\documentclass[12pt]{article}
\pdfoutput=1

\usepackage{putex}
\usepackage{graphicx}
\usepackage{caption}
\usepackage{amsmath, amssymb, amsbsy, mathtools}
\usepackage{array}
\usepackage{subcaption}
\usepackage{cite}
\usepackage[utf8]{inputenc}
\usepackage[colorlinks=true, linkcolor=blue, urlcolor=blue, citecolor=red]{hyperref}
\IfFileExists{dsfont.sty}{\usepackage{dsfont}}{\let\mathds\mathbf}
\usepackage{bm}
\usepackage{booktabs}
\usepackage{placeins}
\usepackage{xcolor}
\usepackage{tikz}
\usetikzlibrary{decorations.pathreplacing}

\newcommand{\pendingfig}[2]{\IfFileExists{#2}{\includegraphics[width=#1]{#2}}{\fbox{\parbox[c][0.3\textwidth][c]{#1}{\centering\color{red}[figure pending: \texttt{\detokenize{#2}}]}}}}

\newcommand{\grp}[1]{\mathrm{#1}}
\newcommand{\grU}{\grp{U}}

\newcommand{\Z}{\mathbb{Z}}

\newcommand{\cO}{{\mathcal O}}
\newcommand{\cW}{{\mathcal W}}
\newcommand{\cP}{{\mathcal P}}
\newcommand{\cM}{{\mathcal M}}
\newcommand{\ket}[1]{\lvert #1\rangle}

\newcommand{\braket}[1]{\langle #1\rangle}
\newcommand{\one}{\mathds{1}}
\DeclareMathOperator{\Ai}{Ai}

\DeclareMathOperator{\Ima}{Im}
\DeclareMathOperator{\Rea}{Re}

\numberwithin{equation}{section}

\begin{document}

\preprint{MIT-CTP/6165}
\institution{MIT}{Center for Theoretical Physics, Massachusetts Institute of Technology, Cambridge, MA 02139, USA}

\title{Hadron structure in $1+1$ dimensions from the infinite Hamiltonian lattice}
\authors{Ross Dempsey\worksat{\MIT}}

\abstract{The internal structure of a hadron is encoded in matrix elements of operators between hadron states of definite momentum. In this paper, we develop a framework for computing such matrix elements in $(1+1)$-dimensional gauge theories by working on an infinite spatial lattice.  We apply the method to the massive Schwinger model. We benchmark our method against discretized lightcone quantization (DLCQ) at $\theta = 0$, and then use the lattice to study hadrons at $\theta\neq 0$, where DLCQ does not apply. We compute parton distribution functions and gravitational form factors, and both show signatures of hadrons becoming large, loosely-bound states when we set $m$ greater than the critical mass $m_c$ and approach $\theta = \pi$ from below. Moreover, we see the gravitational form factor change from that of a free boson to that of a free Majorana fermion when we set $\theta = \pi$ and approach the 2D Ising critical point at $m=m_c$ from below.}
\date{November 2026}

\maketitle
\tableofcontents

\section{Introduction}\label{sec:intro}
 
In a confining gauge theory, the fields that appear in the Lagrangian never appear as asymptotic states; instead, they are bound into hadrons with intricate internal structure. Much of what we know about this structure is encoded in matrix elements
\begin{equation}\label{eq:me_generic}
    \braket{h(p')|\cO|h(p)}
\end{equation}
of local or non-local operators $\cO$ between single-hadron states of definite momentum. Two representative examples are the parton distribution functions (PDFs) \cite{Bjorken:1968dy,Feynman:1969ej,Bjorken:1969ja,Collins:1981uw,Collins:1989gx}, which are forward (i.e., $p=p'$) matrix elements of a bilocal operator containing a lightlike Wilson line, and the gravitational form factors (GFFs) \cite{Kobzarev:1962wt,Pagels:1966zza,Polyakov:2002yz}, which are off-forward (i.e., $p\neq p'$) matrix elements of the stress tensor. PDFs are an essential input for predictions at hadron colliders, and they are determined in QCD by global fits to scattering data \cite{Hou:2019efy,NNPDF:2021njg,PDF4LHCWorkingGroup:2022cjn}. GFFs, which encode the distributions of energy, momentum, and internal forces inside a hadron, have only recently become accessible to experiment \cite{Burkert:2018bqq,Duran:2022xag} and to lattice QCD \cite{Hagler:2003jd,LHPC:2007blg,Alexandrou:2019ali,Shanahan:2018pib,Shanahan:2018nnv,Pefkou:2021fni,Hackett:2023nkr,Hackett:2023rif,Abbott:2025irb}; for reviews, see \cite{Hagler:2009ni,Polyakov:2018zvc,Lorce:2018egm,Burkert:2023wzr}.
 
The massive Schwinger model \cite{Schwinger:1962tp,Lowenstein:1971fc,Coleman:1975pw,Coleman:1976uz} is a natural place to study hadron structure from first principles. Its action is that of quantum electrodynamics in $1+1$ dimensions,
\begin{equation}\label{eq:action}
    S = \int d^2x\left(-\frac{1}{4g^2}F_{\mu\nu}F^{\mu\nu} + \frac{\theta}{4\pi}\epsilon^{\mu\nu}F_{\mu\nu} + \bar\Psi\left(i\gamma^\mu(\partial_\mu - iA_\mu) - m\right)\Psi\right).
\end{equation}
Here $\Psi$ is a Dirac fermion of mass $m$, describing a positron and electron coupled to the $\grU(1)$ gauge field $A_\mu$ with gauge coupling $g$. The dimensionless parameters are $m/g$ and the $\theta$-angle, the latter of which corresponds to a background electric field of $\tfrac{\theta}{2\pi}$ (in units of the fermion charge) \cite{Coleman:1976uz}. We restrict our attention to $m\ge 0$ and $0\leq \theta\leq \pi$; other parameters can be related to this region using discrete symmetries and axial transformations. The theory is super-renormalizable, and so its hadron structure functions are scale-independent and can be compared between methods without any perturbative matching. This is a significant simplification relative to four-dimensional gauge theories.

At $\theta = 0$, the positron and electron are confined into mesons, and the structure of these mesons is simple to calculate. At $m = 0$ (for which the theory is independent of $\theta$ \cite{Coleman:1975pw}), the theory is dual to a free boson \cite{Coleman:1974bu,Mandelstam:1975hb,Coleman:1975pw}, and so there is a unique stable meson state corresponding to that boson, whose structure functions are exactly calculable. At $m>0$, it is straightforward to calculate structure functions of mesons to high accuracy using lightcone quantization \cite{tHooft:1974pnl,Bergknoff:1976xr,Mo:1992sv} and its discretized version (DLCQ) \cite{Eller:1986nt,Pauli:1985ps,Pauli:1985pv}. Furthermore, at $m\gg g$, the heavy fermions bind into nonrelativistic mesons whose structure can be estimated by solving a one-dimensional Schr\"odinger equation.
 
The physics of the Schwinger model changes qualitatively as $\theta$ approaches $\pi$ \cite{Coleman:1976uz}. The background electric field lowers the string tension on one side of each fermion and raises it on the other. Precisely at $\theta = \pi$, in a semiclassical picture the force between an electron and a positron vanishes in one direction, so that they can become deconfined half-asymptotic particles \cite{Coleman:1976uz}. This semiclassical picture is valid when $m\gg g$, and half-asymptotic particles are present at $\theta=\pi$ for any $m>m_c$, with $m_c\approx 0.3335g$ \cite{Byrnes:2002nv,ArguelloCruz:2024xzi,Fujii:2024reh} (see also \cite{Okuda:2026hbg}). For $m>m_c$, we have a first-order phase transition as $\theta$ crosses $\pi$, and this line of first-order phase transitions ends in a second-order critical point at $(m,\theta) = (m_c,\pi)$. 

The approach to $\theta = \pi$ at $m>m_c$ thus offers a controlled setting in which to watch a hadron fall apart. In addition, the approach to $m=m_c$ from below at $\theta=\pi$ interpolates between the massless solvable point and the 2D Ising critical point. However, these regimes are out of reach of DLCQ, which drops the zero modes of the gauge field in which $\theta$ resides \cite{Eller:1986nt}, and of the Euclidean lattice, where a $\theta$ term makes the action complex \cite{Banuls:2019rao}. The Hamiltonian lattice has neither limitation, and in this work we use it to study the structure of Schwinger-model mesons at $\theta\neq 0$, with particular interest in the behavior near and at $\theta = \pi$.

Aside from DLCQ, first-principles calculations of hadron structure functions have mostly been carried out using Euclidean lattice techniques. This presents an immediate difficulty for the PDF even for actions without a sign problem, because the PDF involves a Wilson line at lightlike separation. There are several well-studied schemes for circumventing this problem by reconstructing the PDF from equal-time correlators. One route is large-momentum effective theory \cite{Ji:2013dva,Ji:2014gla,Ji:2020ect,Zhang:2018nsy,Gao:2020ito,Gao:2022iex}, which proceeds by calculating equal-time correlators in a hadron of large momentum, called quasi-PDFs, and then extrapolating to infinite momentum. A related approach calculates the same correlators at fixed Ioffe time (the product of momentum and separation), giving so-called pseudo-PDFs, and then extrapolating to vanishing separation \cite{Ioffe:1969kf,Braun:1994jq,Radyushkin:2017cyf,Orginos:2017kos,Joo:2019jct,Joo:2019bzr,Karpie:2021pap,Joo:2020spy,Radyushkin:2019mye}. Other methods have also been developed \cite{Liu:1993cv,Liu:1999ak,Detmold:2003tm,Detmold:2005gg,Ma:2014jla,Sufian:2019bol,Detmold:2021uru,Alexandrou:2021mmi} (for reviews, see \cite{Lin:2017snn,Cichy:2018mum,Constantinou:2020pek,Constantinou:2020hdm}), and related techniques give access to transverse-momentum-dependent quantities such as the Collins--Soper kernel \cite{Shanahan:2020zxr,Avkhadiev:2024mgd}. 

By contrast, the Hamiltonian lattice uses Lorentzian time, and so the lightlike correlators of the PDF can be calculated directly. This has motivated many proposals for computing structure functions on quantum computers \cite{Lamm:2019uyc,Mueller:2019qqj,Kreshchuk:2020dla,Echevarria:2020wct,Li:2021kcs,Chen:2025zeh,Chen:2026sgc,Zou:2026cfk}. Additionally, in $1+1$ dimensions, tensor network methods allow for very precise calculations in Hamiltonian lattice gauge theory \cite{Banuls:2013jaa,Buyens:2013yza,Banuls:2019rao,Dempsey:2025wia}. These ideas were combined in \cite{Banuls:2025wiq,Schneider:2024yub} to compute the PDF of the lightest meson in the Schwinger model at $\theta = 0$ and $m/g\gtrsim 1.4$ by using real-time simulations of a finite open chain. Other studies have computed quasi-PDFs, quasi-fragmentation functions, and quasi-generalized parton distribution functions \cite{Grieninger:2024axp,Grieninger:2024cdl,Grieninger:2025mbm} and the hadronic tensor \cite{Ikeda:2025bjb}, all using finite chains. The GFF, by contrast, has only been lightly studied in two-dimensional gauge theories~\cite{Burkardt:2000uu,Ji:2020bby,Jia:2024atq}, and to our knowledge it has not been studied at all in the massive Schwinger model.

In this work, we will take advantage of the Hamiltonian lattice and tensor network methods in a different way. We do not perform any real-time evolution. Instead, we employ uniform matrix product states \cite{White:1992zz,Schollwock:2005zz,McCulloch:2008aun,Orus:2013kga,Cirac:2020obd,Zauner-Stauber:2017eqw,Vanderstraeten:2019voi} and the quasiparticle ansatz \cite{Haegeman:2011lcd,Haegeman:2013xcv}, which allow us to represent hadron states directly on an infinite spatial lattice. This approach has been used in past work on the spectrum and phase diagram of the Schwinger model \cite{Buyens:2013yza,Fujii:2024reh}, and recently it was shown how to extend it to an arbitrary $(1+1)$-dimensional gauge theory with abelian or non-abelian gauge group \cite{Dempsey:2025wia}. The novelty of this work is to calculate structure functions in this setting by reducing matrix elements \eqref{eq:me_generic} to efficiently computable contractions of infinite tensor networks.

For an equal-time matrix element, such as the GFF, we strictly benefit by working on an infinite lattice: we do not have to contend with finite-volume effects, and we can work with exact momentum eigenstates for the lightest hadron or excited hadrons. As a consequence, we are able to calculate GFFs to high precision with relatively little computational effort. We compute the GFFs of the lightest three mesons at $\theta = 0$ for many values of $m/g$, and demonstrate excellent agreement with DLCQ. We then map out the gravitational radius of the lightest meson, seeing that it becomes large as we approach $\theta = \pi$. We also check that at $\theta = \pi$, as we vary $m$ from 0 to $m_c$, the GFF interpolates between that of a free boson and a free Majorana fermion.

For a lightlike matrix element, such as the PDF, our approach involves a trade-off with that of \cite{Banuls:2025wiq,Schneider:2024yub}. While that work uses real-time simulations to calculate lightlike correlators directly, we always study hadron states at a fixed time; we thus borrow ideas from the Euclidean lattice in order to reconstruct the PDF from pseudo-PDFs. However, on the infinite lattice, we have the advantage of working with exact momentum eigenstates that we can compute for any hadron; by contrast, on a finite chain, it is more difficult to isolate excited hadrons, which mix with boosted states of lighter ones. Moreover, we do not have to contend with finite-size effects or with entanglement growth in real-time evolution. To demonstrate our approach, we calculate Ioffe-time distributions (ITDs) of the lightest two Schwinger-model mesons, from which we can reconstruct their PDFs. At $\theta = 0$, we again see good agreement with DLCQ, and as we increase $\theta$ at fixed mass we see the PDFs becoming more narrow, consistent with a broader spatial wavefunction. We also map out the third moment of the PDF, which grows as we approach the 2D Ising critical point.
 
The rest of this paper is organized as follows. In Section~\ref{sec:structure}, we define the PDF, the Ioffe-time distribution, and the GFF in a general two-dimensional gauge theory. In Section~\ref{sec:benchmarks}, we review the Schwinger model and collect the benchmarks against which we test our numerics: the exact massless solution, the nonrelativistic limit, lightcone quantization and DLCQ, and the $\theta = \pi$ critical point. In Section~\ref{sec:lattice}, we describe the infinite-lattice construction of hadron states and their matrix elements. We present our results for the GFF in Section~\ref{sec:gff} and for the PDF in Section~\ref{sec:pdf}, and we conclude in Section~\ref{sec:discussion}. Technical details are relegated to the appendices.
\section{Hadron structure in two dimensions}\label{sec:structure}

In this section, we define the two structure functions that we compute in this paper, and explain which of their features in $3+1$ dimensions survive in $1+1$ dimensions. We work in a general $(1+1)$-dimensional gauge theory with Dirac fermions $\Psi$, and specialize to the Schwinger model only in Section~\ref{sec:benchmarks}. 

Our conventions are $\eta_{\mu\nu} = \text{diag}(1,-1)$, $\epsilon^{01} = +1$, and $\gamma^0 = \sigma^z$, $\gamma^1 = i\sigma^y$, so that $\gamma^5 \equiv \gamma^0\gamma^1 = \sigma^x$; we write $\Psi = (\psi_u, \psi_d)^T$ in this basis. Lightcone components of any vector, including $x^\mu$, $A_\mu$, and $\gamma^\mu$, are $v^\pm \equiv (v^0 \pm v^1)/\sqrt{2}$. A hadron $h$ of mass $M$ and spatial momentum $p$ has two-momentum $p^\mu = (E(p), p) = M(\cosh\vartheta, \sinh\vartheta)$, where $E(p) = \sqrt{M^2+p^2}$ and $\vartheta$ is its rapidity. We use the relativistic normalization
\begin{equation}\label{eq:norm}
    \braket{h(p')|h(p)} = 2E(p)\,(2\pi)\,\delta(p-p')
\end{equation}
throughout this section.\footnote{On the lattice, we use a different normalization, which differs from \eqref{eq:norm} by a factor of $2E(p)$ times the cell length; we make this explicit when we write lattice expressions in Sections~\ref{sec:lattice}--\ref{sec:pdf}.}

One further convention is crucial for the phases of matrix elements \eqref{eq:me_generic} to be well-defined: we fix the phases of the single-hadron states by boosts, \begin{equation}\label{eq:h_boost}
    \ket{h(p)} \equiv U(\Lambda_\vartheta)\ket{h(0)},
\end{equation}
where $U(\Lambda_\vartheta)$ is the unitary operator implementing a boost by rapidity $\vartheta$, and its phase is fixed by the condition $U(\Lambda_\vartheta)\ket{0} = \ket{0}$. Equivalently, we require that the boost commutes with an anti-unitary $CPT$ transformation $\Theta$; i.e., $\Theta U(\Lambda_\vartheta) \Theta^{-1} = U(\Lambda_\vartheta)$.\footnote{To see the equivalence, write $U(\Lambda_\vartheta) = e^{-i\vartheta(K+c)}$, where $K = \int dx\, x\, T^{00}(0,x)$ is the boost generator with the vacuum energy subtracted (so that $K\ket{0} = 0$) and $c$ is an undetermined constant. The condition $U(\Lambda_\vartheta)\ket{0} = \ket{0}$ sets $c = 0$. Alternatively, since $\Theta$ reverses $x$ and preserves $T^{00}$, we have $\Theta K\Theta^{-1} = -K$, and since $\Theta$ is antiunitary, $\Theta U(\Lambda_\vartheta)\Theta^{-1} = e^{-i\vartheta(K-c)}$. Thus, requiring the boost to commute with $CPT$ again fixes $c = 0$.}

In Section~\ref{subsec:pdf_def}, we define the parton distribution function and the closely related Ioffe-time distribution. In Section~\ref{subsec:gff_def}, we define the gravitational form factor.

\subsection{Parton distribution functions and Ioffe time}\label{subsec:pdf_def}

The parton distribution function (PDF) of a hadron is defined by a matrix element of a bilocal operator, in which a fermion and an antifermion are separated along a lightlike direction and connected by a Wilson line \cite{Collins:1981uw,Collins:2011zzd}:
\begin{equation}\label{eq:pdf_def}
    f(x) = \int \frac{dz^-}{4\pi}\, e^{-i x P^+ z^-} \braket{h(p)|\bar\Psi(z^-)\,\gamma^+\, \cW(z^-,0)\,\Psi(0)|h(p)}_c\,.
\end{equation}
Here $P^+ = (E(p)+p)/\sqrt 2$, all operators are inserted at $x^+ = 0$, the subscript $c$ denotes subtraction of the vacuum expectation value, and
\begin{equation}\label{eq:wilson_line}
    \cW(z^-,0) = \cP\exp\left(i\int_0^{z^-} dy^-\, A_-(y^-)\right)
\end{equation}
is a lightlike Wilson line. For $0 < x < 1$, $f(x)$ is the density of fermions carrying a fraction $x$ of the hadron's lightcone momentum $P^+$, while for $-1 < x < 0$, $-f(x)$ is the density of antifermions carrying the fraction $-x$. The support is restricted to $|x| < 1$ because every parton carries positive $P^+$. We write $f_q(x) \equiv f(x)$ and $f_{\bar q}(x) \equiv -f(-x)$ for $0<x<1$.

The PDF of a $(3+1)$-dimensional hadron depends on a renormalization scale, because the Wilson line \eqref{eq:wilson_line} is UV divergent in four dimensions and the bilocal operator mixes under renormalization \cite{Gribov:1972ri,Altarelli:1977zs,Dokshitzer:1977sg}. In $1+1$ dimensions, by contrast, the gauge coupling has dimensions of mass, the theory is super-renormalizable, and \eqref{eq:pdf_def} is finite as it stands. Thus, the PDF of a two-dimensional hadron is a single, scale-independent function, and different methods of computing it may be compared directly, without any perturbative matching.

The PDF obeys two exact constraints. First, because the gauge field is non-dynamical, the lightcone momentum of the hadron is carried entirely by the fermions. This implies the momentum sum rule
\begin{equation}\label{eq:momentum_sum}
    F_1 = 1\,, \qquad \text{where}\qquad F_j \equiv \int_{-1}^1 dx\, x^j f(x) = \int_0^1 dx\, x^j\left(f_q(x) + (-1)^{j+1} f_{\bar q}(x)\right)
\end{equation}
are the moments of the PDF; we will use this sum rule as a check of our numerics. Second, for a hadron of zero fermion number, $\int_0^1 dx\,(f_q - f_{\bar q}) = 0$. Better yet, $CPT$ exchanges fermions and antifermions while preserving their lightcone momentum fractions, and so $f_{q/h}(x) = f_{\bar q/\bar h}(x)$, where $\bar h$ is the $CPT$ conjugate of $h$. For a hadron that is its own $CPT$ conjugate, it follows that $f_q = f_{\bar q}$ identically. This is the case for every meson\footnote{It is not true for the solitons between degenerate vacua at $\theta = \pi$. In this paper we will study only the meson states.} of the one-flavor Schwinger model, at any $\theta$, and so in this paper we simply refer to $f \equiv f_q = f_{\bar q}$ on $[0,1]$ as the PDF of the hadron.

Rather than working with $f(x)$ directly, it is convenient to invert the Fourier transform in \eqref{eq:pdf_def} and work with the matrix element directly. Following \cite{Ioffe:1969kf,Braun:1994jq,Radyushkin:2017cyf}, we define the Ioffe-time distribution (ITD) as the forward matrix element of the bilocal operator at a general (i.e., not necessarily null) separation $z^\mu$:
\begin{equation}\label{eq:itd_def}
    \braket{h(p)|\bar\Psi(z)\,\gamma^\mu\,\cW(z,0)\,\Psi(0)|h(p)}_c = 2p^\mu\, \cM(\nu, z_\rho z^\rho) + z^\mu\, \mathcal N(\nu, z_\rho z^\rho)\,.
\end{equation}
Here $\nu \equiv p_\mu z^\mu$ is called the Ioffe time \cite{Ioffe:1969kf}, $\cW(z,0)$ is the straight Wilson line from $0$ to $z$, and $\cM$ and $\mathcal N$ are Lorentz-invariant functions of the two invariants. With $z^\mu$ aligned with the $x^-$ direction (so that $z_\rho z^\rho = 0$), the second term in \eqref{eq:itd_def} does not contribute to the $\mu = +$ component, and so comparing with \eqref{eq:pdf_def} gives
\begin{equation}\label{eq:itd_pdf}
    \cM(\nu, 0) = \int_{-1}^{1} dx\, e^{ix\nu} f(x)\,.
\end{equation}
Since $f_q = f_{\bar q}$, the real part of $\cM(\nu,0)$ vanishes, and the imaginary part defines the ITD $h(\nu)$:
\begin{equation}\label{eq:itd_sine}
    h(\nu) \equiv \Ima \cM(\nu,0) = 2\int_0^1 dx\, f(x)\sin(\nu x)\,, \qquad f(x) = \frac{1}{\pi}\int_0^\infty d\nu\, h(\nu)\sin(\nu x)\,.
\end{equation}
Expanding \eqref{eq:itd_sine} at small $\nu$, we find
\begin{equation}\label{eq:itd_moments}
    h(\nu) = F_1\,\nu - \frac{F_3}{6}\,\nu^3 + \frac{F_5}{120}\,\nu^5 - \cdots\,.
\end{equation}
Thus, the momentum sum rule \eqref{eq:momentum_sum} fixes $h'(0) = 1$ for every hadron. The first nonzero moment, $F_3$, can be read off from the coefficient of $\nu^3$.

The lightcone ITD $\cM(\nu,0)$ can be estimated without ever separating the fermions along a lightlike direction \cite{Radyushkin:2017cyf}. Indeed, we may choose the separation to be purely spatial, $z^\mu = (0, -z)$, so that $\nu = pz$ is controlled by the hadron momentum, and then set $\mu = 0$, removing the second term of \eqref{eq:itd_def}.\footnote{This is the choice usually made in the pseudo-PDF approach \cite{Radyushkin:2017cyf,Orginos:2017kos}; the alternative choice $\mu = 1$, which is natural in the quasi-PDF approach \cite{Ji:2013dva}, receives a contribution from $\mathcal N$.} The price we pay is that a spatial separation has $z_\rho z^\rho = -z^2 \neq 0$, so $\cM(\nu, -z^2)$ differs from its lightcone value $\cM(\nu,0)$. The $z$-dependent corrections, which vanish as $z \to 0$ at fixed $\nu$, are called higher-twist corrections.\footnote{The name is borrowed from four dimensions, where the higher-twist corrections must be treated by perturbative matching \cite{Radyushkin:2017cyf,Ji:2020ect}.} In $1+1$ dimensions we may simply extrapolate them away together with the lattice artifacts.

Thus, our strategy is to compute $\cM(\nu,-z^2)$ at many values of $(p, z)$ that share the same $\nu$, and then extrapolate to $z \to 0$ at fixed $\nu$ to compute the ITD. We can then invert \eqref{eq:itd_sine} to compute the PDF. The central difficulty is that reaching large $\nu$ with small $z$ requires large $p$, and representing a large-momentum state faithfully on the lattice requires very small spacing. We will address this in Section~\ref{subsec:continuum}.

\subsection{Gravitational form factors}\label{subsec:gff_def}

The gravitational form factors (GFFs) of a hadron are defined by the off-forward matrix element of the stress tensor $T^{\mu\nu}$ \cite{Kobzarev:1962wt,Pagels:1966zza}. In four dimensions, a spin-$\tfrac12$ hadron has three such form factors, $A(t)$, $J(t)$, and $D(t)$, which encode the distributions of energy, angular momentum, and pressure or shear forces inside the hadron \cite{Polyakov:2002yz,Shanahan:2018nnv,Polyakov:2018zvc,Burkert:2023wzr}. In two dimensions, there is no spin. Let $p^\mu$ and $p'^\mu$ be the on-shell two-momenta of the ket and the bra, and define
\begin{equation}
    P^\mu \equiv \tfrac12\left(p^\mu+p'^\mu\right), \qquad \Delta^\mu \equiv p'^\mu - p^\mu\,, \qquad t \equiv \Delta_\mu\Delta^\mu\,.
\end{equation}
The most general symmetric tensor built from $P^\mu$, $\Delta^\mu$, and $\eta^{\mu\nu}$ that is conserved, $\Delta_\mu \braket{h(p')|T^{\mu\nu}|h(p)} = 0$, is then
\begin{equation}\label{eq:gff_AD}
    \braket{h(p')|T^{\mu\nu}(0)|h(p)} = 2P^\mu P^\nu A(t) + \frac12\left(\Delta^\mu\Delta^\nu - \eta^{\mu\nu}\Delta^2\right) D(t)\,.
\end{equation}
However, the two tensor structures in \eqref{eq:gff_AD} are not independent. On shell, $P\cdot\Delta = 0$, so in two dimensions $P^\mu$ must be proportional to $\epsilon^{\mu\nu}\Delta_\nu$, and it is straightforward to check that
\begin{equation}
    \Delta^\mu\Delta^\nu - \eta^{\mu\nu}\Delta^2 = \frac{-t}{M^2 - t/4}\,P^\mu P^\nu\,.
\end{equation}
Thus, a two-dimensional hadron has a single gravitational form factor,
\begin{equation}\label{eq:gff_def}
    \braket{h(p')|T^{\mu\nu}(0)|h(p)} = 2P^\mu P^\nu\, G(t)\,, \qquad G(t) = A(t) - \frac{t/4}{M^2 - t/4}\,D(t)\,.
\end{equation}

In the continuum, $G(t)$ is real for any hadron that is its own $CPT$ conjugate. Indeed, we may choose the phase of $\ket{h(0)}$ so that $\Theta\ket{h(0)} = \ket{h(0)}$, and then since we have chosen the boosts in \eqref{eq:h_boost} to commute with the $CPT$ operator $\Theta$, it follows that $\Theta\ket{h(p)} = \ket{h(p)}$ for every $p$. The stress tensor $T^{\mu\nu}(0)$ commutes with $\Theta$, and so, since $\Theta$ is antiunitary,
\begin{equation}
    \braket{h(p')|T^{\mu\nu}(0)|h(p)}^* = \braket{\Theta h(p')|\,\Theta\, T^{\mu\nu}(0)\,\Theta^{-1}|\Theta h(p)} = \braket{h(p')|T^{\mu\nu}(0)|h(p)}\,.
\end{equation}
As we will see in Section~\ref{subsec:gff_lattice}, the lattice counterpart of the reality of $G(t)$ requires some care.

Since $\int dx\, T^{00}$ is the Hamiltonian, translation invariance and the normalization \eqref{eq:norm} give $\braket{h(p)|T^{00}(0)|h(p)} = 2E(p)^2$, and so the forward limit of \eqref{eq:gff_def} gives $G(0) = 1$ for every hadron. We will mostly work in the Breit frame, $p = -p' = -\tfrac12\sqrt{-t}$. In this frame, the two hadrons move back to back and $P^\mu = (E, 0)$, so that
\begin{equation}\label{eq:gff_breit}
    G(t) = \frac{\braket{h(\tfrac12\sqrt{-t})|T^{00}(0)|h(-\tfrac12\sqrt{-t})}}{2E^2}\,, \qquad E \equiv E\big(\tfrac12\sqrt{-t}\big) = \sqrt{M^2 - t/4}\,.
\end{equation}
Note that $t \leq 0$ for the elastic matrix elements we compute, and that $G(t)$ is naturally a function of $-t/M^2$; we will plot it this way throughout. At small $-t$, the slope $G'(0) \equiv dG/dt|_{t=0} > 0$ measures the spatial extent of the energy distribution of the hadron \cite{Polyakov:2018zvc,Burkardt:2000za,Freese:2021mzg}, and we refer to it as the gravitational radius. (More precisely, in the non-relativistic limit we will find that $8G'(0)$ is a mean-square size; see Section~\ref{subsec:weak}.)

There is one subtlety in \eqref{eq:gff_def}: the stress tensor is defined only up to improvement terms $\Delta T^{\mu\nu} = (\partial^\mu\partial^\nu - \eta^{\mu\nu}\partial^2)\,\mathcal U$, where $\mathcal{U}$ is a dimensionless scalar. Such an improvement term leaves the Hamiltonian invariant, but changes $G(t)$ at $t \neq 0$ \cite{Callan:1970ze}. In the Schwinger model, we could take $\mathcal{U}$ to be any function of $F_{01}/g^2$. However, since $\Delta T^{00} = \partial_1^2\,\mathcal U$, an improvement term would appear on the lattice as an explicit second difference added to the energy of a cell. Our lattice energy density operator (see Section~\ref{subsec:gff_lattice}) is the sum of the local terms of the Hamiltonian, with no such addition, and so it corresponds to the canonical stress tensor. And indeed, we check below that our lattice results match exact results calculated using the canonical stress tensor.

\section{The Schwinger model and its benchmarks}\label{sec:benchmarks}

In this section, we review the massive Schwinger model and collect the analytic and semi-analytic results for its PDFs and GFFs against which we will test our lattice method. 

In Section~\ref{subsec:massless} we consider the massless theory, which is solvable, and write the exact structure functions for its single stable meson. In Section~\ref{subsec:weak}, we treat the $m\gg g$ limit using nonrelativistic quantum mechanics. In Section~\ref{subsec:lightcone} we describe lightcone quantization, which, in its discretized form, provides an independent numerical benchmark for both the PDF and the GFF at $\theta = 0$. Finally, in Section~\ref{subsec:ising}, we derive the form factor of the lightest meson at the $\theta = \pi$ critical point.

\subsection{Massless theory}\label{subsec:massless}

Under bosonization \cite{Coleman:1974bu,Mandelstam:1975hb}, the theory \eqref{eq:action} is equivalent to a scalar $\phi$ with Lagrangian
\begin{equation}\label{eq:bosonized}
    \mathcal L = \frac12(\partial_\mu\phi)^2 - \frac{g^2}{2\pi}\phi^2 + c_0\, m g\cos\left(2\sqrt\pi\,\phi - \theta\right),
\end{equation}
where $c_0 = e^{\gamma_E}/(2\pi^{3/2})$ in a suitable normal-ordering scheme, with $\gamma_E$ the Euler--Mascheroni constant. At $m = 0$, \eqref{eq:bosonized} describes a free boson of mass $g/\sqrt\pi$, and $\theta$ can be rotated away; that free boson is dual to the sole stable meson of the massless Schwinger model.

Both the PDF and the GFF of this meson can be computed exactly using \eqref{eq:bosonized}. The stress tensor is that of a free massive scalar, and evaluating \eqref{eq:gff_AD} by Wick contraction gives $A(t) = 1$ and $D(t) = -1$, so that
\begin{equation}\label{eq:gff_massless}
    G(t) = \frac{1}{1 - t/4M^2}\,.
\end{equation}
The PDF is most easily derived in lightcone quantization, as we review in Section~\ref{subsec:lightcone}: the meson is a fermion-antifermion pair whose lightcone wavefunction is constant in the momentum fraction $x$ \cite{Bergknoff:1976xr}, so that
\begin{equation}\label{eq:pdf_massless}
    f(x) = 1 \quad (0 < x < 1)\qquad \text{and}\qquad h(\nu) = \frac{2(1-\cos\nu)}{\nu}\,.
\end{equation}
The moments of this PDF are $F_n = \frac{1-(-1)^n}{2} \frac{2}{n+1}$; in particular, $F_3 = \tfrac12$.

\subsection{Weak coupling}\label{subsec:weak}

When $m \gg g$, the low-lying mesons are non-relativistic bound states of a fermion at position $r_+$ and an anti-fermion at $r_-$, and their structure is governed by the Schr\"odinger equation in the relative coordinate $r \equiv r_+ - r_-$. The two particles are bound by the energy of the electric flux between them \cite{Coleman:1976uz}, which gives a non-relativistic potential of
\begin{equation}\label{eq:VNR}
    V_\text{NR}(r) = \frac{g^2}{2}\left(|r| - \frac{\theta}{\pi}\,r\right),
\end{equation}
so that the string tension is $\tfrac{g^2}{2}(1 - \theta/\pi)$ for $r > 0$ and $\tfrac{g^2}{2}(1 + \theta/\pi)$ for $r < 0$. At $\theta = 0$, with reduced mass $m/2$, the relative wavefunction obeys
\begin{equation}\label{eq:schrodinger}
    -\frac{1}{m}\psi_n''(r) + \frac{g^2}{2}|r|\,\psi_n(r) = \epsilon_n\psi_n(r)\,, \qquad M_n \approx 2m + \epsilon_n\,.
\end{equation}
In terms of $\alpha \equiv (mg^2/2)^{1/3}$ and $u = \alpha r$, the solutions are $\psi_n(r) = \sqrt\alpha\,\hat\psi_n(\alpha r)$ with wavefunctions
\begin{equation}\label{eq:airy}
    \hat\psi_{2k-1}(u) \propto \Ai\left(|u| + \zeta'_k\right)\,, \qquad \hat\psi_{2k}(u) \propto \sgn(u)\Ai\left(|u| + \zeta_k\right)
\end{equation}
and energies
\begin{equation}
    \epsilon_n = -\frac{\alpha^2}{m}\times\begin{cases}\zeta'_k & n = 2k-1\\ \zeta_k & n = 2k\end{cases}\,.
\end{equation}
Here $\zeta_k < 0$ and $\zeta'_k < 0$ are the $k$th zeros of $\Ai$ and $\Ai'$, and $\int du\,|\hat\psi_n|^2 = 1$. The even (odd) wavefunctions are the $C$-odd (even) mesons, because the constituents are fermions.

We can then compute both structure functions using $\psi_n$. To compute the GFF, we note that at leading order the energy density is dominated by the rest masses of the two constituents, $T^{00}(y) \approx m\left(\delta(y - r_+) + \delta(y - r_-)\right)$. Taking matrix elements between the two-body states $e^{ip\bar r}\psi_n(r)$, where $\bar r \equiv \tfrac12(r_+ + r_-)$ is the center of mass, and comparing with \eqref{eq:gff_breit}, we find
\begin{equation}\label{eq:gff_weak}
    G_n(t) = \int dr\,|\psi_n(r)|^2\cos\left(\frac{\sqrt{-t}\,r}{2}\right) = 1 + \frac{t}{8}\braket{r^2}_n + \cdots\,, \qquad \braket{r^2}_n \equiv \int dr\, r^2|\psi_n(r)|^2\,.
\end{equation}
Thus, at weak coupling, $G_n$ is the Fourier transform of the distribution of the separation between the constituents, and $8G_n'(0) = \braket{r^2}_n$ is its mean square radius. The same expression applies at $\theta \neq 0$, with $\psi_n$ the eigenfunctions in the asymmetric potential \eqref{eq:VNR}.

Similarly, boosting the bound state to large momentum, the constituent with relative momentum $q$ carries the momentum fraction $x = \tfrac12 + \tfrac{q}{2m}$ at leading order, and so
\begin{equation}\label{eq:pdf_weak}
    f_n(x) = 2m\left|\tilde\psi_n\big(2m(x - \tfrac12)\big)\right|^2\,, \qquad\text{with } \tilde\psi_n(q) = \int\frac{dr}{\sqrt{2\pi}}\,e^{-iqr}\psi_n(r)\,.
\end{equation}
The PDF is therefore concentrated in a window of width of order $\alpha/m \sim (m/g)^{-2/3}$ around $x = \tfrac12$, and in the limit $m/g \to \infty$ it approaches $\delta(x - \tfrac12)$, for which $F_3 = \tfrac14$.

\subsection{Lightcone quantization}\label{subsec:lightcone}

Since the Wilson line in \eqref{eq:pdf_def} is lightlike, the PDF becomes relatively simple in lightcone quantization, where we let $x^+$ play the role of time. Indeed, in lightcone gauge $A_- = 0$ the Wilson line becomes trivial,\footnote{Note that this step discards the holonomy mode of the gauge field, which is responsible for the restriction of lightcone quantization to $\theta = 0$.} and moreover the field $A_+$ and the $\gamma^5 = -1$ component of $\Psi$ are non-dynamical, so we may integrate them out. The remaining dynamical field is the $\gamma^5 = +1$ component, $\psi \equiv 2^{1/4}\tfrac12(1+\gamma^5)\Psi$; the normalization is chosen so that we have the mode expansion
\begin{equation}\label{eq:lc_modes}
    \psi(x^-) = \int_0^\infty\frac{dk}{\sqrt{2\pi}}\left(c(k)e^{-ikx^-} + d^\dagger(k)e^{ikx^-}\right)
\end{equation}
with $\{c(k),c^\dagger(k')\} = \{d(k),d^\dagger(k')\} = \delta(k-k')$.

We can construct a Fock space by acting with these modes on a vacuum $\ket{0}$ annihilated by $c(k)$ and $d(k)$ for all $k>0$; we restrict to gauge-invariant states, which have equal numbers of $c^\dagger$ and $d^\dagger$ operators. On this space, the lightcone momentum $P^+$ acts diagonally as
\begin{equation}
    P^+ = \int_0^\infty dk\,k \left(c^\dagger(k)c(k) + d^\dagger(k)d(k)\right)\,.
\end{equation}
The lightcone Hamiltonian $P^-$, which consists of the mass term $\tfrac{m^2}{2}\int dk\, k^{-1}(c^\dagger(k) c(k) + d^\dagger(k) d(k))$ and a four-fermion Coulomb interaction, is not diagonal: it connects two-body states to four-body states, and so on. Hadron states are simultaneous eigenstates of $P^+$ and $P^-$, with $M^2 = 2P^+ P^-$. 

If we truncate to the two-body sector (the Tamm--Dancoff approximation; for improvements including higher Fock states, see \cite{Mo:1992sv,Harada:1995au}), with $\ket{h} = \int_0^1 dx\,\varphi(x)\, c^\dagger(xP^+)d^\dagger((1-x)P^+)\ket{0}$, the eigenvalue equation $2P^+P^-\ket{h} = M^2\ket{h}$ becomes the 't~Hooft-like equation \cite{tHooft:1974pnl,Bergknoff:1976xr}
\begin{equation}\label{eq:thooft}
    M^2\varphi(x) = \left(m^2 - \frac{g^2}{\pi}\right)\left(\frac1x + \frac{1}{1-x}\right)\varphi(x) + \frac{g^2}{\pi}\int_0^1 dy\,\varphi(y) - \frac{g^2}{\pi}\,\text{P.V.}\!\int_0^1 dy\,\frac{\varphi(y)}{(x-y)^2}\,,
\end{equation}
where P.V.\ denotes the principal value. At $m = 0$, $\varphi(x) = 1$ solves \eqref{eq:thooft} with $M^2 = g^2/\pi$. One can show that the two-body state with $\varphi(x) = 1$ is in fact an exact eigenstate of the full $P^-$ at $m = 0$.

In a hadron state $\ket{h}$, the PDF \eqref{eq:pdf_def} is simply the number density of fermions \cite{Brodsky:1997de}, 
\begin{equation}
    f(x) \propto \braket{h|c^\dagger(xP^+)c(xP^+)|h}\,.
\end{equation}
Applying this to the two-body eigenstate at $m = 0$ yields \eqref{eq:pdf_massless}.

To go beyond the two-body truncation, we use discretized lightcone quantization (DLCQ) \cite{Eller:1986nt,Pauli:1985pv,Pauli:1985ps}. We compactify $x^- \sim x^- + 2\pi R$ with antiperiodic boundary conditions for $\psi$, so that the momenta in \eqref{eq:lc_modes} become $k = r/2R$ with $r$ a positive odd integer; we label the surviving modes of \eqref{eq:lc_modes} by e.g. $C^\dagger_r$. We then diagonalize $P^-$ exactly in the finite-dimensional space of states of total momentum $P^+ = K/2R$, where $K$ is an even integer called the harmonic resolution. The PDF of an eigenstate $\ket{h_{n,K}}$ is then sampled on the grid $x = r/K$,
\begin{equation}
    f\!\left(\frac rK\right) = \frac K2\braket{h_{n,K}|C^\dagger_r C_r|h_{n,K}}\,,
\end{equation}
and the continuum PDF is reached as $K \to \infty$. The explicit form of the discretized $P^-$ is given in Appendix~\ref{app:dlcq}. In this paper we compute at $K$ up to $106$ and then extrapolate $K\to\infty$; the PDFs are converged well enough to serve as a reliable benchmark.\footnote{The first work to diagonalize the full $P^-$ operator in the Schwinger model was \cite{Eller:1986nt}, which reached $K = 32$ (in our normalization of $K$, which differs from theirs) using periodic boundary conditions. There, the gauge-invariant Hilbert space has a basis of 124 states. For our calculation at $K=106$, the gauge-invariant space has dimension $329{,}931$.}

DLCQ also provides a benchmark for the GFF. In lightcone gauge, $T^{++}$ is a pure fermion bilinear, since the gauge-field contribution to the stress tensor has no $++$ component in two dimensions. Explicitly, it is
\begin{equation}\label{eq:Tpp}
    T^{++} = \frac{i}{2}\left(\psi^\dagger\,\partial_-\psi - \partial_-\psi^\dagger\,\psi\right).
\end{equation}
Substituting the mode expansion \eqref{eq:lc_modes} and discretizing, this becomes
\begin{equation}\label{eq:Tpp_dlcq}
    T^{++}(x^-) = \frac{1}{4\pi R^2}\sum_{\Delta K\text{ even}} e^{i(\Delta K) x^-/2R}\,\mathcal T_{\Delta K}\,,
\end{equation}
where, for $\Delta K \geq 0$,
\begin{equation}
    \mathcal{T}_{\Delta K} = \sum_{r> 0} \frac{2r+\Delta K}{2}\left(C^\dagger_{r+\Delta K}C_r + D^\dagger_{r+\Delta K}D_r\right) + \sum_{0<r<\Delta K}\frac{2r - \Delta K}{2}\, C^\dagger_r D^\dagger_{\Delta K - r}\,,
\end{equation}
and $\mathcal T_{-\Delta K} = \mathcal T_{\Delta K}^\dagger$. 

The matrix element of $T^{++}$ between DLCQ eigenstates $\ket{h_{n,K}}$ and $\ket{h_{n,K'}}$, of total momenta $P^+ = K/2R$ and $P'^+ = K'/2R$, gives an approximation to the GFF. Since $p^- = M^2/2p^+$ for both states, $t = 2\Delta^+\Delta^- = -M^2(p'^+-p^+)^2/(p^+p'^+)$, i.e.
\begin{equation}\label{eq:dlcq_t}
    -\frac{t}{M^2} = \frac{(K' - K)^2}{KK'}\,.
\end{equation}
The relativistically normalized states are $\sqrt{2p^+\cdot 2\pi R}\,\ket{h_{n,K}}$, and only the $\mathcal T_{K'-K}$ term of \eqref{eq:Tpp_dlcq} contributes, so comparing with \eqref{eq:gff_def} at $\mu = \nu = +$ gives
\begin{equation}\label{eq:dlcq_gff}
    G(t) = \frac{\sqrt{KK'}}{\big((K+K')/2\big)^2}\,\braket{h_{n,K'}|\mathcal{T}_{K'-K}|h_{n,K}}\,.
\end{equation}

In the massless theory, the meson is the flat two-body state $\ket{h_{1,K}} = (2/K)^{1/2}\sum_{r<K} C^\dagger_r D^\dagger_{K-r}\ket{0}$, which is an exact eigenstate at every $K$ (see Appendix~\ref{app:dlcq}). The pair-creation term of $\mathcal T_{K'-K}$ maps it to four-body states, which are orthogonal to $\ket{h_{1,K'}}$, while 
\begin{equation}
    C^\dagger_{r+\Delta K}C_r C^\dagger_r D^\dagger_{K-r}\ket0 = C^\dagger_{r+\Delta K} D^\dagger_{K'-(r+\Delta K)}\ket0\,,
\end{equation}
with $\Delta K = K'-K$. Thus the $C^\dagger C$ term contributes
\begin{equation}
    \frac{2}{\sqrt{KK'}}\sum_{r<K}\frac{2r+\Delta K}{2} = \frac{2}{\sqrt{KK'}}\left(\frac{K^2}{4} + \frac{K\,\Delta K}{4}\right) = \frac{\sqrt{KK'}}{2}\,,
\end{equation}
and the $D^\dagger D$ term contributes the same, so that $\braket{h_{1,K'}|\mathcal T_{K'-K}|h_{1,K}} = \sqrt{KK'}$. Then \eqref{eq:dlcq_gff} gives
\begin{equation}
    G = \frac{4KK'}{(K+K')^2} = \frac{1}{1 + (K'-K)^2/4KK'} = \frac{1}{1 - t/4M^2}
\end{equation}
at every $(K,K')$, in agreement with \eqref{eq:gff_massless}.

\subsection{The \texorpdfstring{$\theta = \pi$}{theta = pi} critical point}\label{subsec:ising}

Finally, consider $\theta = \pi$ with $m$ slightly below $m_c$. In this regime, the lightest meson is nearly massless, with a mass $M_1 \propto m_c - m$, and the long-distance physics is that of the Ising field theory (i.e., the Ising conformal field theory perturbed by its dimension-1 energy operator $\varepsilon$). Under fermionization, that theory is in turn equivalent to a free Majorana fermion of mass $M_1$ \cite{Schultz:1964fv,Zuber:1976aa}. The GFF of the lightest meson therefore approaches that of a free Majorana fermion.

Form factors of local operators, including the trace of the stress tensor, are known exactly in the Ising field theory and other integrable theories from the form-factor bootstrap \cite{Karowski:1978vz,Smirnov:1992vz,Yurov:1990kv,Delfino:1996nf}. Here we only need the simplest of them; we show in Appendix~\ref{app:majorana} that the GFF of the free Majorana fermion of mass $M$ is
\begin{equation}\label{eq:gff_majorana}
    G(t) = \frac{1}{\sqrt{1 - t/4M^2}}\,.
\end{equation}
Comparing \eqref{eq:gff_majorana} with \eqref{eq:gff_massless}, we see that the lightest meson at $\theta = \pi$ must interpolate between $G(t) = (1-t/4M^2)^{-1}$ at $m = 0$ and $G(t) = (1-t/4M^2)^{-1/2}$ at $m = m_c$. In particular, $4M^2G'(0)$ must change from $1$ to $\tfrac12$. We will test this prediction in Section~\ref{subsec:gff_results}.

\section{Hadrons on the infinite Hamiltonian lattice}\label{sec:lattice}

Both structure functions of Section~\ref{sec:structure} are matrix elements between single-hadron states of definite momentum. In this section, we explain how such states, and the matrix elements between them, can be constructed directly on an infinite spatial lattice. We begin in Section~\ref{subsec:lattice_H} by reviewing the lattice Hamiltonian of the Schwinger model and its representation as a link-enhanced matrix product operator (LEMPO)~\cite{Dempsey:2025wia}. In Section~\ref{subsec:qp}, we review uniform matrix product states (uMPS) and the quasiparticle ansatz, which we use to write our hadron states, and in Section~\ref{subsec:me} we describe the two kinds of matrix elements we need.

\subsection{The lattice Hamiltonian}\label{subsec:lattice_H}

We use the Kogut--Susskind formulation of the Schwinger model \cite{Kogut:1974ag,Banks:1975gq,Susskind:1976jm} on an infinite chain with lattice spacing $a$. We have a single-component (staggered) fermion $c_n$ on each site $n \in \Z$, with $\{c^\dagger_m, c_n\} = \delta_{mn}$, and a $\grU(1)$ link variable $U_n$ with conjugate electric field $L_n$ on each link $(n,n+1)$, satisfying $[L_m, U_n] = \delta_{mn} U_n$. The Hamiltonian is
\begin{equation}\label{eq:H_lattice}
    H = \sum_{n\in\Z}\left[\frac{g^2a}{2}\left(L_n + \frac{\theta}{2\pi}\right)^2 - \frac{i}{2a}\left(c_n^\dagger U_n c_{n+1} - \text{h.c.}\right) + m_\text{lat}(-1)^n c_n^\dagger c_n\right],
\end{equation}
where $m_\text{lat} = m - g^2a/8$ \cite{Dempsey:2022nys}. This Hamiltonian acts on gauge-invariant states, which obey the Gauss law
\begin{equation}\label{eq:gauss}
    L_n - L_{n-1} = c_n^\dagger c_n - \tfrac12\left(1 - (-1)^n\right).
\end{equation}

The discrete symmetries of \eqref{eq:H_lattice} will play a role in Section~\ref{sec:gff}. Charge conjugation $C$, parity $P$, and time reversal $T$ act as
\begin{equation}\label{eq:CPT_lattice}
\begin{alignedat}{5}
    C&:\quad && c_n \to c_{n+1}^\dagger\,, \qquad && U_n \to U_{n+1}^\dagger\,, \qquad && L_n \to -L_{n+1}\,, \qquad && \theta \to -\theta\,,\\
    P&:\quad && c_n \to (-1)^n c_{-n}\,, \qquad && U_n \to U_{-n-1}^\dagger\,, \qquad && L_n \to -L_{-n-1}\,, \qquad && \theta \to -\theta\,,\\
    T&:\quad && c_n \to (-1)^n c_n\,, \qquad && U_n \to U_n\,, \qquad && L_n \to L_n\,, \qquad && \theta \to \theta\,.
\end{alignedat}
\end{equation}
The operators $C$ and $P$ are unitary, whereas $T$ is anti-unitary.

Each of these transformations maps \eqref{eq:H_lattice} to the Hamiltonian with $\theta$ transformed as indicated, and preserves Gauss's law \eqref{eq:gauss}. Thus, $T$ is a symmetry for any $\theta$, while $C$ and $P$ are symmetries only at $\theta = 0$ or $\theta = \pi$. The products $CP$ and $CPT$ are also symmetries at every $\theta$. Because of the one-site translation contained in $C$, they map a unit cell onto the cell shifted by one site.

For our numerical calculations, we will represent states as uniform matrix product states (see Section~\ref{subsec:qp}). However, the Hilbert space is not a tensor product of local Hilbert spaces, because of the constraint \eqref{eq:gauss}, and the electric field energy term acts on links rather than on sites; both of these features make standard uMPS representations inconvenient. Both problems can be solved by giving the virtual spaces a $\grU(1)$ grading and requiring that the uMPS tensors are $\grU(1)$-symmetric \cite{Singh:2009cd}. As explained in \cite{Dempsey:2025wia}, if we assign the physical charge $c_n^\dagger c_n - \tfrac12(1-(-1)^n)$ to site $n$, then the charge on the virtual bond between sites $n$ and $n+1$ can be identified with the electric field $L_n$, and the symmetry condition on the MPS tensors is equivalent to the Gauss law \eqref{eq:gauss}. The electric term in \eqref{eq:H_lattice} is then an operator acting on the virtual bond, and the Hamiltonian becomes a LEMPO: an MPO whose tensors are interleaved with link tensors acting on virtual bonds \cite{Dempsey:2025wia}. 

To write the explicit LEMPO for \eqref{eq:H_lattice}, we first map the fermions to qubits by a Jordan--Wigner transformation, which we give explicitly in \eqref{eq:jw}. For the precise LEMPO tensors we use, see \cite{Dempsey:2025wia}.

\subsection{Uniform matrix product states and the quasiparticle ansatz}\label{subsec:qp}

After the Jordan--Wigner transformation \eqref{eq:jw}, each site $n$ carries a qubit, with basis states $\ket{s_n}$ labeled by the occupation number $s_n = c_n^\dagger c_n \in \{0, 1\}$. Given the electric field far to the left, Gauss's law \eqref{eq:gauss} determines every $L_n$ in terms of the occupations, and so the product states $\ket{\{s\}} \equiv \bigotimes_n \ket{s_n}$ span the gauge-invariant Hilbert space.

We label the unit cells so that cell $j \in \Z$ consists of the sites $\{2j+1, 2j+2\}$, and we approximate the vacuum of \eqref{eq:H_lattice} by a uniform matrix product state with this two-site unit cell,
\begin{equation}\label{eq:umps}
    \ket{\Omega} = \sum_{\{s\}} \cdots A^{s_{2j+1}}_{1} A^{s_{2j+2}}_{2} A^{s_{2j+3}}_{1} A^{s_{2j+4}}_2 \cdots \ket{\cdots s_{2j+1}s_{2j+2}s_{2j+3}s_{2j+4}\cdots}\,,
\end{equation}
where $A^s_{1,2}$ are $D\times D$ matrices for each value $s \in\{0,1\}$ of the qubit. Here $D$ is the bond dimension, and it is adjusted dynamically as we determine the tensors $A^s_{1,2}$ by minimizing the energy density using the variational uniform MPS (VUMPS) algorithm \cite{Zauner-Stauber:2017eqw,Vanderstraeten:2019voi}, adapted to LEMPOs as in \cite{Dempsey:2025wia}. Since \eqref{eq:umps} is defined directly in the thermodynamic limit, there are no finite-size effects; the only approximation making $\ket{\Omega}$ differ from the true ground state of \eqref{eq:H_lattice} is the finite bond dimension.

We describe excitations using the quasiparticle ansatz \cite{Haegeman:2011lcd,Haegeman:2013xcv} (see also \cite{Pufu:2026kqk} for a continuum analogue), in which the vacuum tensor on a single site of \eqref{eq:umps} is replaced by an excitation tensor, and the result is summed over positions with a plane-wave phase:
\begin{equation}\label{eq:qp}
    \ket{\phi_B(k)} = \sum_{n\in\Z} e^{ikn/2} \sum_{\{s\}}\cdots A^{s_{n-1}}_{n-1}\, B^{s_n}_{n}\, A^{s_{n+1}}_{n+1}\cdots\ket{\{s\}}\,.
\end{equation}
Here $A_n$ is the vacuum tensor of site $n$ (i.e., $A_1$ for odd $n$ and $A_2$ for even $n$, as in \eqref{eq:umps}), and the excitation is carried by two tensors $B_1$ and $B_2$, again for odd and even sites respectively. The state \eqref{eq:qp} has definite lattice momentum $k \in [-\pi,\pi)$ (corresponding to physical momentum $p = k/(2a)$), and it is an exact momentum eigenstate for any choice of $B_{1,2}$. The tensors $B_{1,2}$ are found by diagonalizing the Hamiltonian within the linear space of states of the form \eqref{eq:qp} at fixed $k$~\cite{Haegeman:2011lcd,Vanderstraeten:2019voi}. In practice, we use the implementation of this ansatz in \texttt{MPSKit.jl}~\cite{MPSKit}.

The lowest few eigenvalues at fixed $k$ give lattice approximations to the energies $E_n(p)$ of the mesons $n = 1, 2, \ldots$ above the vacuum, and the corresponding eigenvectors define our lattice approximations to the states $\ket{n, p}$.

There are two caveats. First, the quasiparticle ansatz describes an isolated single-particle state only if that state lies below the two-particle continuum at the given momentum; above threshold, the lowest eigenvalues of the effective Hamiltonian need not converge to anything physical. All the mesons we study lie below the two-particle threshold.\footnote{Note that discrete symmetries can forbid the decay of a hadron into two of the lightest mesons, and so physically, there can be stable states above this threshold. In our numerical approach, however, we are not able to reliably extract such states.} Second, the lattice dispersion relation $E_n(p)$ agrees with $\sqrt{M_n^2 + p^2}$ only up to corrections of order $(ap)^2$, so we have to keep $ap$ small (with $ap \leq 0.4$ a reasonable heuristic).

We do not fix the bond dimension by hand: a two-site version of VUMPS \cite{Zauner-Stauber:2017eqw,MPSKit} adapts the virtual spaces, discarding singular values below a fixed relative tolerance, so that the bond dimension grows as needed toward the continuum. We have checked that tightening this tolerance changes our matrix elements by much less than the uncertainties of the continuum extrapolations.

\subsection{Matrix elements between quasiparticles}\label{subsec:me}

The two structure functions we study require two different kinds of matrix elements between the states $\ket{n,p}$. The PDF requires the forward matrix element $\braket{n,p|\cO|n,p}$ where $\cO$ is an extended Wilson line bilinear. The GFF requires the off-forward matrix element $\braket{n,p'|\cO|n,p}$ of a local operator, the energy density, built from the energies of single unit cells (Section~\ref{subsec:gff_lattice}). In both cases we write the operator as a (LE)MPO, and the matrix element becomes a contraction of the infinite network formed by the two quasiparticle states and the operator.

Such contractions are standard for forward matrix elements \cite{Haegeman:2013vud,Vanderstraeten:2013xda,Zauner:2014iea,Vanderstraeten:2019voi,VanDamme:2021xqk,Osborne:2024oqg}. Substituting \eqref{eq:qp} for both the bra and ket gives a double sum over the positions of the two excitation tensors $B$ and $B'$. Some terms are removed by a gauge condition on the excitation tensors (see \cite{Vanderstraeten:2019voi}), and the surviving terms are geometric series in the transfer matrices of the vacuum, which can be resummed in closed form. 

For the off-forward case, the same logic goes through, except that the phases in the geometric series involve the momentum transfer $p'-p$ as well as the individual momenta; see Appendix~\ref{app:qp_me}. The cost depends on the operator through its length and through the bond dimension of its MPO representation; in Section~\ref{sec:pdf}, we will use Wilson lines spanning $\ell$ sites, represented as MPOs of bond dimension $\ell + 2$. In practice, Wilson lines spanning well over 100 sites pose no difficulty.

We normalize the quasiparticle states as $\braket{n,k'|n,k} = 2\pi\delta(k-k')$; note that this differs from \eqref{eq:norm}. With this normalization, the connected\footnote{Note that every matrix element we compute is the connected one. The vacuum expectation value of the operator appears in every term in which neither excitation tensor overlaps with the operator, and we must subtract this contribution before performing the geometric sums in order for them to converge.} forward matrix element $\braket{n,k|\cO_0|n,k}_c$ of an operator $\cO_0$ supported near cell $0$ gives the change in the expectation value of its sum $\sum_j \cO_j$ when the meson is added to the vacuum. 

\section{Gravitational form factors}\label{sec:gff}

In this section, we compute the gravitational form factors of the lightest mesons of the Schwinger model. We treat the GFF before the PDF because it is the simpler of the two: the operator is local, so we can use the method of Appendix~\ref{app:qp_me} to calculate the matrix elements efficiently, and, as we will see, the lattice artifacts are small. 

In Section~\ref{subsec:gff_lattice}, we define the lattice energy density operator, being careful to make it invariant under $CPT$ (up to artifacts that vanish as $a\to 0$) so that we can follow the argument of Section~\ref{subsec:gff_def} to fix the phases of the quasiparticle states and obtain a real GFF. In Section~\ref{subsec:gff_results}, we present our results: the form factors of the three lightest mesons at $\theta = 0$, which we compare with DLCQ, the gravitational radius of the lightest meson as a function of $m/g$ and $\theta$, and the approach to the Ising critical point at $\theta = \pi$.

\subsection{The energy density and the Breit frame}\label{subsec:gff_lattice}

\begin{figure}[t]
    \centering
    \begin{tikzpicture}[x=2.2cm, y=1cm]
        \foreach \i in {1,2,3,4} {
            \fill (\i,0) circle (2.2pt);
            \node[below=3pt] at (\i,0) {$\i$};
        }
        \draw[thick] (0.55,0) -- (4.45,0);
        \foreach \i in {1,2,3} { \node[above=2pt, font=\small] at (\i+0.5,0) {$L_{\i}$}; }
        \draw[decorate, decoration={brace, amplitude=5pt}, thick] (0.85,0.65) -- (3.15,0.65) node[midway, above=6pt] {$H_0$: sites $1, 2$; links $(1,2)$, $(2,3)$};
        \draw[decorate, decoration={brace, amplitude=5pt, mirror}, thick] (1.85,-0.75) -- (4.15,-0.75) node[midway, below=6pt] {$H'_0$: sites $2, 3$; links $(2,3)$, $(3,4)$};
    \end{tikzpicture}
    \caption{Definitions of the operators $H_0$ and $H'_0$ appearing in \eqref{eq:Fsym_main}. Each site carries a mass term, and each link carries an electric term and a hopping term.}
    \label{fig:cells}
\end{figure}
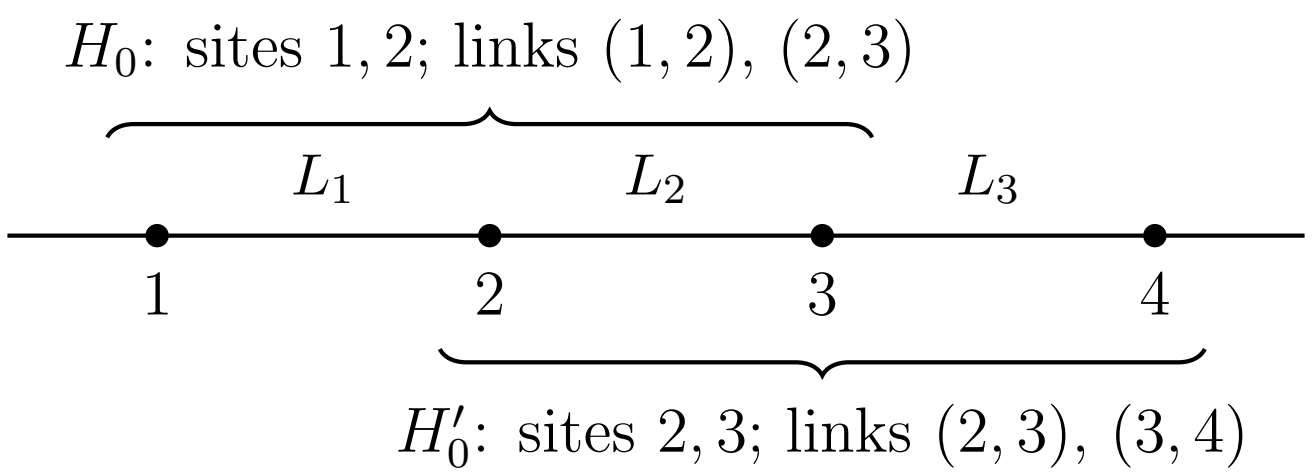

On the lattice, the natural counterpart of $T^{00}$ is the energy of a unit cell. We write the Hamiltonian \eqref{eq:H_lattice} as $H = \sum_{n\in\Z} h_n$, where
\begin{equation}\label{eq:h_n}
    h_n \equiv \frac{g^2a}{2}\left(L_n + \frac{\theta}{2\pi}\right)^2 - \frac{i}{2a}\left(c_n^\dagger U_n c_{n+1} - \text{h.c.}\right) + m_\text{lat}(-1)^n c_n^\dagger c_n
\end{equation}
contains the mass term on site $n$ and the electric and hopping terms on the link $(n, n+1)$. We then define the energy of cell $j$ by
\begin{equation}\label{eq:cell_energy}
    H_j \equiv h_{2j+1} + h_{2j+2}\,,
\end{equation}
and a shifted counterpart by
\begin{equation}\label{eq:cell_energy_prime}
    H_j' \equiv h_{2j+2} + h_{2j+3}\,,
\end{equation}
as shown in Figure~\ref{fig:cells}. Since $\sum_j H_j = H$ exactly, the forward matrix element of $H_0$ in the normalization of Section~\ref{subsec:me} is $E_n(p)$, and so $G(0) = 1$ holds exactly at any lattice spacing, with no renormalization. This is in contrast with lattice QCD, where the gluon and quark parts of the stress tensor must be separately renormalized \cite{Shanahan:2018pib,Pefkou:2021fni,Hackett:2023nkr,Hackett:2023rif}.

As in the continuum (see Section~\ref{subsec:gff_def}), the reality of $G(t)$ rests on $CPT$ symmetry. Let $\Theta$ be the anti-unitary lattice $CPT$ of Section~\ref{subsec:lattice_H}. Since $C$ preserves the momentum while $P$ and $T$ reverse it, $\Theta\ket{n,p}$ is proportional to $\ket{n,p}$, and we may choose the phases of the states so that $\Theta\ket{n,p} = \ket{n,p}$. This fixes each phase up to a sign, and by the argument of Section~\ref{subsec:gff_def}, it agrees with the boost convention as $a \to 0$. Anti-unitarity then gives, for any operator $\cO$,
\begin{equation}\label{eq:cpt_conj}
    \braket{n,p'|\cO|n,p}^* = \braket{n,p'|\Theta\cO\Theta^{-1}|n,p}\,,
\end{equation}
so that the matrix elements of a $\Theta$-invariant operator are real.

For this argument to go through, we need our energy density operator to be invariant under $\Theta$, and we need a prescription for choosing the phases of the states. For the operator, we see from \eqref{eq:CPT_lattice} that $\Theta$ acts on positions as the reflection $n \to 1-n$. This maps the sites of $H_j$ onto those of $H_{-j-1}$, but its links onto those of $H'_{-j-2}$, so neither $H_0$ nor $H'_0$ is $\Theta$-invariant. Instead, we use their average, with $H'_0$ re-centered onto $H_0$ by the phase of a one-site translation:
\begin{equation}\label{eq:Fsym_main}
    \mathcal F_n(p', p) \equiv \frac12\left[\braket{n,p'|H_0|n,p} + e^{i(p'-p)a}\braket{n,p'|H'_0|n,p}\right].
\end{equation}
Provided $\Theta\ket{n,p} = \ket{n,p}$, \eqref{eq:cpt_conj} maps the mass terms in \eqref{eq:Fsym_main} onto themselves and exchanges the link terms of the two cells, up to the offset by one site between the mass terms and the link terms, so that $\mathcal F_n(p',p)$ is real up to lattice artifacts of order $ap$.\footnote{At $\theta = 0$ or $\pi$, $C$ is itself a symmetry, which maps $H_0$ onto $H'_0$, and the two terms in \eqref{eq:Fsym_main} are equal.}

For the states, one could consider directly constructing $\Theta$ and imposing $\Theta\ket{n,p} = \ket{n,p}$, but in practice we use a different approach. We first compute the states with arbitrary phases on a uniform grid of momenta $p_j = j\,\delta p$, $j = -N_p, \ldots, N_p$. Then, starting from $\ket{n,0}$ and moving outward in both directions, we choose the phase of each state so that the near-forward element $\mathcal F_n(p_{j+1}, p_j)$ is real and positive. In the convention $\Theta\ket{n,p} = \ket{n,p}$, these elements are real up to lattice artifacts by the argument above, and positive since they approach $E_n(p) > 0$ as $\delta p \to 0$. This procedure therefore reproduces that convention, including the signs, up to a phase of order $ap$ accumulated from the imaginary lattice artifacts in \eqref{eq:Fsym_main}. The Breit-frame elements $\mathcal F_n(p,-p)$ are then real up to residuals that vanish as $a\to 0$, and their signs are physical. This is important for the heavier mesons, whose form factors change sign as we vary $t$.

To evaluate \eqref{eq:gff_breit} on a lattice of spacing $a$, we take the bra and ket to be the quasiparticle states $\ket{n, \pm p}$, both constructed over the same vacuum, so that
\begin{equation}\label{eq:gff_lat}
    G_n(t;a) = \frac{\Rea\mathcal F_n(p,-p)}{E_n(p)}\,, \qquad t = -4p^2\,.
\end{equation}
To see that this is the lattice counterpart of \eqref{eq:gff_breit}, note that the quasiparticle normalization $\braket{n,k'|n,k} = 2\pi\delta(k-k')$, with $k = 2ap$, differs from \eqref{eq:norm} by a factor $2E\cdot 2a$, and that $T^{00}(0)$ corresponds to $H_0/2a$, the energy of a cell divided by its length. Thus, the numerator of \eqref{eq:gff_breit} is $2E\,\mathcal F_n(p,-p)$, and dividing by $2E^2$ gives \eqref{eq:gff_lat}. The matrix element is computed as described in Appendix~\ref{app:qp_me}.

To obtain the continuum GFF, we extrapolate $G_n(t;a)$ to $a\to 0$.\footnote{The set of spacings depends on the mass and on $\theta$, within $0.05 \leq ga \leq 0.2$, except at $\theta = \pi$, where we use spacings up to $ga = 0.45$ for reasons explained below. We fit a linear function of $(ga)^2$.} We find that the discretization effects are at most a few percent, and are largest for the heaviest meson we study at large $-t$, where we need larger momenta to obtain a given value of $-t/M^2$.

\subsection{Numerical results}\label{subsec:gff_results}

\begin{figure}[t]
    \centering
    \begin{subfigure}{0.49\textwidth}\centering\pendingfig{\linewidth}{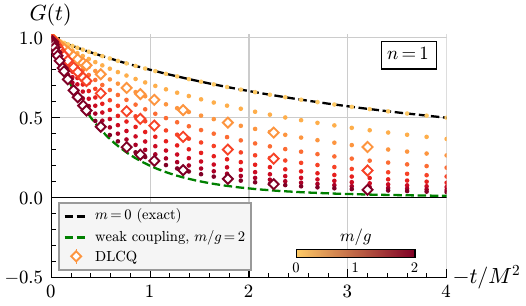}\caption{}\label{fig:gff_1}\end{subfigure}\hfill
    \begin{subfigure}{0.49\textwidth}\centering\pendingfig{\linewidth}{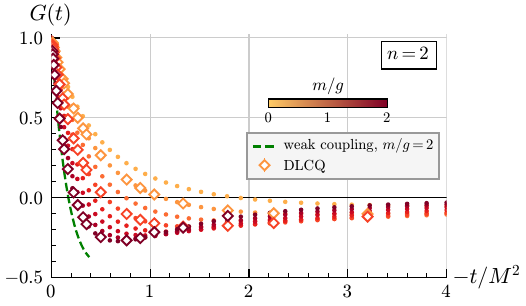}\caption{}\label{fig:gff_2}\end{subfigure}\\[0.5em]
    \begin{subfigure}{0.49\textwidth}\centering\pendingfig{\linewidth}{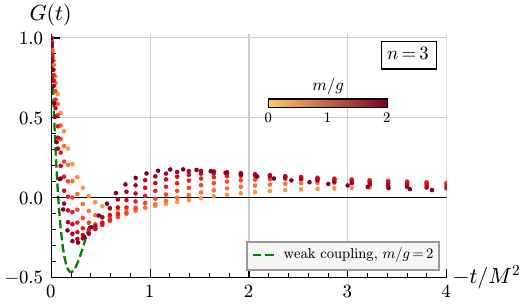}\caption{}\label{fig:gff_3}\end{subfigure}
    \caption{Continuum-extrapolated gravitational form factors of the three lightest mesons at $\theta = 0$ as functions of $-t/M^2$, for $m/g=0,0.25,\ldots,2$. (At the lowest values of $m/g$, there are only one or two stable mesons, so panels (b) and (c) have fewer mass values.) We compare with DLCQ at $m/g = 0.5$, $1$, and $2$ (diamonds) for $n=1,2$, and the exact massless result \eqref{eq:gff_massless} (black dashed line) for $n=1$, and both show excellent agreement with the lattice. We also plot the weak-coupling prediction \eqref{eq:gff_weak} at $m/g = 2$ (gray dashed line), which is most accurate at small $-t$ (for which relativistic corrections are small).}
    \label{fig:gff_bands}
\end{figure}

We begin with the lightest meson at $\theta = 0$, whose continuum form factor is shown in Figure~\ref{fig:gff_1} for $m/g=0,0.25,\ldots,2$. At $m = 0$, the lattice result agrees with the exact free-boson form factor \eqref{eq:gff_massless} to within $0.01\%$ over the entire range $-t \leq 4M^2$. As $m/g$ increases, $G(t)$ falls off faster in units of $M^2$. This is to be expected: the meson becomes a non-relativistic bound state whose size is of order $\alpha^{-1}$, where $\alpha = (mg^2/2)^{1/3}$ is the inverse length scale of the Airy wavefunctions \eqref{eq:airy}, and so it is parametrically larger than the Compton wavelength $1/M \approx 1/2m$. And indeed, at $m/g = 2$, the lattice result is close to the weak-coupling prediction \eqref{eq:gff_weak} at small $-t$, where the relativistic corrections are smallest. 

At intermediate $m/g$, where neither limit applies, DLCQ provides an independent benchmark, as described in Section~\ref{subsec:lightcone}. The two methods agree to within the uncertainty of the $K\to\infty$ extrapolation for DLCQ.

The form factors of the second- and third-lightest mesons, shown in Figure~\ref{fig:gff_bands}(b) and (c), are qualitatively different: $G_2$ has one zero and $G_3$ has two, after which they approach zero from below and above, respectively. This is the structure predicted by \eqref{eq:gff_weak}, where the zeros reflect the nodes of the wavefunctions $\hat\psi_2$ and $\hat\psi_3$. Quantitatively, however, the weak-coupling prediction is not very accurate at $m/g = 2$, except at the smallest values of $-t$. Comparing with DLCQ confirms that this discrepancy is a genuine relativistic effect rather than a lattice artifact; indeed, since the heavier mesons are less tightly bound, one expects their relativistic corrections to be larger.

Unlike DLCQ, which is limited to $\theta = 0$, the lattice computation applies equally well at any $\theta$. We will use it to map out the gravitational radius of the lightest meson, defined by the slope $G'(0) = dG/dt|_{t=0}$ of its form factor (see Section~\ref{subsec:gff_def}), as a function of $m/g$ and $\theta$. We present it in two ways. The dimensionless combination
\begin{equation}\label{eq:s_def}
    s \equiv 4M^2\, G'(0)
\end{equation}
measures the radius in units of the Compton wavelength of the meson. In particular, $s = 1$ for the free boson \eqref{eq:gff_massless} and $s = \tfrac12$ for the free Majorana fermion \eqref{eq:gff_majorana}. The combination $g^2 G'(0)$, on the other hand, measures the physical size; at weak coupling, by \eqref{eq:gff_weak}, it is $g^2\braket{r^2}/8$. At $m = 0$ and any value of $\theta$, we have $g^2 G'(0) = \pi/4$. 

\begin{figure}[t]
    \centering
    \begin{subfigure}{0.48\textwidth}\centering\pendingfig{\linewidth}{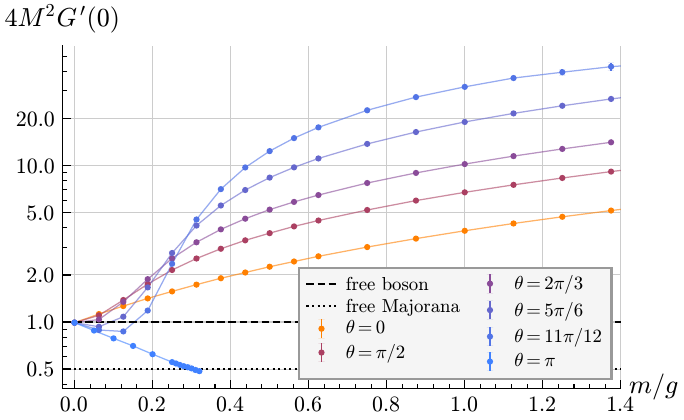}\caption{}\end{subfigure}\hfill
    \begin{subfigure}{0.48\textwidth}\centering\pendingfig{\linewidth}{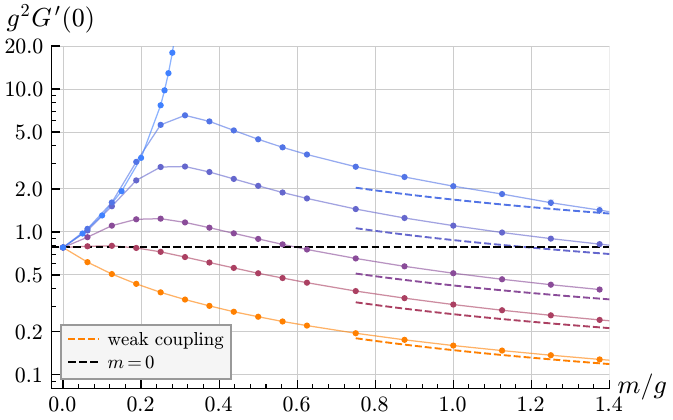}\caption{}\end{subfigure}
    \caption{The gravitational radius of the lightest meson as a function of $m/g$ for several values of $\theta$, in units of its Compton wavelength (a) and in coupling units (b). In (b), the dashed curves are the weak-coupling predictions computed from \eqref{eq:gff_weak} with the potential \eqref{eq:VNR}.}
    \label{fig:gff_radius}
\end{figure}

In Figure~\ref{fig:gff_radius}, we plot the gravitational radius in both units for $0 \leq m/g \leq 1.4$ and $\theta = 0$, $\pi/2$, $2\pi/3$, $5\pi/6$, and $11\pi/12$, and at $\theta = \pi$ for $0.05 \leq m/g \leq 0.32$. The slope is obtained by fitting a polynomial in $t$ to the points with $G(t;a) \geq 0.8$ at each lattice spacing, and extrapolating the linear coefficient to $a\to 0$.

The most striking feature of Figure~\ref{fig:gff_radius} is the rapid growth of the physical radius as we increase $\theta$ towards $\pi$ at fixed mass. We can understand this qualitatively from the weak-coupling picture: at $\theta \neq 0$, the non-relativistic potential \eqref{eq:VNR} becomes asymmetric, and as $\theta \to \pi$ the tension in one direction vanishes, allowing the bound state to swell significantly in that direction.

\begin{figure}[t]
    \centering
    \pendingfig{0.8\linewidth}{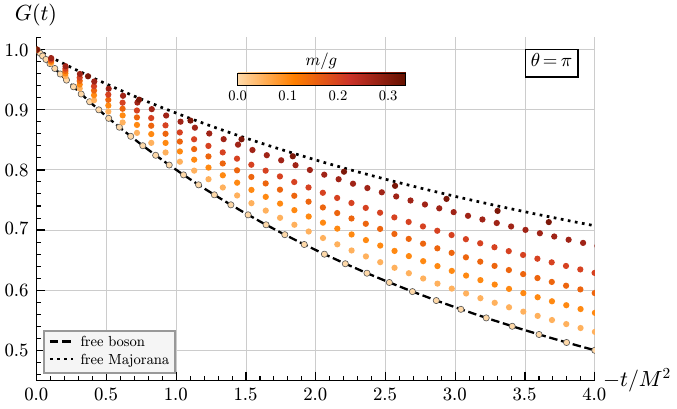}
    \caption{The continuum-extrapolated gravitational form factor of the lightest meson at $\theta = \pi$. We plot $G(t)$ for $m/g = 0, 0.05, \ldots, 0.2, 0.27$, and $0.31$. We compare with the free-boson and free-Majorana form factors, and we see the transition from the free-boson behavior at small $m/g$ to the free-Majorana behavior near the critical point.}
    \label{fig:gff_thetapi}
\end{figure}

Finally, at $\theta = \pi$, we compute the GFF of the lightest meson for $0 \leq m/g \leq 0.31$, approaching the critical point $m_c/g \approx 0.3335$.\footnote{As $m \to m_c$, the correlation length $\xi = 1/M$ diverges, and the bond dimension needed to represent the vacuum and the meson grows with $\xi/a$. We therefore use coarser lattices closer to the critical point, up to $ag = 0.4$ at $m/g = 0.31$, so that $1/(aM) \lesssim 150$ throughout. This is acceptable because the relevant momenta are small, $|p| \lesssim M$, so that $ap\lesssim 0.01$ there; the remaining lattice artifacts, at the scale $g$, are removed by the continuum extrapolation.} At small $m/g$, $G(t)$ is close to the free-boson form factor \eqref{eq:gff_massless}; as $m \to m_c$, it moves steadily toward the Majorana form factor \eqref{eq:gff_majorana}, as shown in Figure~\ref{fig:gff_thetapi}.

Thus, the gravitational form factor shows how the lightest bound state in the Schwinger model begins to resemble that of Ising field theory near the critical point.

\section{Parton distribution functions}\label{sec:pdf}
 
In this section, we compute the PDFs of the two lightest mesons. Following the strategy outlined at the end of Section~\ref{subsec:pdf_def}, we compute the ITD from spatial Wilson lines in quasiparticle states of large momentum, extrapolate to the continuum at fixed Ioffe time, and then invert \eqref{eq:itd_sine}.
 
In Section~\ref{subsec:wilson}, we define the lattice Wilson-line operator. In Section~\ref{subsec:continuum}, we explain how its matrix elements deviate from the continuum ITD, and we show that we can recover the continuum limit using a single extrapolation along a suitable path in the space of lattice spacings and momenta. In Section~\ref{subsec:inversion}, we discuss the inversion to recover the PDF, and in Section~\ref{subsec:pdf_results} we present our results.
 
\subsection{Wilson lines on the lattice}\label{subsec:wilson}
 
The lattice counterpart of the bilocal operator in \eqref{eq:itd_def}, with a spatial separation $z = a\ell$, is a pair of fermion operators separated by $\ell$ sites and connected by the product of the intervening links,
\begin{equation}\label{eq:wilson_lat}
    \cO_\ell = \frac{1}{2}\sum_{j=0,1} c^\dagger_{j}\, U_{j}U_{j+1}\cdots U_{j+\ell-1}\, c_{j+\ell} \qquad (\ell \text{ even})\,.
\end{equation}
This is manifestly gauge invariant. We have taken $\ell$ to be even so that both endpoints lie on the same sublattice; then $\cO_\ell$ connects like components of the Dirac fermion, and the average over the two starting sublattices gives the lattice version of $\Psi^\dagger(z)\cW(z,0)\Psi(0) = \bar\Psi(z)\gamma^0\cW(z,0)\Psi(0)$ (with $z^\mu = (0,-z)$, as in Section~\ref{subsec:pdf_def}). This corresponds to the component $\mu = 0$ of \eqref{eq:itd_def}, with no contamination from $\mathcal N$. We give the MPO representation of $\cO_\ell$ in Appendix~\ref{app:wilson_lempo}.
 
The lattice ITD of the $n$\textsuperscript{th} meson is then defined by the connected forward matrix element
\begin{equation}\label{eq:itd_lattice}
    h_{\text{lat},n}(a, z, ap) \equiv 2\Ima V_n(p,\ell)\,,\qquad \text{with }V_n(p, \ell) \equiv \braket{n, p|\cO_\ell|n, p}_c, \qquad \nu = p z = (ap)\,\ell\,.
\end{equation}
In the normalization of Section~\ref{subsec:me}, $V_n$ is the change in the expectation value of the translation sum of $\cO_\ell$ when the meson $\ket{n,p}$ is added to the vacuum, so that the factor $2p^0 = 2E_n(p)$ of \eqref{eq:itd_def} is divided out automatically. Thus, we find $V_n(p,\ell) \to \tfrac12\cM(\nu,-z^2)$ in the continuum, and $h_{\text{lat},n} = 2\Ima V_n$ approaches $h(\nu) = \Ima\cM$ of \eqref{eq:itd_sine}. Since no normalization is imposed by hand, the momentum sum rule $h'(0) = 1$ is a genuine check of both the normalization and the sign; below we find that it is satisfied to within $0.2\%$.
 
\subsection{The continuum limit at fixed Ioffe time}\label{subsec:continuum}
 
The lattice ITD \eqref{eq:itd_lattice} differs from the continuum ITD $h(\nu)$ in three ways, each controlled by its own small parameter. First, there are discretization effects; these enter at order $(ga)^2$ (rather than $ga$, as a result of the mass shift~\cite{Dempsey:2022nys}). Second, there are higher-twist effects, i.e., the corrections by which $\cM(\nu,-z^2)$ differs from $\cM(\nu,0)$ because the Wilson line is spacelike rather than lightlike. At fixed Ioffe time, $z = \nu/p$, and so these effects are small only if the momentum $p$ is large. Third, there are finite-momentum effects, which arise because boost symmetry is broken on the lattice.
 
To determine what sets the scale for the finite-momentum effects, consider a meson with $p \gg M$. Its energy is $E \approx p + M^2/2p$; the first term is simply shared among the constituents according to their momentum fractions, and so the leading effect of the internal dynamics lives in the subleading term. More precisely, configurations of the constituents with different momentum fractions differ in energy only by amounts of order $M^2/p$, and it is these small differences that determine how the momentum is shared, and hence the PDF. On the lattice, a constituent carrying a momentum $xp$ has an energy that deviates from its continuum value by an amount of order $xp\,(axp)^2$. Since this deviation is not linear in $x$, it changes the relative energies of different configurations by an amount of order $p\,(ap)^2$. The meson therefore has the correct internal structure only if this change is small compared with $M^2/p$, i.e., only if $\varepsilon \ll 1$, where
\begin{equation}\label{eq:eps_def}
    \varepsilon^2 \sim \frac{p\,(ap)^2}{M^2/p}\,, \qquad \varepsilon \equiv \frac{(ap)^2}{aM} = ap\cdot\frac{p}{M}\,.
\end{equation}
 
At fixed $\nu$, the higher-twist effects thus require $p$ to be large, while the finite-momentum effects require $p^2 \ll M/a$. To satisfy both, we must take $a \to 0$, but moreover we must be careful with how we change the momentum as we approach the continuum limit. We will take
\begin{equation}\label{eq:path}
    ap = \kappa\,(aM)^{2/3}\,, \qquad \text{i.e.,}\qquad p = \kappa\, M\,(aM)^{-1/3}\,,
\end{equation}
where $\kappa$ is a constant of order one. Along the path \eqref{eq:path}, $z = \nu/p$ and $\varepsilon = \kappa^2 (aM)^{1/3}$ both vanish as $(aM)^{1/3}$, while $ap$ and $a$ vanish faster. Since the corrections are even in $z$ and in $\varepsilon$, all of the leading corrections are linear in $y^2 \equiv (aM)^{2/3}$, the next ones (including those in $(ap)^2$) enter at order $y^4$, and the discretization effects enter only at order $y^6$. At each $\nu$, we therefore fit the points along the path to
\begin{equation}\label{eq:itd_fit}
    h_{\text{lat},n} \approx h(\nu) + c_1(\nu)\, y^2 + c_2(\nu)\, y^4\,,
\end{equation}
and take the intercept $h(\nu)$ as our estimate of the continuum ITD.\footnote{Note that \eqref{eq:itd_fit} contains no logarithms of $z^2$. In four dimensions, such logarithms appear at leading power and must be treated by perturbative matching \cite{Radyushkin:2017cyf,Ji:2020ect}; in $1+1$ dimensions, where the theory is super-renormalizable (Section~\ref{subsec:pdf_def}), they are absent at leading power, and we assume that any at higher powers are negligible in our fit range.}
 
The constant $\kappa$ sets the balance between the two leading corrections: a smaller $\kappa$ reduces $\varepsilon$ at the cost of larger $z$. The coefficients of $z^2$ and $\varepsilon^2$ are balanced at $\kappa \approx 1.3$ for all of our mesons with $m/g \geq 1$, nearly independently of $\nu$, and at somewhat larger $\kappa$ for lighter mesons. In practice, we compute the ITD on the paths $\kappa = 1.3$, $1.6$, and $2.0$, using four or five spacings with $aM$ between $0.03$ and $0.13$ and Wilson lines of up to $182$ sites.\footnote{For both mesons, we use the same set of lattices with the paths \eqref{eq:path} defined using the mass $M$ of the lightest meson. In units of the mass of the second-lightest meson, the paths therefore lie at somewhat smaller values of $\kappa$, between about $1.0$ and $1.6$.} We take as our central value the path whose leading coefficient $c_1$ is smallest, and we estimate the uncertainty from the spread of the extrapolated values when we change the order of \eqref{eq:itd_fit}, replace it by the line through the two finest spacings, drop the coarsest spacing, or move to the neighboring paths.
 
\begin{figure}[t]
    \centering
    \pendingfig{0.7\linewidth}{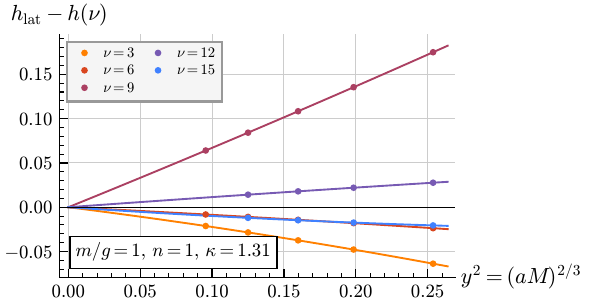}
    \caption{The approach of the lattice ITD to its continuum limit along the central path $\kappa = 1.31$, for the lightest meson at $m/g = 1$ and $\theta = 0$. We plot the deviation $h_\text{lat} - h(\nu)$ against $y^2 = (aM)^{2/3}$ at five values of $\nu$; the curves are the fits \eqref{eq:itd_fit}, whose intercepts define $h(\nu)$.}
    \label{fig:itd_conv}
\end{figure}

We illustrate the extrapolation in Figure~\ref{fig:itd_conv}, for the lightest meson at $m/g = 1$. Along the central path, the deviation $h_{\text{lat}} - h(\nu)$ is nearly linear in $y^2$ at every $\nu$, as anticipated above.
 
\subsection{Inversion}\label{subsec:inversion}
 
The PDF follows from the inverse transform \eqref{eq:itd_sine}, which requires $h(\nu)$ for all $\nu$. (See \cite{Karpie:2019eiq,Grebe:2026qmt} for more discussion of this inverse problem.) In practice, we know the continuum ITD only up to some cutoff $\nu_\text{max}$. However, we have chosen $\nu_\text{max}$ large enough that the ITD has decayed by then: we take $\nu_\text{max} = 20$--$25$ for the lightest meson at $\theta = 0$ and $\nu_\text{max} = 30$ for the second-lightest meson and for $\theta = 0.9\pi$, and in every case $|h(\nu_\text{max})| \lesssim 0.01$. We can therefore simply truncate the integral,
\begin{equation}\label{eq:trunc}
    f(x) = \frac1\pi\int_0^{\nu_\text{max}} d\nu\, h(\nu)\sin(\nu x)\,.
\end{equation}
Since $h(\nu)$ is already small at the cutoff, the truncation does not produce any severe ringing. The uncertainty of the PDF combines the effect of moving the cutoff to $0.9\,\nu_\text{max}$ with the uncertainty in the ITD that we discussed in Section~\ref{subsec:continuum}, which we propagate by transforming each of the variant ITDs in the same way.
 
Note that \eqref{eq:trunc} does not enforce $f(1) = 0$ or the support $[0,1]$, which instead emerge (approximately) from the data. At $m/g = 0.5$, the true PDFs fall off more slowly toward the endpoints; relatedly, the truncated transform of the lattice ITD leaves $f(1)$ slightly positive.

\begin{figure}[t]
    \centering
    \begin{subfigure}{0.48\textwidth}\centering\pendingfig{\linewidth}{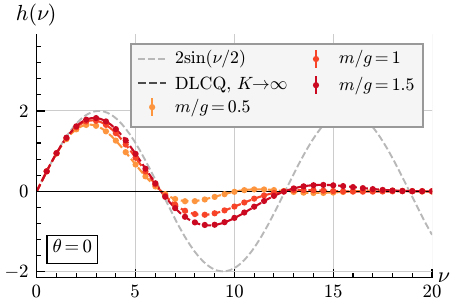}\caption{}\end{subfigure}
    \begin{subfigure}{0.48\textwidth}\centering\pendingfig{\linewidth}{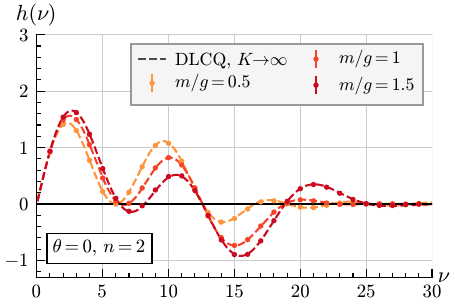}\caption{}\end{subfigure}
    \caption{The ITD at $\theta = 0$. (a) The continuum ITD of the lightest meson at $m/g = 0.5$, $1$, and $1.5$, compared with DLCQ (dashed, matching colors) and with the large-$m/g$ limit $2\sin(\nu/2)$ (gray dashed). (b) The continuum ITD of the second-lightest meson at the same masses, compared with DLCQ (dashed, matching colors).}
    \label{fig:itd}
\end{figure}
 
\subsection{Numerical results}\label{subsec:pdf_results}
 
We first show, in Figure~\ref{fig:itd}(a), the continuum ITD of the lightest meson at $\theta = 0$ for $m/g = 0.5$, $1$, and $1.5$, together with the DLCQ result; we see that we can determine it accurately at all three masses, up to $\nu = 20$. All of the curves leave the origin with unit slope, as required by the momentum sum rule \eqref{eq:momentum_sum}, and lie above the massless ITD \eqref{eq:pdf_massless} at small $\nu$. As $m/g$ increases, the first minimum deepens toward the oscillation $h(\nu) = 2\sin(\nu/2)$ of a PDF concentrated at $\delta\left(x - \tfrac12\right)$, shown as a gray dashed curve. In Figure~\ref{fig:itd}(b), we show that we can also determine the ITD of the second-lightest meson, using the same set of example masses. Since these decay more slowly, we compute them up to $\nu = 30$.
 
\begin{figure}[t]
    \centering
    \begin{subfigure}{0.48\textwidth}\centering\pendingfig{\linewidth}{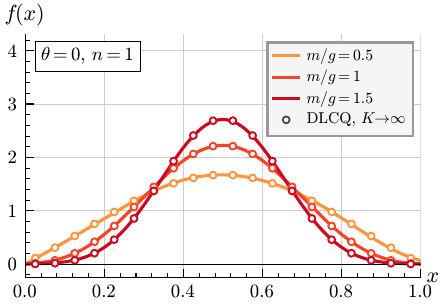}\caption{}\end{subfigure}\hfill
    \begin{subfigure}{0.48\textwidth}\centering\pendingfig{\linewidth}{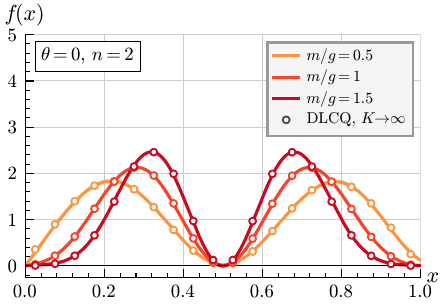}\caption{}\end{subfigure}
    \caption{The PDFs of (a) the lightest meson at $m/g = 0.5$, $1$, and $1.5$ and (b) the second-lightest meson at the same masses. All are at $\theta = 0$, obtained from the continuum ITD by the truncated transform \eqref{eq:trunc}, and compared with DLCQ extrapolated to $K \to\infty$ (open circles, so as to be visually separate from the lattice results which are nearly identical).}
    \label{fig:pdf}
\end{figure}
 
Figure~\ref{fig:pdf} shows the PDFs of the two lightest mesons at the same three masses, all compared with DLCQ. We find excellent agreement at every mass, with root-mean-square deviations less than $0.025$, or about $1\%$ of the peak height.
 
\begin{figure}[t]
    \centering
    \pendingfig{0.7\linewidth}{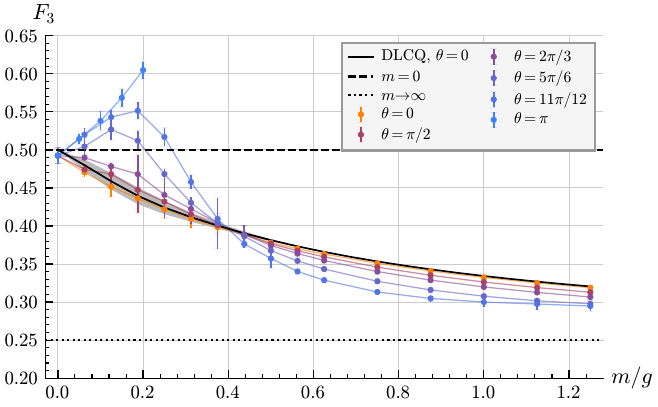}
    \caption{The third moment $F_3$ of the PDF of the lightest meson, as a function of $m/g$ for six values of $\theta$. We compare with DLCQ (extrapolated to $K\to\infty$) at $\theta = 0$, and we indicate the massless and nonrelativistic limits $F_3 = \tfrac12$ and $\tfrac14$ respectively. The gray band is the uncertainty of the DLCQ extrapolation to $K \to \infty$.}
    \label{fig:f3}
\end{figure}

We now turn to the moments of the PDF. The low moments are relatively inexpensive to compute, since by \eqref{eq:itd_moments} they are determined by the ITD at small $\nu$ \cite{Karpie:2018zaz}, where the continuum extrapolation is most precise. We estimate $F_3$ by fitting $h(\nu) - \nu = -\tfrac16F_3\nu^3 + \tfrac1{120}F_5\nu^5$ to the continuum ITD at small $\nu$, which only requires the ITD up to $\nu \approx 3.5$, and hence short Wilson lines. We show our estimates of $F_3$ in Figure~\ref{fig:f3} as a function of $m/g$ for six values of $\theta$. At $m = 0$, we find $F_3 = 0.49(1)$, consistent with the exact value $\tfrac12$ of \eqref{eq:pdf_massless}. At $\theta = 0$, we can also compare with $F_3$ from DLCQ at every mass. All of our values at $\theta = 0$ agree with DLCQ within the uncertainty of its $K\to\infty$ extrapolation. We see that for $\theta = 0$, $F_3$ decreases monotonically with $m/g$, on its way to the limiting value $\tfrac14$ at $m\gg g$ (see Section~\ref{subsec:weak}).
 
At $\theta = \pi$, $F_3$ exceeds its massless value $\tfrac12$ as $m/g$ grows toward the critical point, so our data predict a PDF with more support toward the endpoints than the flat massless one. It would be interesting to understand the enhancement at the endpoints from the point of view of the Ising field theory.
 
\begin{figure}[t]
    \centering
    \begin{subfigure}{0.48\textwidth}\centering\pendingfig{\linewidth}{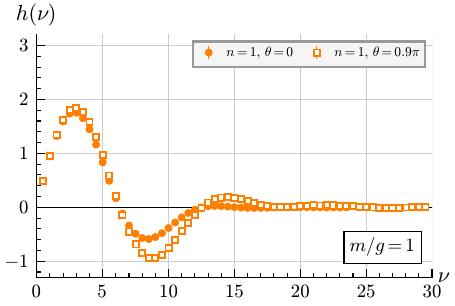}\caption{}\end{subfigure}\hfill
    \begin{subfigure}{0.48\textwidth}\centering\pendingfig{\linewidth}{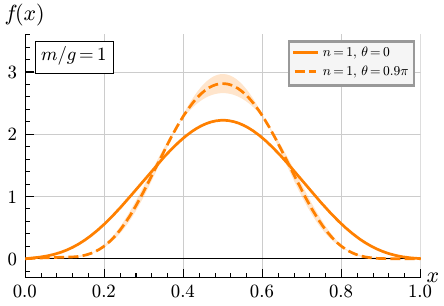}\caption{}\end{subfigure}
    \caption{The ITD (a) and the PDF (b) of the lightest meson at $m/g = 1$ for $\theta = 0$ and $\theta = 0.9\pi$. The bands are the uncertainties described in Sections~\ref{subsec:continuum} and~\ref{subsec:inversion}. At $\theta = 0.9\pi$, the ITD of the lightest meson decays more slowly, and its PDF is narrower.}
    \label{fig:pdf_theta}
\end{figure}

Away from $\theta = 0$, we can also reconstruct the full PDF. In Figure~\ref{fig:pdf_theta}, we compare the ITD and the PDF of the lightest meson at $m/g = 1$ for $\theta = 0$ and $\theta = 0.9\pi$. At $\theta = 0.9\pi$, the ITD of the lightest meson decays more slowly, and its PDF is correspondingly narrower, with a peak about $25\%$ higher. This narrowing in momentum fraction accompanies the spatial broadening seen in the GFF (Figure~\ref{fig:gff_radius}), but it is much weaker: between $\theta = 0$ and $0.9\pi$, $g^2G'(0)$ grows by an order of magnitude, while the width $\braket{(x-\tfrac12)^2}$ of the PDF only decreases by a factor of about $1.6$. 

The non-relativistic picture of Section~\ref{subsec:weak} explains why. As $\theta \to \pi$, we see from \eqref{eq:VNR} that the string tension on one side of the meson nearly vanishes, and so the bound state becomes lopsided, with a nonzero mean displacement $\langle r\rangle$. The GFF measures the mean square separation $\braket{r^2}$, which includes this displacement, whereas the width of the PDF is set by the momentum spread $\langle q^2\rangle$, which is sensitive only to the growth in the variance of $r$.\footnote{For the non-relativistic ground state at $m/g = 1$, $\braket{r^2}$ grows by a factor of about $10$ between $\theta = 0$ and $0.9\pi$, but the variance of $r$ grows only by a factor of $3$, and $\braket{q^2}$ decreases by a factor of about $2.6$.}
 
\FloatBarrier
\section{Discussion}\label{sec:discussion}

Confinement is a double-edged sword in the study of strongly-coupled gauge theories. On one hand, it prevents us from studying partons directly; on the other hand, it turns every asymptotic state into a rich laboratory for exploring the dynamics of the theory. Hadron structure functions are our windows into these laboratories, allowing us to characterize the internal structure of bound states in a way that connects to experimental observables.

In this paper, we have demonstrated how to calculate hadron structure functions using Hamiltonian lattice gauge theory on an infinite spatial lattice. As an example, we have studied mesons in the massive Schwinger model, for which infinite lattice techniques have long been understood \cite{Buyens:2013yza,Fujii:2024reh}. However, we have used LEMPOs~\cite{Dempsey:2025wia} as the tensor network backend for our calculation, and this construction generalizes immediately to non-abelian gauge theories in $1+1$ dimensions. Thus, our approach applies more generally to any such gauge theory (although the computational difficulty is likely greater in other cases).

In the particular case of the Schwinger model, we benchmark our infinite lattice method by comparing with discretized lightcone quantization, which gives very accurate estimates of structure functions but only works at $\theta = 0$. We then move to $\theta\neq 0$ and use the infinite lattice to calculate the gravitational form factor and the parton distribution function for mesons. As $\theta\to \pi$ for $m>m_c\approx 0.3335g$ \cite{Byrnes:2002nv,ArguelloCruz:2024xzi,Fujii:2024reh}, we approach a line of first-order phase transitions at which hadrons fall apart into charged solitons. The structure functions show quantitative signatures of the approach to that deconfinement transition.

Several questions remain open in the massive Schwinger model. It would be interesting to study other hadron structure functions for the mesons using a similar approach. Indeed, we have demonstrated in this paper how to compute off-forward matrix elements (for the GFF) and nonlocal matrix elements (for the PDF); by combining the two techniques used here, one can in principle calculate any matrix element. In addition, it would be valuable to generalize our methods slightly in order to calculate the structure functions of the solitons between the two degenerate vacua at $\theta = \pi$ and $m>m_c$. One could also consider how the structure functions relate to real-time scattering in the Schwinger model, which has also been studied using tensor networks \cite{Rigobello:2021fxw,Papaefstathiou:2024zsu}, including for solitons \cite{Belyansky:2023rgh,Dempsey:2026yut}.

While we have focused only on the massive Schwinger model in this paper, nothing in our construction is specific to that case. It applies equally to non-abelian gauge theories, to theories with several flavors (e.g., the two-flavor Schwinger model, studied using infinite lattices in \cite{Cuomo:2026kmy}), and to matter in various representations~\cite{Dempsey:2026hlf}. The internal structure of hadrons should be richer in these more general settings.

Finally, a natural long-term goal of this program is to study hadron structure in higher dimensions, aiming towards QCD\@. At present, tensor network methods are much less powerful in more than one space dimension, and so results are quite limited. Nevertheless, it would be interesting to explore how infinite-lattice techniques can help.

\section*{Acknowledgments}
We thank Phiala Shanahan for useful initial suggestions, collaboration on related projects, and comments on the manuscript. We also thank Brian Jiang and Patrick Oare for collaboration on related projects, and Silviu Pufu, Benjamin S\o{}gaard, and Manuel Schneider for helpful discussions. RD is supported by a Pappalardo Fellowship in Physics at MIT.

We also acknowledge the use of Claude (primarily Opus 4.8 and 5.5) for assistance with setting up code, preparing plots, and drafting text.

\appendix
\section{Off-forward matrix elements between quasiparticles}\label{app:qp_me}
\tikzset{
    tn/.style={draw, rounded corners=1.5pt, fill=white, minimum width=7mm, minimum height=6mm, inner sep=1pt, font=\small},
    op/.style={draw, rounded corners=1.5pt, fill=white, minimum height=6mm, inner sep=1pt, font=\small},
}

In this appendix, we explain how we evaluate the matrix element of a local operator between two quasiparticle states of different momenta, which we need in order to calculate the gravitational form factor via \eqref{eq:gff_lat}. The forward case is described in \cite{Vanderstraeten:2019voi,Osborne:2024oqg}; the off-forward case follows the same strategy, with phases that depend on both momenta. 

In Appendix~\ref{subsec:qp_me_states}, we review the mixed canonical form of the vacuum and the gauge freedom of the quasiparticle ansatz. In Appendix~\ref{subsec:qp_me_families}, we use the gauge condition to reduce the matrix element to four families of terms, and in Appendix~\ref{subsec:qp_me_resum}, we resum the infinite families as geometric series in the transfer matrices of the vacuum.

\subsection{Vacuum and quasiparticle states}\label{subsec:qp_me_states}

We describe the vacuum $\ket{\Omega}$ by a uniform MPS in the mixed canonical form~\cite{Vanderstraeten:2019voi}. It is specified by left- and right-gauged tensors $A_L$ and $A_R$, related by a matrix $C$ on the bond between them:
\begin{equation}\label{eq:AL_C_AR}
    \begin{tikzpicture}[baseline=-0.6ex, x=1cm, y=1cm]
        \draw (-0.65,0) -- (1.85,0);
        \draw (0,0) -- (0,-0.7);
        \node[tn] at (0,0) {$A_L$};
        \node[tn] at (1.2,0) {$C$};
    \end{tikzpicture}
    \;=\;
    \begin{tikzpicture}[baseline=-0.6ex, x=1cm, y=1cm]
        \draw (-0.65,0) -- (1.85,0);
        \draw (1.2,0) -- (1.2,-0.7);
        \node[tn] at (0,0) {$C$};
        \node[tn] at (1.2,0) {$A_R$};
    \end{tikzpicture}\,,
\end{equation}
where horizontal lines are virtual indices and vertical lines are physical indices. The gauge conditions defining $A_L$ and $A_R$ are
\begin{equation}\label{eq:isometry}
    \begin{tikzpicture}[baseline=-0.6ex, x=1cm, y=1cm]
        \draw[rounded corners=4pt] (0.65,0.6) -- (-0.6,0.6) -- (-0.6,-0.6) -- (0.65,-0.6);
        \draw (0,0.6) -- (0,-0.6);
        \node[tn] at (0,0.6) {$A_L$};
        \node[tn] at (0,-0.6) {$\bar A_L$};
    \end{tikzpicture}
    \;=\;
    \begin{tikzpicture}[baseline=-0.6ex, x=1cm, y=1cm]
        \draw[rounded corners=4pt] (0.6,0.6) -- (0,0.6) -- (0,-0.6) -- (0.6,-0.6);
    \end{tikzpicture}\,,
    \qquad\qquad
    \begin{tikzpicture}[baseline=-0.6ex, x=1cm, y=1cm]
        \draw[rounded corners=4pt] (-0.65,0.6) -- (0.6,0.6) -- (0.6,-0.6) -- (-0.65,-0.6);
        \draw (0,0.6) -- (0,-0.6);
        \node[tn] at (0,0.6) {$A_R$};
        \node[tn] at (0,-0.6) {$\bar A_R$};
    \end{tikzpicture}
    \;=\;
    \begin{tikzpicture}[baseline=-0.6ex, x=1cm, y=1cm]
        \draw[rounded corners=4pt] (-0.6,0.6) -- (0,0.6) -- (0,-0.6) -- (-0.6,-0.6);
    \end{tikzpicture}\,,
\end{equation}
where the upper (lower) layer is the ket (bra), and the caps denote identity maps on virtual spaces. Since our unit cell has two sites, each of these tensors carries a sublattice index, $A_{L,j}$ and $A_{R,j}$, with $C_j$ on the bond between sites $j$ and $j+1$; we will mostly suppress the sublattice index.

We write the quasiparticle ansatz \eqref{eq:qp} as
\begin{equation}\label{eq:qp_sites}
    \ket{\phi(k)} = \sum_{n\in\Z} e^{ikn/2}\ket{B;n}\,,
\end{equation}
where $\ket{B;n}$ is the state with $A_L$ on every site to the left of site $n$, the excitation tensor $B_1$ or $B_2$ on site $n$, according to its sublattice, and $A_R$ on every site to the right. The tensors $B_j$ are not unique: shifting them by $B \to B + A_L X - e^{-ik/2}X A_R$, for any matrix $X$ on the virtual bond, leaves $\ket{\phi(k)}$ unchanged~\cite{Haegeman:2013vud}. We remove this redundancy, together with the overlap with the vacuum at $k = 0$, by imposing the left gauge condition
\begin{equation}\label{eq:B_gauge}
    \begin{tikzpicture}[baseline=-0.6ex, x=1cm, y=1cm]
        \draw[rounded corners=4pt] (0.65,0.6) -- (-0.6,0.6) -- (-0.6,-0.6) -- (0.65,-0.6);
        \draw (0,0.6) -- (0,-0.6);
        \node[tn] at (0,0.6) {$B$};
        \node[tn] at (0,-0.6) {$\bar A_L$};
    \end{tikzpicture}
    \;=\; 0
\end{equation}
on each site of the unit cell. 

A key consequence of \eqref{eq:B_gauge} will be important below. Consider a contraction $\braket{B',n'|B,n}$ with $n\neq n'$; for concreteness, take $n<n'$. Then each site to the left of $n$ has $A_L$ in both layers, and so applying \eqref{eq:isometry} repeatedly, starting from the left, contracts all of them to the identity. The identity then annihilates the site by \eqref{eq:B_gauge}:
\begin{equation}\label{eq:B_gauge_consequence}
    \begin{tikzpicture}[baseline=-0.6ex, x=1cm, y=1cm]
        \draw[rounded corners=4pt] (0.45,0.6) -- (0,0.6) -- (0,-0.6) -- (0.45,-0.6);
        \draw (0.45,0.6) -- (1.3,0.6) (2.3,0.6) -- (4.45,0.6);
        \draw (0.45,-0.6) -- (1.3,-0.6) (2.3,-0.6) -- (4.45,-0.6);
        \draw (0.8,0.6) -- (0.8,-0.6);
        \draw (2.8,0.6) -- (2.8,-0.6);
        \draw (3.8,0.6) -- (3.8,-0.6);
        \node[tn] at (0.8,0.6) {$A_L$};
        \node[tn] at (0.8,-0.6) {$\bar A_L$};
        \node at (1.8,0.6) {$\cdots$};
        \node at (1.8,-0.6) {$\cdots$};
        \node[tn] at (2.8,0.6) {$A_L$};
        \node[tn] at (2.8,-0.6) {$\bar A_L$};
        \node[tn] at (3.8,0.6) {$B$};
        \node[tn] at (3.8,-0.6) {$\bar A_L$};
    \end{tikzpicture}
    \;\overset{\eqref{eq:isometry}}{=}\;
    \begin{tikzpicture}[baseline=-0.6ex, x=1cm, y=1cm]
        \draw[rounded corners=4pt] (0.65,0.6) -- (-0.6,0.6) -- (-0.6,-0.6) -- (0.65,-0.6);
        \draw (0,0.6) -- (0,-0.6);
        \node[tn] at (0,0.6) {$B$};
        \node[tn] at (0,-0.6) {$\bar A_L$};
    \end{tikzpicture}
    \;\overset{\eqref{eq:B_gauge}}{=}\; 0\,.
\end{equation}
The same holds for a site on which only the bra carries an excitation tensor. As a result, the overlap $\braket{B';n'|B;n}$ vanishes unless $n' = n$, and the norm of $\ket{\phi(k)}$ per unit cell is
\begin{equation}\label{eq:qp_norm}
    \mathcal N(k) \;=\; \sum_{j=1,2}\;
    \begin{tikzpicture}[baseline=-0.6ex, x=1cm, y=1cm]
        \draw[rounded corners=4pt] (0.45,0.6) -- (0,0.6) -- (0,-0.6) -- (0.45,-0.6);
        \draw[rounded corners=4pt] (5.55,0.6) -- (6.0,0.6) -- (6.0,-0.6) -- (5.55,-0.6);
        \draw (1.45,0.6) -- (4.55,0.6);
        \draw (1.45,-0.6) -- (4.55,-0.6);
        \draw (2.0,0.6) -- (2.0,-0.6);
        \draw (3.0,0.6) -- (3.0,-0.6);
        \draw (4.0,0.6) -- (4.0,-0.6);
        \node at (0.95,0.6) {$\cdots$};
        \node at (0.95,-0.6) {$\cdots$};
        \node[tn] at (2.0,0.6) {$A_L$};
        \node[tn] at (2.0,-0.6) {$\bar A_L$};
        \node[tn] at (3.0,0.6) {$B_j$};
        \node[tn] at (3.0,-0.6) {$\bar B_j$};
        \node[tn] at (4.0,0.6) {$A_R$};
        \node[tn] at (4.0,-0.6) {$\bar A_R$};
        \node at (5.05,0.6) {$\cdots$};
        \node at (5.05,-0.6) {$\cdots$};
    \end{tikzpicture}
    \;\overset{\eqref{eq:isometry}}{=}\; \sum_{j=1,2}\;
    \begin{tikzpicture}[baseline=-0.6ex, x=1cm, y=1cm]
        \draw[rounded corners=4pt] (-0.6,0.6) rectangle (0.6,-0.6);
        \draw (0,0.6) -- (0,-0.6);
        \node[tn] at (0,0.6) {$B_j$};
        \node[tn] at (0,-0.6) {$\bar B_j$};
    \end{tikzpicture}\,.
\end{equation}

\subsection{Reduction to four families}\label{subsec:qp_me_families}

Let the bra carry momentum $k'$ and tensors $B'_{1,2}$, both in the gauge \eqref{eq:B_gauge}, and let $\cO$ be a local operator supported on a window of $W$ sites, $n = 1, \ldots, W$. Then
\begin{equation}\label{eq:me_double}
    \braket{\phi(k')|\cO|\phi(k)} = \sum_{n',n\in\Z} e^{i(kn - k'n')/2}\braket{B';n'|\cO|B;n}_c\,,
\end{equation}
where the subscript $c$ indicates that we subtract $\braket{\Omega|\cO|\Omega}\braket{B';n'|B;n}$ from each term. This subtraction only affects the terms with $n' = n$, because of \eqref{eq:B_gauge_consequence}. Without it, the terms with $n' = n$ far from the window would sum to a term proportional to $\delta(k-k')$.

Again by \eqref{eq:B_gauge_consequence}, reading any term from the left, the first site on which the bra and ket differ from the vacuum cannot be one with a single excitation tensor and no operator. We may thus restrict to terms for which the leftmost nontrivial site is either in the window where the operator sits or is a site where $n' = n$. Sorting the surviving terms by the positions $n'$ and $n$ relative to the window, we are left with four families:
\begin{enumerate}
    \item[(i)] $n'$ and $n$ both in the window, which is a finite sum;
    \item[(ii)] one of $n'$, $n$ in the window and the other to the right of the window;
    \item[(iii)] $n' = n$ to the left of the window;
    \item[(iv)] $n'$ and $n$ both to the right of the window.
\end{enumerate}

\subsection{Resummation}\label{subsec:qp_me_resum}

The infinite families (ii)--(iv) are sums over the distance between the window and an excitation tensor, or between the two excitation tensors. Over such a stretch, each site contributes a transfer matrix of the vacuum,
\begin{equation}\label{eq:T_diagram}
    T_{XY} \;\equiv\; \sum_{s} X^{s}\otimes\bar Y^{s} \;=\;
    \begin{tikzpicture}[baseline=-0.6ex, x=1cm, y=1cm]
        \draw (-0.65,0.6) -- (0.65,0.6);
        \draw (-0.65,-0.6) -- (0.65,-0.6);
        \draw (0,0.6) -- (0,-0.6);
        \node[tn] at (0,0.6) {$X$};
        \node[tn] at (0,-0.6) {$\bar Y$};
    \end{tikzpicture}\,,
\end{equation}
where $X, Y \in \{A_L, A_R\}$ are the tensors of the ket and the bra on that site, and we write, e.g., $T_{RL}$ for $X = A_R$, $Y = A_L$. Each of these transfer matrices has an eigenvalue of unit modulus \cite{Fannes:1990ur}. By \eqref{eq:isometry}, the identity is the left fixed point of $T_{LL}$ and the right fixed point of $T_{RR}$, and by \eqref{eq:AL_C_AR}, the fixed points of the mixed transfer matrices $T_{LR}$ and $T_{RL}$ are given by $C$. Each infinite family is then built from geometric series of the form
\begin{equation}\label{eq:geometric}
    \sum_{l\geq 0} v\left(e^{i\kappa}\,T_{XY}\right)^l = v\left(1 - e^{i\kappa}\,T_{XY}\right)^{-1}\,,
\end{equation}
where $v$ is a left environment (i.e., the partial contraction of the network up to some bond \cite{Vanderstraeten:2019voi}, on which $T_{XY}$ acts from the right), and $\kappa$ is the phase per site.\footnote{Since $T_{XY}$ depends on the sublattice, in practice we resum each series over full cells, $\big(1 - e^{2i\kappa}\,T_\text{cell}\big)^{-1}$, where $T_\text{cell}$ is the product of the transfer matrices of the two sites in a cell.} In practice, acting with the operator \eqref{eq:geometric} means solving a linear system involving the matrix $1-e^{i\kappa} T_{XY}$. 

In family (ii), only one position is summed. If the ket is in the window and the bra to its right, then $T_{XY} = T_{RL}$ and $\kappa = -k'/2$, and in the other case $T_{XY} = T_{LR}$ and $\kappa = k/2$. In family (iii), and in the part of family (iv) with $n' = n$, the two positions are summed together, with $T_{XY} = T_{RR}$ and $\kappa = -(k-k')/2$, or $T_{XY} = T_{LL}$ and $\kappa = (k-k')/2$, respectively; in family (iii), $v$ is the site $n' = n$, and the series sums over its distance from the window.\footnote{Note that when $e^{i\kappa}$ is the inverse of the unit-modulus eigenvalue of $T_{XY}$, the operator \eqref{eq:geometric} is singular when it acts along the corresponding fixed point vector. However, this component is annihilated either by the seed $v$ or by the tensor that closes the network \cite{Vanderstraeten:2019voi}. Projecting the fixed point out of $T_{XY}$ therefore removes the singularity and leaves the result unchanged.}

The part of family (iv) with $n' \neq n$ requires the most care, since the two positions must be summed independently. Consider first $W < n' < n$. Splitting the phase as $e^{i(kn-k'n')/2} = e^{i(k-k')n'/2}\,e^{ik(n-n')/2}$, we find
\begin{equation}\label{eq:double_geometric}
\begin{split}
    &\sum_{W<n'<n} e^{i(kn-k'n')/2}\braket{B';n'|\cO|B;n}\\
    &\qquad = e^{i(k-k')(W+1)/2 + ik/2}\; L_c\,\big(1 - e^{i(k-k')/2}\,T_{LL}\big)^{-1}\,\mathcal P_{B'}\,\big(1 - e^{ik/2}\,T_{LR}\big)^{-1}\,\mathcal C_{B}\,,
\end{split}
\end{equation}
where the subscript $c$ may be omitted on the left-hand side since $n' \neq n$, and the objects on the right-hand side are defined by
\begin{equation}\label{eq:env_diagrams}
\begin{split}
    L_c \;&=\;
    \begin{tikzpicture}[baseline=-0.6ex, x=1cm, y=1cm]
        \draw[rounded corners=4pt] (0.45,1.0) -- (0,1.0) -- (0,-1.0) -- (0.45,-1.0);
        \draw (0.45,1.0) -- (1.3,1.0) (2.3,1.0) -- (3.35,1.0);
        \draw (0.45,-1.0) -- (1.3,-1.0) (2.3,-1.0) -- (3.35,-1.0);
        \draw (0.8,1.0) -- (0.8,-1.0);
        \draw (2.8,1.0) -- (2.8,-1.0);
        \node[op, minimum width=27mm] at (1.8,0) {$\cO$};
        \node[tn] at (0.8,1.0) {$A_L$};
        \node[tn] at (2.8,1.0) {$A_L$};
        \node[tn] at (0.8,-1.0) {$\bar A_L$};
        \node[tn] at (2.8,-1.0) {$\bar A_L$};
        \node at (1.8,1.0) {$\cdots$};
        \node at (1.8,-1.0) {$\cdots$};
    \end{tikzpicture}
    \;-\; \braket{\Omega|\cO|\Omega}
    \begin{tikzpicture}[baseline=-0.6ex, x=1cm, y=1cm]
        \draw[rounded corners=4pt] (0.6,1.0) -- (0,1.0) -- (0,-1.0) -- (0.6,-1.0);
    \end{tikzpicture}\,,\\[1ex]
    \mathcal P_{B'} \;&=\;
    \begin{tikzpicture}[baseline=-0.6ex, x=1cm, y=1cm]
        \draw (-0.65,0.6) -- (0.65,0.6);
        \draw (-0.65,-0.6) -- (0.65,-0.6);
        \draw (0,0.6) -- (0,-0.6);
        \node[tn] at (0,0.6) {$A_L$};
        \node[tn] at (0,-0.6) {$\bar B'$};
    \end{tikzpicture}\,,
    \qquad
    \mathcal C_{B} \;=\;
    \begin{tikzpicture}[baseline=-0.6ex, x=1cm, y=1cm]
        \draw[rounded corners=4pt] (-0.65,0.6) -- (0.6,0.6) -- (0.6,-0.6) -- (-0.65,-0.6);
        \draw (0,0.6) -- (0,-0.6);
        \node[tn] at (0,0.6) {$B$};
        \node[tn] at (0,-0.6) {$\bar A_R$};
    \end{tikzpicture}\,.
\end{split}
\end{equation}
Here the box labeled $\cO$ is the operator on the $W$ sites of the window, and the caps are the identity, as in \eqref{eq:isometry}. Thus, $L_c$ is the connected left environment at the right edge of the window, $\mathcal P_{B'}$ is the site $n'$ with $B'$ in the bra, and $\mathcal C_B$ is the site $n$ with $B$ in the ket, which closes the network. The first resolvent sums over $n'$, and the second over $n - n'$. Since the sum over $n - n'$ depends on $n'$ only through the environment it acts on, evaluating the double series reduces to solving two successive linear systems coming from \eqref{eq:geometric}, the first with $T_{LL}$ and the second with $T_{LR}$. The terms with $W < n < n'$ are given by the same expression with the roles of the bra and the ket exchanged, i.e., with $T_{RL}$ and $\kappa = -k'/2$ in the second solve, and with $B$ placed first.

Summing the four families with the phases of \eqref{eq:me_double} and dividing by $\sqrt{\mathcal N(k)\,\mathcal N(k')}$ gives the matrix element that enters \eqref{eq:gff_lat}. We have checked that our implementation agrees with a na\"ive version in which we evaluate \eqref{eq:me_double} by truncating the sums at increasingly large distances.

\section{The Wilson line as an MPO}\label{app:wilson_lempo}

In this appendix, we give the MPO representation of the Wilson-line operator \eqref{eq:wilson_lat}. We map the fermions to qubits by the Jordan--Wigner transformation
\begin{equation}\label{eq:jw}
    c_n = \Big(\prod_{k<n}\left(-i\sigma^z_k\right)\Big)\sigma^-_n\,, \qquad c_n^\dagger = \Big(\prod_{k<n}\left(i\sigma^z_k\right)\Big)\sigma^+_n\,,
\end{equation}
where $\sigma^z_n = +1$ ($-1$) if site $n$ is occupied (empty), and $\sigma^\pm_n$ raise and lower the occupation. The factors of $i$ in the string are chosen so that the hopping term of \eqref{eq:H_lattice} becomes real, $-i\left(c_n^\dagger c_{n+1} - \text{h.c.}\right) = \sigma^+_n\sigma^-_{n+1} + \sigma^-_n\sigma^+_{n+1}$, with the link $U_n$ implemented as explained below. Using $\sigma^+_n\sigma^z_n = -\sigma^+_n$, we find
\begin{equation}\label{eq:jw_bilinear}
    c_j^\dagger c_{j+\ell} = i^\ell\,\sigma^+_j\Big(\prod_{k=j+1}^{j+\ell-1}\left(-\sigma^z_k\right)\Big)\sigma^-_{j+\ell}\,,
\end{equation}
so that the operator $c_j^\dagger U_j\cdots U_{j+\ell-1}c_{j+\ell}$ becomes the right-hand side of \eqref{eq:jw_bilinear} times links; the phase $i^\ell$ is real for even $\ell$.

We represent the sum over $j$ in \eqref{eq:wilson_lat} as an MPO with the same tensor $\mathcal W$ on every site, and an $(\ell+2)$-dimensional virtual space with basis $\{\mathrm{i}, \mathrm{o}_1, \ldots, \mathrm{o}_\ell, \mathrm{f}\}$, where $\mathrm{i}$ is the initial state before the line has begun, $\mathrm{f}$ is the final state after it has ended, and $\mathrm{o}_k$ records that the line was opened $k$ sites ago, as in the finite-automaton construction of MPOs \cite{Crosswhite:2008ucd,Crosswhite:2008rml}. The only nonzero entries of $\mathcal W$, viewed as a matrix of operators on the site qubit, are
\begin{equation}\label{eq:W}
    \mathcal W_{\mathrm{i}\mathrm{i}} = \mathcal W_{\mathrm{f}\mathrm{f}} = \one\,, \qquad \mathcal W_{\mathrm{i}\mathrm{o}_1} = \sigma^+\,, \qquad \mathcal W_{\mathrm{o}_k\mathrm{o}_{k+1}} = -\sigma^z \ \ (1\leq k<\ell)\,, \qquad \mathcal W_{\mathrm{o}_\ell\mathrm{f}} = \sigma^-\,,
\end{equation}
and the translation sum of the operator \eqref{eq:wilson_lat} is $\tfrac{i^\ell}{2}$ times the infinite contraction of $\mathcal W$'s, whose expectation value in the uMPS or quasiparticle state is the sum over the two sublattice starting points per unit cell. Reading $\mathcal W$ as an automaton, the line is opened by $\sigma^+$ on one site, carried across the $\ell - 1$ intermediate sites by the Jordan--Wigner string, and closed by $\sigma^-$ on the final site.

Note that, unlike the Hamiltonian, this operator is an ordinary MPO: it requires no link tensors acting on the virtual bonds, even though it contains the links $U_n$. Indeed, in the symmetric-MPS representation of Section~\ref{subsec:lattice_H}, the electric field $L_n$ is the charge on a virtual bond, and the link $U_n$ raises it by one unit. The states $\mathrm{o}_k$ of the automaton therefore carry a $\grU(1)$ charge of $+1$, created by $\sigma^+$ and absorbed by $\sigma^-$, and as this charge is transported along the virtual bond of the MPO it shifts the electric field on every link between the endpoints. This is exactly the action of $U_j\cdots U_{j+\ell-1}$, and so the links do not require any explicit insertion of the $U_j$ operators.

\section{Discretized lightcone quantization}\label{app:dlcq}

In this appendix, we give the discretized lightcone Hamiltonian used in Section~\ref{subsec:lightcone}, following \cite{Eller:1986nt,Brodsky:1997de}, together with the expressions for the PDF and the GFF. Compactifying $x^-\sim x^- + 2\pi R$ with antiperiodic boundary conditions, the mode expansion \eqref{eq:lc_modes} becomes
\begin{equation}
    \psi(x^-) = \frac{1}{\sqrt{2\pi R}}\sum_{r>0,\ r\text{ odd}}\left(C_r e^{-irx^-/2R} + D_r^\dagger e^{irx^-/2R}\right),
\end{equation}
with $\{C_r, C^\dagger_{r'}\} = \{D_r, D^\dagger_{r'}\} = \delta_{rr'}$, so that a parton with label $r$ carries $k = r/2R$, and
\begin{equation}
    P^+ = \frac{1}{2R}\sum_r r\left(C_r^\dagger C_r + D_r^\dagger D_r\right) \equiv \frac{K}{2R}\,.
\end{equation}
Here and below, every momentum label $r$ runs over the positive odd integers, subject only to the constraints indicated under the sums. The Fourier modes of the current $J^+ = {:}\psi^\dagger\psi{:}$ are, for even $q > 0$,
\begin{equation}\label{eq:dlcq_J}
    J_q \equiv \int_{-\pi R}^{\pi R} dx^-\, e^{-iqx^-/2R}\, J^+ = \sum_r\left(C^\dagger_{r+q}C_r - D^\dagger_{r+q}D_r\right) + \sum_{r<q} C^\dagger_r D^\dagger_{q-r}\,, \qquad J_{-q} = J_q^\dagger\,.
\end{equation}
In lightcone gauge, eliminating $A_+$ gives an instantaneous Coulomb interaction expressed as $\tfrac{g^2}{2}\int dx^-\, J^+ (-\partial_-^2)^{-1} J^+$, and so the lightcone Hamiltonian is $P^- = P^-_\text{mass} + P^-_\text{int}$, with
\begin{equation}\label{eq:dlcq_Pminus_compact}
    P^-_\text{mass} = m^2 R\sum_r \frac1r\left(C_r^\dagger C_r + D_r^\dagger D_r\right)\,, \qquad P^-_\text{int} = \frac{g^2 R}{\pi}\sum_{q\neq 0,\ q\text{ even}}\frac{1}{q^2}\,J_qJ_{-q}\,.
\end{equation}
The $q = 0$ term is omitted because $J_0 = \sum_r(C^\dagger_r C_r - D^\dagger_r D_r)$ vanishes on gauge-invariant states.

The product $J_qJ_{-q}$ of the normal-ordered currents is not itself normal-ordered. Bringing it to normal order gives four-fermion terms, together with two-fermion self-energy terms:
\begin{equation}\label{eq:dlcq_Pint}
\begin{split}
    P^-_\text{int} = \frac{g^2R}{\pi}\Bigg\{&\sum_r \sigma_r\left(C_r^\dagger C_r + D_r^\dagger D_r\right)\\
    &+ \sum_{\substack{r_1+r_2 = r_3+r_4\\ r_3\neq r_1}}\frac{1}{(r_1-r_3)^2}\left(C^\dagger_{r_3}C^\dagger_{r_4}C_{r_2}C_{r_1} + D^\dagger_{r_3}D^\dagger_{r_4}D_{r_2}D_{r_1}\right)\\
    &- 2\sum_{r_1+r_2 = r_3+r_4}\left[\frac{1-\delta_{r_1r_3}}{(r_1-r_3)^2} - \frac{1}{(r_1+r_2)^2}\right] C^\dagger_{r_3}D^\dagger_{r_4}D_{r_2}C_{r_1}\\
    &- \Bigg(\sum_{r_1 = r_2+r_3+r_4}\left[\frac{1}{(r_2+r_3)^2} - \frac{1}{(r_2+r_4)^2}\right]\\
    &\hphantom{- \Bigg(}\times\left(C^\dagger_{r_3}C^\dagger_{r_4}D^\dagger_{r_2}C_{r_1} + D^\dagger_{r_3}D^\dagger_{r_4}C^\dagger_{r_2}D_{r_1}\right) + \text{h.c.}\Bigg)\Bigg\}\,,
\end{split}
\end{equation}
where the self-energy coefficient is
\begin{equation}\label{eq:dlcq_self}
    \sigma_r = 2\sum_{\substack{0<q<r\\ q\text{ even}}}\frac{1}{q^2}\,.
\end{equation}

The operator $P^-$ commutes with $P^+$ (and with $J_0$), and so we can diagonalize it in the finite-dimensional space spanned by the neutral Fock states $C^\dagger_{r_1}\cdots C^\dagger_{r_N}\,D^\dagger_{r'_1}\cdots D^\dagger_{r'_N}\ket{0}$ with $\sum_i (r_i + r'_i) = K$. Since $P^+ = K/2R$, the invariant mass is
\begin{equation}\label{eq:dlcq_M2}
    M^2 = 2P^+P^- = \frac{K}{R}\,P^-\,,
\end{equation}
which is independent of $R$. At $m = 0$, the flat state $\ket{h_{1,K}} = (2/K)^{1/2}\sum_{r<K} C^\dagger_r D^\dagger_{K-r}\ket{0}$ is an exact eigenstate with $M^2 = g^2/\pi$ at every $K$.

The PDF of the eigenstate $\ket{h_{n,K}}$, normalized to unity, is sampled on the grid $x = r/K$ (with spacing $2/K$):
\begin{equation}\label{eq:dlcq_pdf}
    f\!\left(\frac rK\right) = \frac K2\braket{h_{n,K}|C^\dagger_r C_r|h_{n,K}}\,, \qquad \bar f\!\left(\frac rK\right) = \frac K2\braket{h_{n,K}|D^\dagger_r D_r|h_{n,K}}\,.
\end{equation}
The momentum sum rule $\sum_r \tfrac2K\,\tfrac rK\,(f + \bar f) = 1$ holds exactly. Moreover, the lightcone Hamiltonian has a charge conjugation symmetry exchanging $C^\dagger_r$ with $D^\dagger_r$, and so we have $\bar f = f$ exactly.

The third moment of the PDF can be estimated by
\begin{equation}\label{eq:dlcq_F3}
    F_3 = \frac{4}{K}\sum_r \left(\frac rK\right)^3 f\!\left(\frac rK\right)\,.
\end{equation}
At $m = 0$, the flat eigenstate $\ket{h_{1,K}}$ has $F_3 = \tfrac12 - 1/K^2$, so it converges to the correct value as $K\to\infty$.

\begin{figure}[t]
    \centering
    \begin{subfigure}{0.48\textwidth}\centering\pendingfig{\linewidth}{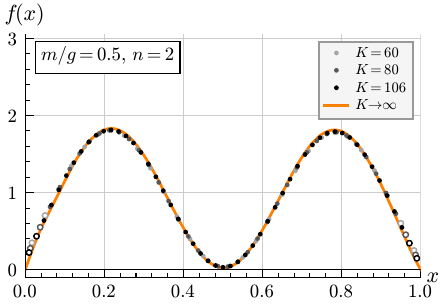}\caption{}\end{subfigure}\hfill
    \begin{subfigure}{0.48\textwidth}\centering\pendingfig{\linewidth}{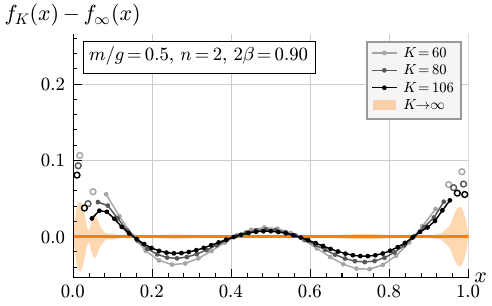}\caption{}\end{subfigure}
    \caption{Convergence of DLCQ with the harmonic resolution $K$, for the second-lightest meson at $m/g = 0.5$ and $\theta = 0$. (a) The PDF at $K = 60$, $80$, and $106$, sampled at $x = r/K$, and its extrapolation to $K\to\infty$ (orange), with the uncertainty of the extrapolation as a band. (b) The deviation of the sampled PDF at each $K$ from the extrapolated one. Open markers are the lowest modes at each end, which we omit from the extrapolation. The PDFs extrapolated to $K\to\infty$ are the benchmarks in Figures~\ref{fig:itd} and~\ref{fig:pdf}.}
    \label{fig:dlcq_kconv}
\end{figure}

The rate at which the DLCQ PDF converges is set by the behavior of the lightcone wavefunction near the endpoints, $\varphi(x)\sim x^\beta$, where \eqref{eq:thooft} requires $\pi\beta\cot\pi\beta = 1 - \pi m^2/g^2$ \cite{tHooft:1974pnl,Bergknoff:1976xr}. The corrections then fall off as $K^{-2\beta}$, and so we extrapolate to $K\to\infty$ by fitting $f_K(x) = f_\infty(x) + c(x)\,K^{-2\beta}$ at each $x$. To interpolate $f_K$ between the sampling points, we use a cubic spline of $f_K(x)/[x(1-x)]^{2\beta}$, which is smooth at the endpoints, and we omit the two lowest modes at each end ($r = 1, 3$ and $K-1, K-3$), which are not samples of a smooth function. We estimate the uncertainty of $f_\infty$ by comparing with a fit in $1/K$ and with a fit omitting only the lowest mode. Figure~\ref{fig:dlcq_kconv} shows the approach to $K\to\infty$ for the second-lightest meson at $m/g = 0.5$, where the convergence is slowest.

\section{The gravitational form factor of a free Majorana fermion}\label{app:majorana}

In this appendix, we derive \eqref{eq:gff_majorana}. As explained in Section~\ref{subsec:ising}, at $\theta = \pi$ and $m \to m_c$ the lightest meson becomes the particle of the Ising field theory, which is a free Majorana fermion of mass $M$. Since this theory is free, we can compute its GFF directly from the definition \eqref{eq:gff_def}: we write down the stress tensor of the free field, expand the field in creation and annihilation operators, and evaluate the one-particle matrix element.

A Majorana fermion is a two-component spinor field $\chi$ that is its own antiparticle. This condition is simplest to impose in a basis in which the Dirac operator $i\gamma^\mu\partial_\mu$ is real, since then it reads $\chi^* = \chi$, and the field equation preserves it. The basis of Section~\ref{sec:structure}, in which $\gamma^0$ and $\gamma^1$ are real, does not have this property. In this appendix only, we therefore use the Majorana basis
\begin{equation}
    \gamma^0 = \sigma^y\,, \qquad \gamma^1 = i\sigma^x\,,
\end{equation}
which satisfies the same algebra $\{\gamma^\mu, \gamma^\nu\} = 2\eta^{\mu\nu}$ and is related to the basis of Section~\ref{sec:structure} by a unitary change of basis, and in which $i\gamma^0$ and $i\gamma^1$ are real. Since $G(t)$ is defined by a matrix element of $T^{\mu\nu}$, it does not depend on this choice. For a real field, $\bar\chi \equiv \chi^\dagger\gamma^0 = \chi^T\gamma^0$.

The action of a free Majorana fermion is
\begin{equation}\label{eq:majorana_action}
    S = \int d^2x\left[\frac{i}{2}\,\bar\chi\gamma^\mu\partial_\mu\chi - \frac{M}{2}\,\bar\chi\chi\right],
\end{equation}
where the factors of $\tfrac12$, appropriate for a real field, give the canonical anticommutator $\{\chi_a(t,x), \chi_b(t,y)\} = \delta_{ab}\,\delta(x-y)$. The field equation is the Dirac equation,
\begin{equation}\label{eq:majorana_eom}
    \left(i\gamma^\mu\partial_\mu - M\right)\chi = 0\,.
\end{equation}
The symmetric stress tensor, obtained by varying \eqref{eq:majorana_action} with respect to the metric, is
\begin{equation}\label{eq:majorana_T}
    T^{\mu\nu} = \frac{i}{4}\,\bar\chi\left(\gamma^\mu\partial^\nu + \gamma^\nu\partial^\mu\right)\chi - \eta^{\mu\nu}\,\mathcal{L}\,,
\end{equation}
and the Lagrangian term vanishes on shell. As a check of the normalization, $T^{00} = \tfrac{i}{2}\,\chi^T\partial_0\chi$, whose integral is the Hamiltonian. Using \eqref{eq:majorana_eom}, the trace is
\begin{equation}\label{eq:majorana_trace}
    T^\mu{}_\mu = \frac{i}{2}\,\bar\chi\gamma^\mu\partial_\mu\chi = \frac{M}{2}\,\bar\chi\chi\,.
\end{equation}
Note that the trace vanishes when $M = 0$, as it must at the critical point.

To evaluate the matrix element of \eqref{eq:majorana_T}, we expand $\chi$ in plane waves. Because $\chi$ obeys the Dirac equation \eqref{eq:majorana_eom}, each plane wave $e^{-ip\cdot x}$ with $p^2 = M^2$ must come with a definite spinor $u(p)$, the positive-energy solution of
\begin{equation}\label{eq:majorana_u}
    \left(\gamma^\mu p_\mu - M\right)u(p) = 0\,.
\end{equation}
The reality condition $\chi^* = \chi$ then fixes the negative-frequency part as the complex conjugate of the positive-frequency part, so that
\begin{equation}
    \chi(x) = \int\frac{dp}{2\pi\,\cdot\,2E(p)}\left[u(p)\,a(p)\,e^{-ip\cdot x} + u(p)^*\,a^\dagger(p)\,e^{ip\cdot x}\right],
\end{equation}
with $\{a(p), a^\dagger(p')\} = 2E(p)(2\pi)\delta(p-p')$, so that $\ket{p} = a^\dagger(p)\ket{0}$ is normalized as in \eqref{eq:norm}, and $\braket{0|\chi(x)|p} = u(p)\,e^{-ip\cdot x}$.

To solve \eqref{eq:majorana_u}, we parametrize $p^\mu = M(\cosh\vartheta, \sinh\vartheta)$, so that $E \pm p = Me^{\pm\vartheta}$. In the Majorana basis,
\begin{equation}
    \gamma^\mu p_\mu = E\sigma^y - ip\,\sigma^x = \begin{pmatrix} 0 & -iMe^{\vartheta} \\ iMe^{-\vartheta} & 0\end{pmatrix},
\end{equation}
and \eqref{eq:majorana_u} gives $u_1 = -ie^{\vartheta}u_2$. Normalizing so that $\bar uu = 2M$ (equivalently, $u^\dagger u = 2E$), we find
\begin{equation}
    u(\vartheta) = \sqrt{M}\begin{pmatrix} -ie^{\vartheta/2} \\ e^{-\vartheta/2}\end{pmatrix}.
\end{equation}
Writing $\overline{\vartheta} = \tfrac12(\vartheta + \vartheta')$ and $\delta\vartheta = \vartheta - \vartheta'$ for the incoming and outgoing rapidities $\vartheta$ and $\vartheta'$, the bilinears we will need are
\begin{equation}\label{eq:majorana_bilinears}
    \bar u(\vartheta')\,u(\vartheta) = 2M\cosh\frac{\delta\vartheta}{2}\,, \qquad \bar u(\vartheta')\,\gamma^\mu u(\vartheta) = 2M\left(\cosh\overline{\vartheta}, \sinh\overline{\vartheta}\right).
\end{equation}

In the one-particle matrix element of a bilinear $\bar\chi\Gamma\partial^\nu\chi = \chi^T\gamma^0\Gamma\,\partial^\nu\chi$, either of the two fields can annihilate the incoming particle while the other creates the outgoing one. For $\Gamma = \gamma^\mu$, the matrix $\gamma^0\gamma^\mu$ is symmetric, and the two contractions combine into
\begin{equation}
    \braket{p'|\bar\chi\gamma^\mu\partial^\nu\chi(0)|p} = -i p^\nu\,\bar u(p')\gamma^\mu u(p) - ip'^\nu\,\bar u(p')\gamma^\mu u(p) = -2iP^\nu\,\bar u(p')\gamma^\mu u(p)\,,
\end{equation}
where the second term comes from the contraction in which the differentiated field creates the outgoing particle, and the relative sign from anticommuting the two fields is compensated by transposing $\gamma^0\gamma^\mu$. Substituting into \eqref{eq:majorana_T}, we find
\begin{equation}
    \braket{p'|T^{\mu\nu}(0)|p} = \frac12\left[P^\nu\,\bar u(p')\gamma^\mu u(p) + P^\mu\,\bar u(p')\gamma^\nu u(p)\right].
\end{equation}
Finally, $P^\mu = \tfrac12(p+p')^\mu = M\cosh\tfrac{\delta\vartheta}{2}\,(\cosh\overline{\vartheta}, \sinh\overline{\vartheta})$, so that by \eqref{eq:majorana_bilinears}, $\bar u(p')\gamma^\mu u(p) = 2P^\mu/\cosh\tfrac{\delta\vartheta}{2}$. Therefore,
\begin{equation}
    \braket{p'|T^{\mu\nu}(0)|p} = \frac{2P^\mu P^\nu}{\cosh(\delta\vartheta/2)}\,,
\end{equation}
which has the form \eqref{eq:gff_def}, with $G = 1/\cosh(\delta\vartheta/2)$. Since $t = (p'-p)^2 = -4M^2\sinh^2\tfrac{\delta\vartheta}{2}$, we have $\cosh\tfrac{\delta\vartheta}{2} = \sqrt{1 - t/4M^2}$, and hence
\begin{equation}
    G(t) = \frac{1}{\sqrt{1 - t/4M^2}}\,.
\end{equation}

\bibliographystyle{JHEP}
\bibliography{refs}

@article{Okuda:2026hbg,
    author = "Okuda, Takuya",
    title = "{Critical behavior of the bosonized Schwinger model with uniform matrix product states}",
    eprint = "2611.16092",
    archivePrefix = "arXiv",
    primaryClass = "hep-lat",
    month = "11",
    year = "2026"
}

@article{Fannes:1990ur,
    author = "Fannes, M. and Nachtergaele, B. and Werner, R. F.",
    title = "{FINITELY CORRELATED STATES ON QUANTUM SPIN CHAINS}",
    reportNumber = "DIAS-STP-90-11",
    doi = "10.1007/BF02099178",
    journal = "Commun. Math. Phys.",
    volume = "144",
    pages = "443--490",
    year = "1992"
}

@article{Fujii:2024reh,
    author = "Fujii, Hirotsugu and Fujikura, Kohei and Kikukawa, Yoshio and Okuda, Takuya and Pedersen, Juan W.",
    title = "{Critical behavior of the Schwinger model via gauge-invariant variational uniform matrix product states}",
    eprint = "2412.03569",
    archivePrefix = "arXiv",
    primaryClass = "hep-lat",
    reportNumber = "UT-Komaba/24-10",
    doi = "10.1103/PhysRevD.111.094505",
    journal = "Phys. Rev. D",
    volume = "111",
    number = "9",
    pages = "094505",
    year = "2025"
}

@article{ArguelloCruz:2024xzi,
    author = "Arguello Cruz, Erick and Tarnopolsky, Grigory and Xin, Yuan",
    title = "{Precision study of the massive Schwinger model near quantum criticality}",
    eprint = "2412.01902",
    archivePrefix = "arXiv",
    primaryClass = "hep-th",
    doi = "10.1103/vf7m-kzk2",
    journal = "Phys. Rev. D",
    volume = "112",
    number = "3",
    pages = "034023",
    year = "2025"
}

@article{Schwinger:1962tp,
    author = "Schwinger, Julian S.",
    title = "{Gauge Invariance and Mass. 2.}",
    doi = "10.1103/PhysRev.128.2425",
    journal = "Phys. Rev.",
    volume = "128",
    pages = "2425--2429",
    year = "1962"
}

@article{Lowenstein:1971fc,
    author = "Lowenstein, J. H. and Swieca, J. A.",
    title = "{Quantum electrodynamics in two-dimensions}",
    doi = "10.1016/0003-4916(71)90246-6",
    journal = "Annals Phys.",
    volume = "68",
    pages = "172--195",
    year = "1971"
}

@article{Coleman:1975pw,
    author = "Coleman, Sidney R. and Jackiw, R. and Susskind, Leonard",
    title = "{Charge Shielding and Quark Confinement in the Massive Schwinger Model}",
    reportNumber = "MIT-CTP-470",
    doi = "10.1016/0003-4916(75)90212-2",
    journal = "Annals Phys.",
    volume = "93",
    pages = "267",
    year = "1975"
}

@article{Coleman:1976uz,
    author = "Coleman, Sidney R.",
    title = "{More About the Massive Schwinger Model}",
    reportNumber = "Print-76-0357 (HARVARD)",
    doi = "10.1016/0003-4916(76)90280-3",
    journal = "Annals Phys.",
    volume = "101",
    pages = "239",
    year = "1976"
}

@article{Kogut:1974ag,
    author = "Kogut, John B. and Susskind, Leonard",
    title = "{Hamiltonian Formulation of Wilson's Lattice Gauge Theories}",
    reportNumber = "Print-74-1186 (CORNELL)",
    doi = "10.1103/PhysRevD.11.395",
    journal = "Phys. Rev. D",
    volume = "11",
    pages = "395--408",
    year = "1975"
}

@article{Banks:1975gq,
    author = "Banks, Tom and Susskind, Leonard and Kogut, John B.",
    title = "{Strong Coupling Calculations of Lattice Gauge Theories: (1+1)-Dimensional Exercises}",
    reportNumber = "CLNS-318",
    doi = "10.1103/PhysRevD.13.1043",
    journal = "Phys. Rev. D",
    volume = "13",
    pages = "1043",
    year = "1976"
}

@article{Susskind:1976jm,
    author = "Susskind, Leonard",
    title = "{Lattice Fermions}",
    reportNumber = "PTENS-76-1",
    doi = "10.1103/PhysRevD.16.3031",
    journal = "Phys. Rev. D",
    volume = "16",
    pages = "3031--3039",
    year = "1977"
}

@article{Byrnes:2002nv,
    author = "Byrnes, T. and Sriganesh, P. and Bursill, R. J. and Hamer, C. J.",
    title = "{Density matrix renormalization group approach to the massive Schwinger model}",
    eprint = "hep-lat/0202014",
    archivePrefix = "arXiv",
    doi = "10.1103/PhysRevD.66.013002",
    journal = "Phys. Rev. D",
    volume = "66",
    pages = "013002",
    year = "2002"
}

@article{Banuls:2013jaa,
    author = "Ba{\~n}uls, M. C. and Cichy, K. and Jansen, K. and Cirac, J. I.",
    title = "{The mass spectrum of the Schwinger model with Matrix Product States}",
    eprint = "1305.3765",
    archivePrefix = "arXiv",
    primaryClass = "hep-lat",
    reportNumber = "DESY-13-084, SFB-CPP-13-31, DESY 13-084, SFB/CPP-13-31",
    doi = "10.1007/JHEP11(2013)158",
    journal = "JHEP",
    volume = "11",
    pages = "158",
    year = "2013"
}

@article{Buyens:2013yza,
    author = "Buyens, Boye and Haegeman, Jutho and Van Acoleyen, Karel and Verschelde, Henri and Verstraete, Frank",
    title = "{Matrix product states for gauge field theories}",
    eprint = "1312.6654",
    archivePrefix = "arXiv",
    primaryClass = "hep-lat",
    doi = "10.1103/PhysRevLett.113.091601",
    journal = "Phys. Rev. Lett.",
    volume = "113",
    pages = "091601",
    year = "2014"
}

@article{Dempsey:2022nys,
    author = "Dempsey, Ross and Klebanov, Igor R. and Pufu, Silviu S. and Zan, Bernardo",
    title = "{Discrete chiral symmetry and mass shift in the lattice Hamiltonian approach to the Schwinger model}",
    eprint = "2206.05308",
    archivePrefix = "arXiv",
    primaryClass = "hep-th",
    reportNumber = "PUPT-2634",
    doi = "10.1103/PhysRevResearch.4.043133",
    journal = "Phys. Rev. Res.",
    volume = "4",
    number = "4",
    pages = "043133",
    year = "2022"
}

@article{Cuomo:2026kmy,
    author = "Cuomo, Gabriel and Dempsey, Ross and Katsevich, Andrei and Klebanov, Igor R. and Kochergin, Ilia V. and Pufu, Silviu S. and S{\o}gaard, Benjamin T.",
    title = "{The two-flavor Schwinger model at 50: Solving Coleman's puzzles}",
    eprint = "2605.08042",
    archivePrefix = "arXiv",
    primaryClass = "hep-th",
    reportNumber = "PUPT-2662, MIT-CTP/6036",
    month = "5",
    year = "2026"
}

@article{White:1992zz,
    author = "White, Steven R.",
    title = "{Density matrix formulation for quantum renormalization groups}",
    doi = "10.1103/PhysRevLett.69.2863",
    journal = "Phys. Rev. Lett.",
    volume = "69",
    pages = "2863--2866",
    year = "1992"
}

@article{Schollwock:2005zz,
    author = "Schollwock, U.",
    title = "{The density-matrix renormalization group}",
    eprint = "cond-mat/0409292",
    archivePrefix = "arXiv",
    doi = "10.1103/RevModPhys.77.259",
    journal = "Rev. Mod. Phys.",
    volume = "77",
    pages = "259--315",
    year = "2005"
}

@article{Orus:2013kga,
    author = "Orus, Roman",
    title = "{A Practical Introduction to Tensor Networks: Matrix Product States and Projected Entangled Pair States}",
    eprint = "1306.2164",
    archivePrefix = "arXiv",
    primaryClass = "cond-mat.str-el",
    doi = "10.1016/j.aop.2014.06.013",
    journal = "Annals Phys.",
    volume = "349",
    pages = "117--158",
    year = "2014"
}

@article{Cirac:2020obd,
    author = "Cirac, J. Ignacio and Perez-Garcia, David and Schuch, Norbert and Verstraete, Frank",
    title = "{Matrix product states and projected entangled pair states: Concepts, symmetries, theorems}",
    eprint = "2011.12127",
    archivePrefix = "arXiv",
    primaryClass = "quant-ph",
    doi = "10.1103/RevModPhys.93.045003",
    journal = "Rev. Mod. Phys.",
    volume = "93",
    number = "4",
    pages = "045003",
    year = "2021"
}

@article{McCulloch:2008aun,
    author = "McCulloch, I. P.",
    title = "{Infinite size density matrix renormalization group, revisited}",
    eprint = "0804.2509",
    archivePrefix = "arXiv",
    primaryClass = "cond-mat.str-el",
    month = "4",
    year = "2008"
}

@article{Zauner-Stauber:2017eqw,
    author = "Zauner-Stauber, V. and Vanderstraeten, L. and Fishman, M. T. and Verstraete, F. and Haegeman, J.",
    title = "{Variational optimization algorithms for uniform matrix product states}",
    eprint = "1701.07035",
    archivePrefix = "arXiv",
    primaryClass = "quant-ph",
    doi = "10.1103/PhysRevB.97.045145",
    journal = "Phys. Rev. B",
    volume = "97",
    number = "4",
    pages = "045145",
    year = "2018"
}

@article{Vanderstraeten:2019voi,
    author = "Vanderstraeten, Laurens and Haegeman, Jutho and Verstraete, Frank",
    title = "{Tangent-space methods for uniform matrix product states}",
    eprint = "1810.07006",
    archivePrefix = "arXiv",
    primaryClass = "cond-mat.str-el",
    doi = "10.21468/SciPostPhysLectNotes.7",
    journal = "SciPost Phys. Lect. Notes",
    volume = "7",
    pages = "1",
    year = "2019"
}

@article{Haegeman:2011lcd,
    author = "Haegeman, Jutho and Pirvu, Bogdan and Weir, David J. and Cirac, J. Ignacio and Osborne, Tobias J. and Verschelde, Henri and Verstraete, Frank",
    title = "{Variational matrix product ansatz for dispersion relations}",
    eprint = "1103.2286",
    archivePrefix = "arXiv",
    primaryClass = "quant-ph",
    doi = "10.1103/PhysRevB.85.100408",
    journal = "Phys. Rev. B",
    volume = "85",
    number = "10",
    pages = "100408",
    year = "2012"
}

@article{Haegeman:2013xcv,
    author = "Haegeman, Jutho and Michalakis, Spyridon and Nachtergaele, Bruno and Osborne, Tobias J. and Schuch, Norbert and Verstraete, Frank",
    title = "{Elementary Excitations in Gapped Quantum Spin Systems}",
    eprint = "1305.2176",
    archivePrefix = "arXiv",
    primaryClass = "quant-ph",
    doi = "10.1103/PhysRevLett.111.080401",
    journal = "Phys. Rev. Lett.",
    volume = "111",
    number = "8",
    pages = "080401",
    year = "2013"
}

@article{Haegeman:2013vud,
    author = "Haegeman, Jutho and Osborne, Tobias J. and Verstraete, Frank",
    title = "{Post-matrix product state methods: To tangent space and beyond}",
    eprint = "1305.1894",
    archivePrefix = "arXiv",
    primaryClass = "quant-ph",
    doi = "10.1103/PhysRevB.88.075133",
    journal = "Phys. Rev. B",
    volume = "88",
    number = "7",
    pages = "075133",
    year = "2013"
}

@article{Zauner:2014iea,
    author = "Zauner, V. and Draxler, D. and Vanderstraeten, L. and Degroote, M. and Haegeman, J. and Rams, M. M. and Stojevic, V. and Schuch, N. and Verstraete, F.",
    title = "{Transfer Matrices and Excitations with Matrix Product States}",
    eprint = "1408.5140",
    archivePrefix = "arXiv",
    primaryClass = "quant-ph",
    doi = "10.1088/1367-2630/17/5/053002",
    journal = "New J. Phys.",
    volume = "17",
    number = "5",
    pages = "053002",
    year = "2015"
}

@article{Vanderstraeten:2013xda,
    author = "Vanderstraeten, Laurens and Haegeman, Jutho and Osborne, Tobias J. and Verstraete, Frank",
    title = "{S-matrix from matrix product states}",
    eprint = "1312.6793",
    archivePrefix = "arXiv",
    primaryClass = "cond-mat.str-el",
    doi = "10.1103/PhysRevLett.112.257202",
    journal = "Phys. Rev. Lett.",
    volume = "112",
    number = "25",
    pages = "257202",
    year = "2014"
}

@article{VanDamme:2021xqk,
    author = "Van Damme, Maarten and Vanhove, Robijn and Haegeman, Jutho and Verstraete, Frank and Vanderstraeten, Laurens",
    title = "{Efficient MPS methods for extracting spectral information on rings and cylinders}",
    eprint = "2102.10982",
    archivePrefix = "arXiv",
    primaryClass = "cond-mat.str-el",
    doi = "10.1103/PhysRevB.104.115142",
    journal = "Phys. Rev. B",
    volume = "104",
    pages = "115142",
    year = "2021"
}

@article{Osborne:2024oqg,
    author = "Osborne, Jesse J. and McCulloch, Ian P.",
    title = "{Efficient and systematic calculation of arbitrary observables for the matrix product state excitation ansatz}",
    eprint = "2408.17117",
    archivePrefix = "arXiv",
    primaryClass = "cond-mat.str-el",
    doi = "10.1103/PhysRevResearch.7.023018",
    journal = "Phys. Rev. Res.",
    volume = "7",
    number = "2",
    pages = "023018",
    year = "2025"
}

@article{Singh:2009cd,
    author = "Singh, Sukhwinder and Pfeifer, Robert N. C. and Vidal, Guifre",
    title = "{Tensor network decompositions in the presence of a global symmetry}",
    eprint = "0907.2994",
    archivePrefix = "arXiv",
    primaryClass = "cond-mat.str-el",
    doi = "10.1103/PhysRevA.82.050301",
    journal = "Phys. Rev. A",
    volume = "82",
    pages = "050301",
    year = "2010"
}

@article{Banuls:2019rao,
    author = "Ba{\~n}uls, Mari Carmen and Cichy, Krzysztof",
    title = "{Review on Novel Methods for Lattice Gauge Theories}",
    eprint = "1910.00257",
    archivePrefix = "arXiv",
    primaryClass = "hep-lat",
    doi = "10.1088/1361-6633/ab6311",
    journal = "Rept. Prog. Phys.",
    volume = "83",
    number = "2",
    pages = "024401",
    year = "2020"
}

@article{Dempsey:2025wia,
    author = {Dempsey, Ross and Gl{\"u}ck, Anna-Maria E. and Pufu, Silviu S. and S{\o}gaard, Benjamin T.},
    title = "{Infinite matrix product states for (1 + 1)-dimensional gauge theories}",
    eprint = "2508.16363",
    archivePrefix = "arXiv",
    primaryClass = "hep-th",
    reportNumber = "PUPT-2657",
    doi = "10.1007/JHEP03(2026)181",
    journal = "JHEP",
    volume = "03",
    pages = "181",
    year = "2026"
}

@article{Dempsey:2026hlf,
    author = {Dempsey, Ross and Gl{\"u}ck, Anna-Maria E. and Pufu, Silviu S. and S{\o}gaard, Benjamin T.},
    title = "{Vacuum structure of gapped QCD$_2$ theories from the infinite Hamiltonian lattice}",
    eprint = "2601.16262",
    archivePrefix = "arXiv",
    primaryClass = "hep-th",
    reportNumber = "PUPT-2660, MIT-CTP/5983",
    month = "1",
    year = "2026"
}

@article{Bjorken:1968dy,
    author = "Bjorken, J. D.",
    title = "{Asymptotic Sum Rules at Infinite Momentum}",
    reportNumber = "SLAC-PUB-0510",
    doi = "10.1103/PhysRev.179.1547",
    journal = "Phys. Rev.",
    volume = "179",
    pages = "1547--1553",
    year = "1969"
}

@article{Feynman:1969ej,
    author = "Feynman, Richard P.",
    editor = "Brown, L. M.",
    title = "{Very high-energy collisions of hadrons}",
    reportNumber = "PRINT-69-2817",
    doi = "10.1103/PhysRevLett.23.1415",
    journal = "Phys. Rev. Lett.",
    volume = "23",
    pages = "1415--1417",
    year = "1969"
}

@article{Bjorken:1969ja,
    author = "Bjorken, J. D. and Paschos, Emmanuel A.",
    title = "{Inelastic Electron Proton and gamma Proton Scattering, and the Structure of the Nucleon}",
    reportNumber = "SLAC-PUB-0572",
    doi = "10.1103/PhysRev.185.1975",
    journal = "Phys. Rev.",
    volume = "185",
    pages = "1975--1982",
    year = "1969"
}

@article{Collins:1981uw,
    author = "Collins, John C. and Soper, Davison E.",
    title = "{Parton Distribution and Decay Functions}",
    reportNumber = "OITS-166",
    doi = "10.1016/0550-3213(82)90021-9",
    journal = "Nucl. Phys. B",
    volume = "194",
    pages = "445--492",
    year = "1982"
}

@article{Collins:1989gx,
    author = "Collins, John C. and Soper, Davison E. and Sterman, George F.",
    title = "{Factorization of Hard Processes in QCD}",
    eprint = "hep-ph/0409313",
    archivePrefix = "arXiv",
    reportNumber = "ITP-SB-89-31",
    doi = "10.1142/9789814503266_0001",
    journal = "Adv. Ser. Direct. High Energy Phys.",
    volume = "5",
    pages = "1--91",
    year = "1989"
}

@book{Collins:2011zzd,
    author = "Collins, John",
    title = "{Foundations of Perturbative QCD}",
    doi = "10.1017/9781009401845",
    isbn = "978-1-009-40184-5, 978-1-009-40183-8, 978-1-009-40182-1",
    publisher = "Cambridge University Press",
    volume = "32",
    year = "2011"
}

@article{Gribov:1972ri,
    author = "Gribov, V. N. and Lipatov, L. N.",
    title = "{Deep inelastic e p scattering in perturbation theory}",
    reportNumber = "IPTI-381-71",
    journal = "Sov. J. Nucl. Phys.",
    volume = "15",
    pages = "438--450",
    year = "1972"
}

@article{Altarelli:1977zs,
    author = "Altarelli, Guido and Parisi, G.",
    title = "{Asymptotic Freedom in Parton Language}",
    reportNumber = "LPTENS-77-6",
    doi = "10.1016/0550-3213(77)90384-4",
    journal = "Nucl. Phys. B",
    volume = "126",
    pages = "298--318",
    year = "1977"
}

@article{Dokshitzer:1977sg,
    author = "Dokshitzer, Yuri L.",
    title = "{Calculation of the Structure Functions for Deep Inelastic Scattering and e+ e- Annihilation by Perturbation Theory in Quantum Chromodynamics}",
    journal = "Sov. Phys. JETP",
    volume = "46",
    pages = "641--653",
    year = "1977"
}

@article{PDF4LHCWorkingGroup:2022cjn,
    author = "Ball, Richard D. and others",
    collaboration = "PDF4LHC Working Group",
    title = "{The PDF4LHC21 combination of global PDF fits for the LHC Run III}",
    eprint = "2203.05506",
    archivePrefix = "arXiv",
    primaryClass = "hep-ph",
    reportNumber = "Edinburgh 2021/31, FERMILAB-PUB-22-121-QIS-SCD-T, MSUHEP-22-010, SMU-HEP-22-01, Nikhef 2021-033",
    doi = "10.1088/1361-6471/ac7216",
    journal = "J. Phys. G",
    volume = "49",
    number = "8",
    pages = "080501",
    year = "2022"
}

@article{NNPDF:2021njg,
    author = "Ball, Richard D. and others",
    collaboration = "NNPDF",
    title = "{The path to proton structure at 1{\%} accuracy}",
    eprint = "2109.02653",
    archivePrefix = "arXiv",
    primaryClass = "hep-ph",
    reportNumber = "Edinburgh 2021/12, Nikhef-2021-013, TIF-UNIMI-2021-11",
    doi = "10.1140/epjc/s10052-022-10328-7",
    journal = "Eur. Phys. J. C",
    volume = "82",
    number = "5",
    pages = "428",
    year = "2022"
}

@article{Hou:2019efy,
    author = "Hou, Tie-Jiun and others",
    title = "{New CTEQ global analysis of quantum chromodynamics with high-precision data from the LHC}",
    eprint = "1912.10053",
    archivePrefix = "arXiv",
    primaryClass = "hep-ph",
    reportNumber = "MSUHEP-19-025, PITT-PACC-1911, SMU-HEP-19-03",
    doi = "10.1103/PhysRevD.103.014013",
    journal = "Phys. Rev. D",
    volume = "103",
    number = "1",
    pages = "014013",
    year = "2021"
}

@article{Ioffe:1969kf,
    author = "Ioffe, B. L.",
    title = "{Space-time picture of photon and neutrino scattering and electroproduction cross-section asymptotics}",
    doi = "10.1016/0370-2693(69)90415-8",
    journal = "Phys. Lett. B",
    volume = "30",
    pages = "123--125",
    year = "1969"
}

@article{Braun:1994jq,
    author = "Braun, V. and Gornicki, P. and Mankiewicz, L.",
    title = "{Ioffe - time distributions instead of parton momentum distributions in description of deep inelastic scattering}",
    eprint = "hep-ph/9410318",
    archivePrefix = "arXiv",
    reportNumber = "DESY-94-181",
    doi = "10.1103/PhysRevD.51.6036",
    journal = "Phys. Rev. D",
    volume = "51",
    pages = "6036--6051",
    year = "1995"
}

@article{Ji:2013dva,
    author = "Ji, Xiangdong",
    title = "{Parton Physics on a Euclidean Lattice}",
    eprint = "1305.1539",
    archivePrefix = "arXiv",
    primaryClass = "hep-ph",
    doi = "10.1103/PhysRevLett.110.262002",
    journal = "Phys. Rev. Lett.",
    volume = "110",
    pages = "262002",
    year = "2013"
}

@article{Ji:2014gla,
    author = "Ji, Xiangdong",
    title = "{Parton Physics from Large-Momentum Effective Field Theory}",
    eprint = "1404.6680",
    archivePrefix = "arXiv",
    primaryClass = "hep-ph",
    doi = "10.1007/s11433-014-5492-3",
    journal = "Sci. China Phys. Mech. Astron.",
    volume = "57",
    pages = "1407--1412",
    year = "2014"
}

@article{Ji:2020ect,
    author = "Ji, Xiangdong and Liu, Yu-Sheng and Liu, Yizhuang and Zhang, Jian-Hui and Zhao, Yong",
    title = "{Large-momentum effective theory}",
    eprint = "2004.03543",
    archivePrefix = "arXiv",
    primaryClass = "hep-ph",
    doi = "10.1103/RevModPhys.93.035005",
    journal = "Rev. Mod. Phys.",
    volume = "93",
    number = "3",
    pages = "035005",
    year = "2021"
}

@article{Radyushkin:2017cyf,
    author = "Radyushkin, A. V.",
    title = "{Quasi-parton distribution functions, momentum distributions, and pseudo-parton distribution functions}",
    eprint = "1705.01488",
    archivePrefix = "arXiv",
    primaryClass = "hep-ph",
    reportNumber = "JLAB-THY-17-2455",
    doi = "10.1103/PhysRevD.96.034025",
    journal = "Phys. Rev. D",
    volume = "96",
    number = "3",
    pages = "034025",
    year = "2017"
}

@article{Radyushkin:2019mye,
    author = "Radyushkin, A. V.",
    title = "{Theory and applications of parton pseudodistributions}",
    eprint = "1912.04244",
    archivePrefix = "arXiv",
    primaryClass = "hep-ph",
    reportNumber = "JLAB-THY-19-3118",
    doi = "10.1142/S0217751X20300021",
    journal = "Int. J. Mod. Phys. A",
    volume = "35",
    number = "05",
    pages = "2030002",
    year = "2020"
}

@article{Orginos:2017kos,
    author = "Orginos, Kostas and Radyushkin, Anatoly and Karpie, Joseph and Zafeiropoulos, Savvas",
    title = "{Lattice QCD exploration of parton pseudo-distribution functions}",
    eprint = "1706.05373",
    archivePrefix = "arXiv",
    primaryClass = "hep-ph",
    reportNumber = "JLAB-THY-17-2494",
    doi = "10.1103/PhysRevD.96.094503",
    journal = "Phys. Rev. D",
    volume = "96",
    number = "9",
    pages = "094503",
    year = "2017"
}

@article{Joo:2019jct,
    author = "Jo{\'o}, B{\'a}lint and Karpie, Joseph and Orginos, Kostas and Radyushkin, Anatoly and Richards, David and Zafeiropoulos, Savvas",
    title = "{Parton Distribution Functions from Ioffe time pseudo-distributions}",
    eprint = "1908.09771",
    archivePrefix = "arXiv",
    primaryClass = "hep-lat",
    doi = "10.1007/JHEP12(2019)081",
    journal = "JHEP",
    volume = "12",
    pages = "081",
    year = "2019"
}

@article{Joo:2019bzr,
    author = "Jo{\'o}, B{\'a}lint and Karpie, Joseph and Orginos, Kostas and Radyushkin, Anatoly V. and Richards, David G. and Sufian, Raza Sabbir and Zafeiropoulos, Savvas",
    title = "{Pion valence structure from Ioffe-time parton pseudodistribution functions}",
    eprint = "1909.08517",
    archivePrefix = "arXiv",
    primaryClass = "hep-lat",
    reportNumber = "JLAB-THY-19-3038",
    doi = "10.1103/PhysRevD.100.114512",
    journal = "Phys. Rev. D",
    volume = "100",
    number = "11",
    pages = "114512",
    year = "2019"
}

@article{Karpie:2021pap,
    author = "Karpie, Joseph and Orginos, Kostas and Radyushkin, Anatoly and Zafeiropoulos, Savvas",
    collaboration = "HadStruc",
    title = "{The continuum and leading twist limits of parton distribution functions in lattice QCD}",
    eprint = "2105.13313",
    archivePrefix = "arXiv",
    primaryClass = "hep-lat",
    reportNumber = "JLAB-THY-21-3409",
    doi = "10.1007/JHEP11(2021)024",
    journal = "JHEP",
    volume = "11",
    pages = "024",
    year = "2021"
}

@article{Constantinou:2020pek,
    author = "Constantinou, Martha",
    title = "{The x-dependence of hadronic parton distributions: A review on the progress of lattice QCD}",
    eprint = "2010.02445",
    archivePrefix = "arXiv",
    primaryClass = "hep-lat",
    doi = "10.1140/epja/s10050-021-00353-7",
    journal = "Eur. Phys. J. A",
    volume = "57",
    number = "2",
    pages = "77",
    year = "2021"
}

@article{Cichy:2018mum,
    author = "Cichy, Krzysztof and Constantinou, Martha",
    title = "{A guide to light-cone PDFs from Lattice QCD: an overview of approaches, techniques and results}",
    eprint = "1811.07248",
    archivePrefix = "arXiv",
    primaryClass = "hep-lat",
    doi = "10.1155/2019/3036904",
    journal = "Adv. High Energy Phys.",
    volume = "2019",
    pages = "3036904",
    year = "2019"
}

@article{Lin:2017snn,
    author = "Lin, Huey-Wen and others",
    title = "{Parton distributions and lattice QCD calculations: a community white paper}",
    eprint = "1711.07916",
    archivePrefix = "arXiv",
    primaryClass = "hep-ph",
    reportNumber = "DESY-17-185, IFJPAN-IV-2017-19, INT-PUB-17-042, JLAB-THY-17-2604, MSUHEP-17-017, OUTP-17-15P, SMU-HEP-17-08, NIKHEF-2017-047",
    doi = "10.1016/j.ppnp.2018.01.007",
    journal = "Prog. Part. Nucl. Phys.",
    volume = "100",
    pages = "107--160",
    year = "2018"
}

@article{Constantinou:2020hdm,
    author = "Constantinou, Martha and others",
    title = "{Parton distributions and lattice-QCD calculations: Toward 3D structure}",
    eprint = "2006.08636",
    archivePrefix = "arXiv",
    primaryClass = "hep-ph",
    reportNumber = "Nikhef 2020-018, MIT-CTP/5213, MSUHEP-20-012, IFJPAN-IV-2020-3, SMU-HEP-20-03, DESY-20-109, FERMILAB-PUB-20-787-SCD",
    doi = "10.1016/j.ppnp.2021.103908",
    journal = "Prog. Part. Nucl. Phys.",
    volume = "121",
    pages = "103908",
    year = "2021"
}

@article{Ma:2014jla,
    author = "Ma, Yan-Qing and Qiu, Jian-Wei",
    title = "{Extracting Parton Distribution Functions from Lattice QCD Calculations}",
    eprint = "1404.6860",
    archivePrefix = "arXiv",
    primaryClass = "hep-ph",
    doi = "10.1103/PhysRevD.98.074021",
    journal = "Phys. Rev. D",
    volume = "98",
    number = "7",
    pages = "074021",
    year = "2018"
}

@article{Detmold:2005gg,
    author = "Detmold, William and Lin, C. J. David",
    title = "{Deep-inelastic scattering and the operator product expansion in lattice QCD}",
    eprint = "hep-lat/0507007",
    archivePrefix = "arXiv",
    reportNumber = "INT-PUB-05-16, UW-PT-05-17, NT-UW-05-05",
    doi = "10.1103/PhysRevD.73.014501",
    journal = "Phys. Rev. D",
    volume = "73",
    pages = "014501",
    year = "2006"
}

@article{Liu:1993cv,
    author = "Liu, Keh-Fei and Dong, Shao-Jing",
    title = "{Origin of difference between anti-d and anti-u partons in the nucleon}",
    eprint = "hep-ph/9306299",
    archivePrefix = "arXiv",
    reportNumber = "UK-93-02",
    doi = "10.1103/PhysRevLett.72.1790",
    journal = "Phys. Rev. Lett.",
    volume = "72",
    pages = "1790--1793",
    year = "1994"
}

@article{Liu:1999ak,
    author = "Liu, Keh-Fei",
    title = "{Parton degrees of freedom from the path integral formalism}",
    eprint = "hep-ph/9910306",
    archivePrefix = "arXiv",
    reportNumber = "UK-99-09",
    doi = "10.1103/PhysRevD.62.074501",
    journal = "Phys. Rev. D",
    volume = "62",
    pages = "074501",
    year = "2000"
}

@article{Karpie:2018zaz,
    author = "Karpie, Joseph and Orginos, Kostas and Zafeiropoulos, Savvas",
    title = "{Moments of Ioffe time parton distribution functions from non-local matrix elements}",
    eprint = "1807.10933",
    archivePrefix = "arXiv",
    primaryClass = "hep-lat",
    doi = "10.1007/JHEP11(2018)178",
    journal = "JHEP",
    volume = "11",
    pages = "178",
    year = "2018"
}

@article{Karpie:2019eiq,
    author = "Karpie, Joseph and Orginos, Kostas and Rothkopf, Alexander and Zafeiropoulos, Savvas",
    title = "{Reconstructing parton distribution functions from Ioffe time data: from Bayesian methods to Neural Networks}",
    eprint = "1901.05408",
    archivePrefix = "arXiv",
    primaryClass = "hep-lat",
    reportNumber = "JLAB-THY-19-2898",
    doi = "10.1007/JHEP04(2019)057",
    journal = "JHEP",
    volume = "04",
    pages = "057",
    year = "2019"
}

@article{Grieninger:2024cdl,
    author = "Grieninger, Sebastian and Ikeda, Kazuki and Zahed, Ismail",
    title = "{Quasiparton distributions in massive QED2: Toward quantum computation}",
    eprint = "2404.05112",
    archivePrefix = "arXiv",
    primaryClass = "hep-ph",
    doi = "10.1103/PhysRevD.110.076008",
    journal = "Phys. Rev. D",
    volume = "110",
    number = "7",
    pages = "076008",
    year = "2024"
}

@article{Grieninger:2024axp,
    author = "Grieninger, Sebastian and Zahed, Ismail",
    title = "{Quasifragmentation functions in the massive Schwinger model}",
    eprint = "2406.01891",
    archivePrefix = "arXiv",
    primaryClass = "hep-ph",
    doi = "10.1103/PhysRevD.110.116009",
    journal = "Phys. Rev. D",
    volume = "110",
    number = "11",
    pages = "116009",
    year = "2024"
}

@article{Grieninger:2025mbm,
    author = "Grieninger, Sebastian and Montgomery, Jake and Ringer, Felix and Zahed, Ismail",
    title = "{Tensor network simulations of quasi-GPDs in the massive Schwinger model}",
    eprint = "2511.17752",
    archivePrefix = "arXiv",
    primaryClass = "hep-lat",
    reportNumber = "IQuS@UW-21-115, NT@UW-25-17",
    doi = "10.1103/sz3z-8cyx",
    journal = "Phys. Rev. D",
    volume = "113",
    number = "7",
    pages = "074509",
    year = "2026"
}

@article{Banuls:2025wiq,
    author = "Ba{\~n}uls, Mari Carmen and Cichy, Krzysztof and Lin, C. -J. David and Schneider, Manuel",
    title = "{Parton distribution functions in the Schwinger model from tensor network states}",
    eprint = "2504.07508",
    archivePrefix = "arXiv",
    primaryClass = "hep-lat",
    doi = "10.1103/lcyh-rxf3",
    journal = "Phys. Rev. D",
    volume = "113",
    number = "1",
    pages = "L011502",
    year = "2026"
}

@article{Schneider:2024yub,
    author = "Schneider, Manuel and Ba{\~n}uls, Mari Carmen and Cichy, Krzysztof and Lin, C. -J. David",
    title = "{Parton Distribution Functions in the Schwinger Model with Tensor Networks}",
    eprint = "2409.16996",
    archivePrefix = "arXiv",
    primaryClass = "hep-lat",
    doi = "10.22323/1.466.0024",
    journal = "PoS",
    volume = "LATTICE2024",
    pages = "024",
    year = "2025"
}

@article{Lamm:2019uyc,
    author = "Lamm, Henry and Lawrence, Scott and Yamauchi, Yukari",
    title = "{Parton physics on a quantum computer}",
    eprint = "1908.10439",
    archivePrefix = "arXiv",
    primaryClass = "hep-lat",
    doi = "10.1103/PhysRevResearch.2.013272",
    journal = "Phys. Rev. Res.",
    volume = "2",
    number = "1",
    pages = "013272",
    year = "2020"
}

@article{Mueller:2019qqj,
    author = "Mueller, Niklas and Tarasov, Andrey and Venugopalan, Raju",
    title = "{Deeply inelastic scattering structure functions on a hybrid quantum computer}",
    eprint = "1908.07051",
    archivePrefix = "arXiv",
    primaryClass = "hep-th",
    doi = "10.1103/PhysRevD.102.016007",
    journal = "Phys. Rev. D",
    volume = "102",
    number = "1",
    pages = "016007",
    year = "2020"
}

@article{Kreshchuk:2020dla,
    author = "Kreshchuk, Michael and Kirby, William M. and Goldstein, Gary and Beauchemin, Hugo and Love, Peter J.",
    title = "{Quantum simulation of quantum field theory in the light-front formulation}",
    eprint = "2002.04016",
    archivePrefix = "arXiv",
    primaryClass = "quant-ph",
    doi = "10.1103/PhysRevA.105.032418",
    journal = "Phys. Rev. A",
    volume = "105",
    number = "3",
    pages = "032418",
    year = "2022"
}

@article{Echevarria:2020wct,
    author = "Echevarria, M. G. and Egusquiza, I. L. and Rico, E. and Schnell, G.",
    title = "{Quantum simulation of light-front parton correlators}",
    eprint = "2011.01275",
    archivePrefix = "arXiv",
    primaryClass = "quant-ph",
    doi = "10.1103/PhysRevD.104.014512",
    journal = "Phys. Rev. D",
    volume = "104",
    number = "1",
    pages = "014512",
    year = "2021"
}

@article{Li:2021kcs,
    author = "Li, Tianyin and Guo, Xingyu and Lai, Wai Kin and Liu, Xiaohui and Wang, Enke and Xing, Hongxi and Zhang, Dan-Bo and Zhu, Shi-Liang",
    collaboration = "QuNu",
    title = "{Partonic collinear structure by quantum computing}",
    eprint = "2106.03865",
    archivePrefix = "arXiv",
    primaryClass = "hep-ph",
    doi = "10.1103/PhysRevD.105.L111502",
    journal = "Phys. Rev. D",
    volume = "105",
    number = "11",
    pages = "L111502",
    year = "2022"
}

@article{Chen:2025zeh,
    author = "Chen, Jiunn-Wei and Chen, Yu-Ting and Meher, Ghanashyam",
    title = "{Parton Distributions on a Quantum Computer}",
    eprint = "2506.16829",
    archivePrefix = "arXiv",
    primaryClass = "hep-lat",
    month = "6",
    year = "2025"
}

@article{Ikeda:2025bjb,
    author = "Ikeda, Kazuki and Kang, Zhong-Bo and Kharzeev, Dmitri E. and Qian, Wenyang",
    title = "{Quantum simulation of real-time current correlators and DIS-inspired observables in the Schwinger model}",
    eprint = "2512.18062",
    archivePrefix = "arXiv",
    primaryClass = "hep-ph",
    doi = "10.1007/JHEP07(2026)242",
    journal = "JHEP",
    volume = "07",
    pages = "242",
    year = "2026"
}

@article{Jia:2024atq,
    author = "Jia, Yu and Mo, Zhewen and Xiong, Xiaonu and Yu, Rui",
    title = "{Light-cone and quasigeneralized parton distributions in the {\textquoteright}t Hooft model}",
    eprint = "2401.12786",
    archivePrefix = "arXiv",
    primaryClass = "hep-ph",
    doi = "10.1103/PhysRevD.109.014042",
    journal = "Phys. Rev. D",
    volume = "109",
    number = "1",
    pages = "014042",
    year = "2024"
}

@article{Burkardt:2000uu,
    author = "Burkardt, Matthias",
    title = "{Off forward parton distributions in (1+1)-dimensional QCD}",
    eprint = "hep-ph/0005209",
    archivePrefix = "arXiv",
    doi = "10.1103/PhysRevD.62.094003",
    journal = "Phys. Rev. D",
    volume = "62",
    pages = "094003",
    year = "2000"
}

@article{tHooft:1974pnl,
    author = "'t Hooft, Gerard",
    title = "{A Two-Dimensional Model for Mesons}",
    reportNumber = "CERN-TH-1820",
    doi = "10.1016/0550-3213(74)90088-1",
    journal = "Nucl. Phys. B",
    volume = "75",
    pages = "461--470",
    year = "1974"
}

@article{Bergknoff:1976xr,
    author = "Bergknoff, Hugh",
    title = "{Physical Particles of the Massive Schwinger Model}",
    reportNumber = "CLNS-341",
    doi = "10.1016/0550-3213(77)90204-8",
    journal = "Nucl. Phys. B",
    volume = "122",
    pages = "215--236",
    year = "1977"
}

@article{Eller:1986nt,
    author = "Eller, Thomas and Pauli, Hans Christian and Brodsky, Stanley J.",
    title = "{Discretized Light Cone Quantization: The Massless and the Massive Schwinger Model}",
    reportNumber = "SLAC-PUB-3835",
    doi = "10.1103/PhysRevD.35.1493",
    journal = "Phys. Rev. D",
    volume = "35",
    pages = "1493",
    year = "1987"
}

@article{Pauli:1985pv,
    author = "Pauli, Hans Christian and Brodsky, Stanley J.",
    title = "{Solving Field Theory in One Space One Time Dimension}",
    reportNumber = "SLAC-PUB-3701",
    doi = "10.1103/PhysRevD.32.1993",
    journal = "Phys. Rev. D",
    volume = "32",
    pages = "1993",
    year = "1985"
}

@article{Pauli:1985ps,
    author = "Pauli, Hans Christian and Brodsky, Stanley J.",
    title = "{Discretized Light Cone Quantization: Solution to a Field Theory in One Space One Time Dimensions}",
    reportNumber = "SLAC-PUB-3714",
    doi = "10.1103/PhysRevD.32.2001",
    journal = "Phys. Rev. D",
    volume = "32",
    pages = "2001",
    year = "1985"
}

@article{Brodsky:1997de,
    author = "Brodsky, Stanley J. and Pauli, Hans-Christian and Pinsky, Stephen S.",
    title = "{Quantum chromodynamics and other field theories on the light cone}",
    eprint = "hep-ph/9705477",
    archivePrefix = "arXiv",
    reportNumber = "SLAC-PUB-7484, MPIH-V1-1997",
    doi = "10.1016/S0370-1573(97)00089-6",
    journal = "Phys. Rept.",
    volume = "301",
    pages = "299--486",
    year = "1998"
}

@article{Mo:1992sv,
    author = "Mo, Yi-zhang and Perry, Robert J.",
    title = "{Basis function calculations for the massive Schwinger model in the light front Tamm-Dancoff approximation}",
    reportNumber = "OSU-NT-92-128",
    doi = "10.1006/jcph.1993.1171",
    journal = "J. Comput. Phys.",
    volume = "108",
    pages = "159--174",
    year = "1993"
}

@article{Harada:1995au,
    author = "Harada, Koji and Okazaki, Atsushi and Taniguchi, Masa-aki",
    title = "{Mesons in the massive Schwinger model on the light cone}",
    eprint = "hep-th/9509136",
    archivePrefix = "arXiv",
    reportNumber = "KYUSHU-HET-28",
    doi = "10.1103/PhysRevD.54.7656",
    journal = "Phys. Rev. D",
    volume = "54",
    pages = "7656--7663",
    year = "1996"
}

@article{Ji:2020bby,
    author = "Ji, Xiangdong and Liu, Yizhuang and Zahed, Ismail",
    title = "{Mass structure of hadrons and light-front sum rules in the $'t$ Hooft model}",
    eprint = "2010.06665",
    archivePrefix = "arXiv",
    primaryClass = "hep-ph",
    doi = "10.1103/PhysRevD.103.074002",
    journal = "Phys. Rev. D",
    volume = "103",
    number = "7",
    pages = "074002",
    year = "2021"
}

@article{Kobzarev:1962wt,
    author = "Kobzarev, I. Yu. and Okun', L. B.",
    title = "{Gravitational Interaction of Fermions}",
    journal = "Zh. Eksp. Teor. Fiz.",
    volume = "43",
    pages = "1904--1909",
    year = "1962"
}

@article{Pagels:1966zza,
    author = "Pagels, Heinz",
    title = "{Energy-Momentum Structure Form Factors of Particles}",
    doi = "10.1103/PhysRev.144.1250",
    journal = "Phys. Rev.",
    volume = "144",
    pages = "1250--1260",
    year = "1966"
}

@article{Polyakov:2002yz,
    author = "Polyakov, M. V.",
    title = "{Generalized parton distributions and strong forces inside nucleons and nuclei}",
    eprint = "hep-ph/0210165",
    archivePrefix = "arXiv",
    reportNumber = "RUB-TP2-14-02",
    doi = "10.1016/S0370-2693(03)00036-4",
    journal = "Phys. Lett. B",
    volume = "555",
    pages = "57--62",
    year = "2003"
}

@article{Polyakov:2018zvc,
    author = "Polyakov, Maxim V. and Schweitzer, Peter",
    title = "{Forces inside hadrons: pressure, surface tension, mechanical radius, and all that}",
    eprint = "1805.06596",
    archivePrefix = "arXiv",
    primaryClass = "hep-ph",
    doi = "10.1142/S0217751X18300259",
    journal = "Int. J. Mod. Phys. A",
    volume = "33",
    number = "26",
    pages = "1830025",
    year = "2018"
}

@article{Burkert:2018bqq,
    author = "Burkert, V. D. and Elouadrhiri, L. and Girod, F. X.",
    title = "{The pressure distribution inside the proton}",
    doi = "10.1038/s41586-018-0060-z",
    journal = "Nature",
    volume = "557",
    number = "7705",
    pages = "396--399",
    year = "2018"
}

@article{Hackett:2023rif,
    author = "Hackett, Daniel C. and Pefkou, Dimitra A. and Shanahan, Phiala E.",
    title = "{Gravitational Form Factors of the Proton from Lattice QCD}",
    eprint = "2310.08484",
    archivePrefix = "arXiv",
    primaryClass = "hep-lat",
    reportNumber = "MIT-CTP/5630, FERMILAB-PUB-23-592-T",
    doi = "10.1103/PhysRevLett.132.251904",
    journal = "Phys. Rev. Lett.",
    volume = "132",
    number = "25",
    pages = "251904",
    year = "2024"
}

@article{Hackett:2023nkr,
    author = "Hackett, Daniel C. and Oare, Patrick R. and Pefkou, Dimitra A. and Shanahan, Phiala E.",
    title = "{Gravitational form factors of the pion from lattice QCD}",
    eprint = "2307.11707",
    archivePrefix = "arXiv",
    primaryClass = "hep-lat",
    reportNumber = "MIT-CTP/5585",
    doi = "10.1103/PhysRevD.108.114504",
    journal = "Phys. Rev. D",
    volume = "108",
    number = "11",
    pages = "114504",
    year = "2023"
}

@article{Duran:2022xag,
    author = "Duran, B. and others",
    title = "{Determining the gluonic gravitational form factors of the proton}",
    eprint = "2207.05212",
    archivePrefix = "arXiv",
    primaryClass = "nucl-ex",
    doi = "10.1038/s41586-023-05730-4",
    journal = "Nature",
    volume = "615",
    number = "7954",
    pages = "813--816",
    year = "2023"
}

@article{Burkardt:2000za,
    author = "Burkardt, Matthias",
    title = "{Impact parameter dependent parton distributions and off forward parton distributions for zeta ---{\ensuremath{>}} 0}",
    eprint = "hep-ph/0005108",
    archivePrefix = "arXiv",
    doi = "10.1103/PhysRevD.62.071503",
    journal = "Phys. Rev. D",
    volume = "62",
    pages = "071503",
    year = "2000",
    note = "[Erratum: Phys.Rev.D 66, 119903 (2002)]"
}

@article{Callan:1970ze,
    author = "Callan, Jr., Curtis G. and Coleman, Sidney R. and Jackiw, Roman",
    title = "{A New improved energy - momentum tensor}",
    doi = "10.1016/0003-4916(70)90394-5",
    journal = "Annals Phys.",
    volume = "59",
    pages = "42--73",
    year = "1970"
}

@article{Freese:2021mzg,
    author = "Freese, Adam and Miller, Gerald A.",
    title = "{Unified formalism for electromagnetic and gravitational probes: Densities}",
    eprint = "2108.03301",
    archivePrefix = "arXiv",
    primaryClass = "hep-ph",
    reportNumber = "NT@UW-21-09",
    doi = "10.1103/PhysRevD.105.014003",
    journal = "Phys. Rev. D",
    volume = "105",
    number = "1",
    pages = "014003",
    year = "2022"
}

@article{Lorce:2018egm,
    author = "Lorc{\'e}, C{\'e}dric and Moutarde, Herv{\'e} and Trawi{\'n}ski, Arkadiusz P.",
    title = "{Revisiting the mechanical properties of the nucleon}",
    eprint = "1810.09837",
    archivePrefix = "arXiv",
    primaryClass = "hep-ph",
    doi = "10.1140/epjc/s10052-019-6572-3",
    journal = "Eur. Phys. J. C",
    volume = "79",
    number = "1",
    pages = "89",
    year = "2019"
}

@article{Burkert:2023wzr,
    author = "Burkert, V. D. and Elouadrhiri, L. and Girod, F. X. and Lorc{\'e}, C. and Schweitzer, P. and Shanahan, P. E.",
    title = "{Colloquium: Gravitational form factors of the proton}",
    eprint = "2303.08347",
    archivePrefix = "arXiv",
    primaryClass = "hep-ph",
    reportNumber = "JLAB-PHY-23-3774",
    doi = "10.1103/RevModPhys.95.041002",
    journal = "Rev. Mod. Phys.",
    volume = "95",
    number = "4",
    pages = "041002",
    year = "2023"
}

@article{Coleman:1974bu,
    author = "Coleman, Sidney R.",
    editor = "Stone, M.",
    title = "{The Quantum Sine-Gordon Equation as the Massive Thirring Model}",
    reportNumber = "Print-74-1722 (HARVARD)",
    doi = "10.1103/PhysRevD.11.2088",
    journal = "Phys. Rev. D",
    volume = "11",
    pages = "2088",
    year = "1975"
}

@article{Mandelstam:1975hb,
    author = "Mandelstam, S.",
    editor = "Stone, M.",
    title = "{Soliton Operators for the Quantized Sine-Gordon Equation}",
    reportNumber = "Print-75-0198 (UC,BERKELEY)",
    doi = "10.1103/PhysRevD.11.3026",
    journal = "Phys. Rev. D",
    volume = "11",
    pages = "3026",
    year = "1975"
}

@article{Chen:2026sgc,
    author = "Chen, Jiunn-Wei and Chen, Yu-Ting and Meher, Ghanashyam",
    title = "{Parton distribution function in 1+1D Schwinger Model using Quantum Computing}",
    doi = "10.22323/1.518.0135",
    journal = "PoS",
    volume = "LATTICE2025",
    pages = "135",
    year = "2026"
}

@misc{MPSKit,
  doi = {10.5281/ZENODO.10654900},
  url = {https://zenodo.org/doi/10.5281/zenodo.10654900},
  author = {Devos, Lukas and Van Damme, Maarten and Haegeman, Jutho},
  title = {MPSKit},
  publisher = {Zenodo},
  year = {2026},
  copyright = {MIT License}
}

@article{Yurov:1990kv,
    author = "Yurov, V. P. and Zamolodchikov, A. B.",
    title = "{Correlation functions of integrable 2-D models of relativistic field theory. Ising model}",
    reportNumber = "SISSA-36-90-EP-A",
    doi = "10.1142/S0217751X91001660",
    journal = "Int. J. Mod. Phys. A",
    volume = "6",
    pages = "3419--3440",
    year = "1991"
}

@article{Shanahan:2018pib,
    author = "Shanahan, P. E. and Detmold, W.",
    title = "{Gluon gravitational form factors of the nucleon and the pion from lattice QCD}",
    eprint = "1810.04626",
    archivePrefix = "arXiv",
    primaryClass = "hep-lat",
    reportNumber = "MIT-CTP/5069",
    doi = "10.1103/PhysRevD.99.014511",
    journal = "Phys. Rev. D",
    volume = "99",
    number = "1",
    pages = "014511",
    year = "2019"
}

@article{Shanahan:2018nnv,
    author = "Shanahan, P. E. and Detmold, W.",
    title = "{Pressure Distribution and Shear Forces inside the Proton}",
    eprint = "1810.07589",
    archivePrefix = "arXiv",
    primaryClass = "nucl-th",
    reportNumber = "MIT-CTP/5071",
    doi = "10.1103/PhysRevLett.122.072003",
    journal = "Phys. Rev. Lett.",
    volume = "122",
    number = "7",
    pages = "072003",
    year = "2019"
}

@article{Pefkou:2021fni,
    author = "Pefkou, Dimitra A. and Hackett, Daniel C. and Shanahan, Phiala E.",
    title = "{Gluon gravitational structure of hadrons of different spin}",
    eprint = "2107.10368",
    archivePrefix = "arXiv",
    primaryClass = "hep-lat",
    reportNumber = "MIT-CTP/5318",
    doi = "10.1103/PhysRevD.105.054509",
    journal = "Phys. Rev. D",
    volume = "105",
    number = "5",
    pages = "054509",
    year = "2022"
}

@article{Abbott:2025irb,
    author = "Abbott, Ryan and Hackett, Daniel C. and Pefkou, Dimitra A. and Romero-L{\'o}pez, Fernando and Shanahan, Phiala E.",
    title = "{Lattice Evidence That Scalar Glueballs are Small}",
    eprint = "2508.21821",
    archivePrefix = "arXiv",
    primaryClass = "hep-lat",
    reportNumber = "MIT-CTP/5907, FERMILAB-PUB-25-0621-T, INT-PUB-25-021",
    doi = "10.1103/67xg-qxhz",
    journal = "Phys. Rev. Lett.",
    volume = "136",
    number = "4",
    pages = "041901",
    year = "2026"
}

@article{Grebe:2026qmt,
    author = "Grebe, Anthony V. and Hackett, Daniel C. and Wagman, Michael L. and Zhang, Rui and Zhao, Yong",
    title = "{Direct calculation of parton distributions in momentum space from lattice QCD}",
    eprint = "2606.16877",
    archivePrefix = "arXiv",
    primaryClass = "hep-lat",
    reportNumber = "UMD-PP-026-05, FERMILAB-PUB-26-0392-T, MIT-CTP/6043",
    month = "6",
    year = "2026"
}

@article{Shanahan:2020zxr,
    author = "Shanahan, Phiala and Wagman, Michael and Zhao, Yong",
    title = "{Collins-Soper kernel for TMD evolution from lattice QCD}",
    eprint = "2003.06063",
    archivePrefix = "arXiv",
    primaryClass = "hep-lat",
    reportNumber = "FERMILAB-PUB-20-102-T, MIT/CTP-5184",
    doi = "10.1103/PhysRevD.102.014511",
    journal = "Phys. Rev. D",
    volume = "102",
    number = "1",
    pages = "014511",
    year = "2020"
}

@article{Avkhadiev:2024mgd,
    author = "Avkhadiev, Artur and Shanahan, Phiala E. and Wagman, Michael L. and Zhao, Yong",
    title = "{Determination of the Collins-Soper Kernel from Lattice QCD}",
    eprint = "2402.06725",
    archivePrefix = "arXiv",
    primaryClass = "hep-lat",
    reportNumber = "FERMILAB-PUB-24-0037-T, MIT-CTP/5677",
    doi = "10.1103/PhysRevLett.132.231901",
    journal = "Phys. Rev. Lett.",
    volume = "132",
    number = "23",
    pages = "231901",
    year = "2024"
}

@article{Detmold:2021uru,
    author = "Detmold, William and Grebe, Anthony V. and Kanamori, Issaku and Lin, C. -J. David and Perry, Robert J. and Zhao, Yong",
    collaboration = "HOPE",
    title = "{Parton physics from a heavy-quark operator product expansion: Formalism and Wilson coefficients}",
    eprint = "2103.09529",
    archivePrefix = "arXiv",
    primaryClass = "hep-lat",
    reportNumber = "MIT-CTP/5294",
    doi = "10.1103/PhysRevD.104.074511",
    journal = "Phys. Rev. D",
    volume = "104",
    number = "7",
    pages = "074511",
    year = "2021"
}

@article{Detmold:2003tm,
    author = "Detmold, William and Melnitchouk, W. and Thomas, Anthony William",
    title = "{Parton distribution functions in the pion from lattice QCD}",
    eprint = "hep-lat/0303015",
    archivePrefix = "arXiv",
    reportNumber = "NT-UW-03-04, JLAB-THY-03-28, ADP-03-110-T548",
    doi = "10.1103/PhysRevD.68.034025",
    journal = "Phys. Rev. D",
    volume = "68",
    pages = "034025",
    year = "2003"
}

@article{Pufu:2026kqk,
    author = "Pufu, Silviu S. and S{\o}gaard, Benjamin T. and Terlemes, Spyros",
    title = "{The Quasiparticle Ansatz for Relativistic Continuous Matrix Product States}",
    eprint = "2610.13824",
    archivePrefix = "arXiv",
    primaryClass = "hep-th",
    reportNumber = "PUPT-2664",
    month = "10",
    year = "2026"
}

@article{Hagler:2003jd,
    author = "Hagler, Philipp and Negele, John W. and Renner, Dru Bryant and Schroers, W. and Lippert, T. and Schilling, K.",
    collaboration = "LHPC, SESAM",
    title = "{Moments of nucleon generalized parton distributions in lattice QCD}",
    eprint = "hep-lat/0304018",
    archivePrefix = "arXiv",
    reportNumber = "MIT-CPT-3353",
    doi = "10.1103/PhysRevD.68.034505",
    journal = "Phys. Rev. D",
    volume = "68",
    pages = "034505",
    year = "2003"
}

@article{LHPC:2007blg,
    author = "Hagler, Ph. and others",
    collaboration = "LHPC",
    title = "{Nucleon Generalized Parton Distributions from Full Lattice QCD}",
    eprint = "0705.4295",
    archivePrefix = "arXiv",
    primaryClass = "hep-lat",
    reportNumber = "DESY-07-077, JLAB-THY-07-651, TUM-T39-07-09",
    doi = "10.1103/PhysRevD.77.094502",
    journal = "Phys. Rev. D",
    volume = "77",
    pages = "094502",
    year = "2008"
}

@article{Hagler:2009ni,
    author = "Hagler, Ph.",
    title = "{Hadron structure from lattice quantum chromodynamics}",
    eprint = "0912.5483",
    archivePrefix = "arXiv",
    primaryClass = "hep-lat",
    reportNumber = "TUM-T39-09-12",
    doi = "10.1016/j.physrep.2009.12.008",
    journal = "Phys. Rept.",
    volume = "490",
    pages = "49--175",
    year = "2010"
}

@article{Alexandrou:2019ali,
    author = "Alexandrou, C. and others",
    title = "{Moments of nucleon generalized parton distributions from lattice QCD simulations at physical pion mass}",
    eprint = "1908.10706",
    archivePrefix = "arXiv",
    primaryClass = "hep-lat",
    doi = "10.1103/PhysRevD.101.034519",
    journal = "Phys. Rev. D",
    volume = "101",
    number = "3",
    pages = "034519",
    year = "2020"
}

@article{Zhang:2018nsy,
    author = {Zhang, Jian-Hui and Chen, Jiunn-Wei and Jin, Luchang and Lin, Huey-Wen and Sch{\"a}fer, Andreas and Zhao, Yong},
    title = "{First direct lattice-QCD calculation of the $x$-dependence of the pion parton distribution function}",
    eprint = "1804.01483",
    archivePrefix = "arXiv",
    primaryClass = "hep-lat",
    doi = "10.1103/PhysRevD.100.034505",
    journal = "Phys. Rev. D",
    volume = "100",
    number = "3",
    pages = "034505",
    year = "2019"
}

@article{Sufian:2019bol,
    author = "Sufian, Raza Sabbir and Karpie, Joseph and Egerer, Colin and Orginos, Kostas and Qiu, Jian-Wei and Richards, David G.",
    title = "{Pion Valence Quark Distribution from Matrix Element Calculated in Lattice QCD}",
    eprint = "1901.03921",
    archivePrefix = "arXiv",
    primaryClass = "hep-lat",
    reportNumber = "JLAB-THY-19-2847",
    doi = "10.1103/PhysRevD.99.074507",
    journal = "Phys. Rev. D",
    volume = "99",
    number = "7",
    pages = "074507",
    year = "2019"
}

@article{Gao:2020ito,
    author = "Gao, Xiang and Jin, Luchang and Kallidonis, Christos and Karthik, Nikhil and Mukherjee, Swagato and Petreczky, Peter and Shugert, Charles and Syritsyn, Sergey and Zhao, Yong",
    title = "{Valence parton distribution of the pion from lattice QCD: Approaching the continuum limit}",
    eprint = "2007.06590",
    archivePrefix = "arXiv",
    primaryClass = "hep-lat",
    doi = "10.1103/PhysRevD.102.094513",
    journal = "Phys. Rev. D",
    volume = "102",
    number = "9",
    pages = "094513",
    year = "2020"
}

@article{Gao:2022iex,
    author = "Gao, Xiang and Hanlon, Andrew D. and Karthik, Nikhil and Mukherjee, Swagato and Petreczky, Peter and Scior, Philipp and Shi, Shuzhe and Syritsyn, Sergey and Zhao, Yong and Zhou, Kai",
    title = "{Continuum-extrapolated NNLO valence PDF of the pion at the physical point}",
    eprint = "2208.02297",
    archivePrefix = "arXiv",
    primaryClass = "hep-lat",
    doi = "10.1103/PhysRevD.106.114510",
    journal = "Phys. Rev. D",
    volume = "106",
    number = "11",
    pages = "114510",
    year = "2022"
}

@article{Alexandrou:2021mmi,
    author = {Alexandrou, Constantia and Bacchio, Simone and Clo{\"e}t, Ian and Constantinou, Martha and Hadjiyiannakou, Kyriakos and Koutsou, Giannis and Lauer, Colin},
    collaboration = "ETM",
    title = "{Pion and kaon {\ensuremath{\langle}}x3{\ensuremath{\rangle}} from lattice QCD and PDF reconstruction from Mellin moments}",
    eprint = "2104.02247",
    archivePrefix = "arXiv",
    primaryClass = "hep-lat",
    doi = "10.1103/PhysRevD.104.054504",
    journal = "Phys. Rev. D",
    volume = "104",
    number = "5",
    pages = "054504",
    year = "2021"
}

@article{Joo:2020spy,
    author = "Jo{\'o}, B{\'a}lint and Karpie, Joseph and Orginos, Kostas and Radyushkin, Anatoly V. and Richards, David G. and Zafeiropoulos, Savvas",
    title = "{Parton Distribution Functions from Ioffe Time Pseudodistributions from Lattice Calculations: Approaching the Physical Point}",
    eprint = "2004.01687",
    archivePrefix = "arXiv",
    primaryClass = "hep-lat",
    reportNumber = "JLAB-THY-20-3161",
    doi = "10.1103/PhysRevLett.125.232003",
    journal = "Phys. Rev. Lett.",
    volume = "125",
    number = "23",
    pages = "232003",
    year = "2020"
}

@article{Karowski:1978vz,
    author = "Karowski, M. and Weisz, P.",
    title = "{Exact Form-Factors in (1+1)-Dimensional Field Theoretic Models with Soliton Behavior}",
    reportNumber = "FUB-HEP 78/2",
    doi = "10.1016/0550-3213(78)90362-0",
    journal = "Nucl. Phys. B",
    volume = "139",
    pages = "455--476",
    year = "1978"
}

@book{Smirnov:1992vz,
    author = "Smirnov, F. A.",
    title = "{Form-factors in completely integrable models of quantum field theory}",
    volume = "14",
    year = "1992"
}

@article{Delfino:1996nf,
    author = "Delfino, G. and Simonetti, P. and Cardy, John L.",
    title = "{Asymptotic factorization of form-factors in two-dimensional quantum field theory}",
    eprint = "hep-th/9607046",
    archivePrefix = "arXiv",
    reportNumber = "OUTP-96-25-S, SWAT-95-96-123",
    doi = "10.1016/0370-2693(96)01035-0",
    journal = "Phys. Lett. B",
    volume = "387",
    pages = "327--333",
    year = "1996"
}

@article{Belyansky:2023rgh,
    author = "Belyansky, Ron and Whitsitt, Seth and Mueller, Niklas and Fahimniya, Ali and Bennewitz, Elizabeth R. and Davoudi, Zohreh and Gorshkov, Alexey V.",
    title = "{High-Energy Collision of Quarks and Mesons in the Schwinger Model: From Tensor Networks to Circuit QED}",
    eprint = "2307.02522",
    archivePrefix = "arXiv",
    primaryClass = "quant-ph",
    reportNumber = "UMD-PP-023-02, IQuS@UW-21-050",
    doi = "10.1103/PhysRevLett.132.091903",
    journal = "Phys. Rev. Lett.",
    volume = "132",
    number = "9",
    pages = "091903",
    year = "2024"
}

@article{Papaefstathiou:2024zsu,
    author = "Papaefstathiou, Irene and Knolle, Johannes and Ba{\~n}uls, Mari Carmen",
    title = "{Real-time scattering in the lattice Schwinger model}",
    eprint = "2402.18429",
    archivePrefix = "arXiv",
    primaryClass = "hep-lat",
    doi = "10.1103/PhysRevD.111.014504",
    journal = "Phys. Rev. D",
    volume = "111",
    number = "1",
    pages = "014504",
    year = "2025"
}

@article{Zou:2026cfk,
    author = "Zou, Dairui and Li, Tianyin and Liang, Jian and Wang, Enke and Xing, Hongxi",
    collaboration = "QuNu",
    title = "{Hadronic tensor in lattice gauge theories by quantum computing}",
    eprint = "2606.17003",
    archivePrefix = "arXiv",
    primaryClass = "hep-ph",
    reportNumber = "RIKEN-iTHEMS-Report-26",
    doi = "10.1103/sdvz-1ltn",
    journal = "Phys. Rev. D",
    volume = "114",
    number = "7",
    pages = "076001",
    year = "2026"
}

@article{Crosswhite:2008ucd,
    author = "Crosswhite, Gregory M. and Bacon, Dave",
    title = "{Finite automata for caching in matrix product algorithms}",
    eprint = "0708.1221",
    archivePrefix = "arXiv",
    primaryClass = "quant-ph",
    doi = "10.1103/PhysRevA.78.012356",
    journal = "Phys. Rev. A",
    volume = "78",
    number = "1",
    pages = "012356",
    year = "2008"
}

@article{Crosswhite:2008rml,
    author = "Crosswhite, Gregory M. and Doherty, Andrew C. and Vidal, Guifre",
    title = "{Applying matrix product operators to model systems with long-range interactions}",
    eprint = "0804.2504",
    archivePrefix = "arXiv",
    primaryClass = "cond-mat.other",
    doi = "10.1103/PhysRevB.78.035116",
    journal = "Phys. Rev. B",
    volume = "78",
    pages = "035116",
    year = "2008"
}

@article{Zuber:1976aa,
    author = "Zuber, J. B. and Itzykson, C.",
    title = "{Quantum Field Theory and the Two-Dimensional Ising Model}",
    reportNumber = "PRINT-76-0271 (HARVARD), FERMILAB-PUB-75-057-T",
    doi = "10.1103/PhysRevD.15.2875",
    journal = "Phys. Rev. D",
    volume = "15",
    pages = "2875",
    year = "1977"
}

@article{Schultz:1964fv,
    author = "Schultz, T. D. and Mattis, D. C. and Lieb, Elliott H.",
    title = "{Two-dimensional Ising model as a soluble problem of many fermions}",
    doi = "10.1103/RevModPhys.36.856",
    journal = "Rev. Mod. Phys.",
    volume = "36",
    pages = "856--871",
    year = "1964"
}

@article{Rigobello:2021fxw,
    author = "Rigobello, Marco and Notarnicola, Simone and Magnifico, Giuseppe and Montangero, Simone",
    title = "{Entanglement generation in (1+1)D QED scattering processes}",
    eprint = "2105.03445",
    archivePrefix = "arXiv",
    primaryClass = "hep-lat",
    doi = "10.1103/PhysRevD.104.114501",
    journal = "Phys. Rev. D",
    volume = "104",
    number = "11",
    pages = "114501",
    year = "2021"
}

@article{Dempsey:2026yut,
    author = {Dempsey, Ross and Gl{\"u}ck, Anna-Maria E. and Pufu, Silviu S. and S{\o}gaard, Benjamin T.},
    title = "{String breaking in a (1+1)-dimensional $e^+e^-$ collider}",
    eprint = "2611.27134",
    archivePrefix = "arXiv",
    primaryClass = "hep-th",
    reportNumber = "PUPT-2667, MIT-CTP/6160",
    month = "11",
    year = "2026"
}

\end{document}